\documentclass[12pt]{article}
 \pdfoutput=1

\usepackage[margin=1in,footskip=0.25in]{geometry}
\usepackage{amssymb}
\usepackage{amsmath}
\usepackage{graphicx}
\usepackage{subfig}
\usepackage{hyperref}
\usepackage{caption}
\usepackage{color}
\usepackage{setspace}
\usepackage[squaren]{SIunits}
\usepackage{lineno}

\usepackage{soul}
\usepackage{xparse}
\usepackage[normalem]{ulem}

\newcommand{\ra}[1]{#1}

\newcommand{\rda}[2]{#2}
\newcommand{\rd}[1]{}

\title{Common Lines Modeling for Reference Free\\Ab-Initio Reconstruction in Cryo-EM\protect\footnote{\url{https://doi.org/10.1016/j.jsb.2017.09.007}}}

\author{
  Ido Greenberg\textsuperscript{a,b}, Yoel Shkolnisky\textsuperscript{a}\\
  \textsuperscript{a}Department of Applied Mathematics, School of Mathematical Sciences,\\
  Tel-Aviv University, Israel\\
  \textsuperscript{b}Corresponding author. E-mail address: idogreenberg@mail.tau.ac.il.
}

\begin{document}

\maketitle

\begin{abstract}
We consider the problem of estimating an unbiased and reference-free ab-inito model for non-symmetric molecules from images generated by single-particle cryo-electron microscopy. The proposed algorithm finds the globally optimal assignment of orientations that simultaneously respects all common lines between all images.
The contribution of each common line to the estimated orientations is weighted according to a statistical model for common lines' detection errors. The key property of the proposed algorithm is that it finds the global optimum for the orientations given the common lines. In particular, any local optima in the common lines energy landscape do not affect the proposed algorithm. As a result, it is applicable to thousands of images at once, very robust to noise, completely reference free, and not biased towards any initial model. A byproduct of the algorithm is a set of measures that allow to asses the reliability of the obtained ab-initio model. We demonstrate the algorithm using class averages from two experimental data sets, resulting in ab-initio models with resolutions of 20{\AA} or better, even from class averages consisting of as few as three raw images per class.

\bigskip

\end{abstract}


\section{Introduction}
\label{intro}
\label{sec:background}

One of the primary tasks in single particle reconstruction (SPR) using cryo-electron microscopy (cryo-EM) is to determine the three-dimensional density map of a molecule given only its two-dimensional projection-images. Each projection-image corresponds to a tomographic projection of a randomly oriented copy of the molecule, and a typical data set consists of tens of thousands of raw projection-images.

Existing algorithms for recovering high-resolution density maps are based on an iterative process, which gradually refines an initial low resolution density map using a data set of raw projection-images~\cite{Walz2015,Scheres2012}. The refinement process is intended to optimize some objective function, such as the likelihood of the data with relation to the density map.
Robust methods to such optimization became very popular in recent years as a result of both algorithmic and hardware improvements. Among the optimization methods applied in cryo-EM, one can find the expectation maximization~\cite{Scheres2012}, the stochastic hill climbing~\cite{Elmlund2012420,viper}, and the stochastic gradient descent~\cite{cryoSPARC}.
Such methods achieve in many cases near-atomic resolutions~\cite{Subramaniam2016}.

All existing cryo-EM refinement algorithms require an initial density map to initialize the refinement process.
While in certain cases they may converge successfully even from completely random initialization, they are guaranteed to converge to the correct solution only if initialized with a sufficiently accurate map.

In certain cases, an initial density map for the refinement process can be generated using random conical tilt~\cite{Radermacher1}, electron tomography~\cite{subtomo2016}, or using prior knowledge~\cite{SCHERES2016125}.
In other cases, it can be generated algorithmically using common lines approaches~\cite{VanHeel1987,LYUMKIS2013417}, or
using stochastic optimization procedures~\cite{PRIME13,cryoSPARC,viper}. Stochastic optimization approaches have gained popularity recently, with various attempts of making them as insensitive to the initial model as possible.
These algorithms still require some initialization in order to produce an initial model for refinement, yet, in some cases they converge successfully even from a heuristic or random initialization. In such cases they are said to provide an ``ab-initio reconstruction'', in the sense of requiring only projection-images, without an external reference density map. However, since they essentially rely on some internally-chosen initial map, they are susceptible to similar drawbacks as other optimization algorithms. In particular, their outcome may correspond to a local optimum of the underlying optimization, making it biased towards the initial map.
This problem is well known in computational optimization as sensitivity to starting point and convergence to a local optimum. In the context of cryo-EM, this problem is known as model bias and was demonstrated in~\cite{Henderson2013b}, where it is shown that an arbitrary initial density map along with images of pure noise can converge into a refined map similar to the initial one. This also raises the issue of assessing the reliability of a reconstructed map. Such an assessment usually requires either high-resolution reconstruction exhibiting indicative features, or performing a tilt-pair validation. Unfortunately both these methods are not always applicable. Thus, a reliable method for estimating an unbiased initial density map is likely to be advantageous.

A classical unbiased algorithm for estimating a low-resolution density map is the angular reconstitution~\cite{VanHeel1987,Goncharov}. Unfortunately, this method is very sensitive to noise in the images, and is thus applicable only to a small number of intensively-averaged, manually-verified class averages. As a result, the angular reconstitution method is still not effective enough in practice. The limited robustness of the angular reconstitution is due to its inefficient exploitation of the common lines information~\cite{sync2N}. Specifically, $N$ class averages share $\binom{N}{2}$ common lines, yet only $\mathcal{O}(N)$ of them are used by the angular reconstitution method.

To improve the robustness of the angular reconstitution to noise,~\cite{sync2N} showed how to use all the common lines between all class averages to estimate the orientations of all class averages simultaneously, resulting in an algorithm which is much more robust to noise than~\cite{VanHeel1987,Goncharov}, and turning the classical angular reconstitution method into a more practical tool for ab-initio modeling. The simultaneous estimation of the orientations relies on the surprising mathematical fact that all unknown imaging orientations (up to handedness) are given explicitly in closed form by (the eigenvectors of) an appropriately constructed matrix.

The idea of simultaneous synchronization was further improved in~\cite{sync3N}, resulting in even further noise robustness. Nevertheless, even the algorithm of~\cite{sync3N} fails if the projection-images are too noisy (due to high rate of misidentified common lines). However, if we knew which common lines are wrong, we could ignore them to achieve superior robustness to noise. Indeed, according to~\cite{sync3N}, the contribution of each common line to the computation of the orientations can be controlled by incorporating weights in the computations.

In order to find appropriate weights, we propose in this paper a model for the errors in the estimated common lines between pairs of class averages (Section~{\ref{sec:errors_model}}), use this model to estimate the reliability of each common line, and derive an algorithm which uses these estimated reliabilities to improve the estimation of the orientations (Section~{\ref{sec:errors_indications}}).
We demonstrate the algorithm in Section~\ref{sec:examples} using experimental data sets, \rda{such}{so} that it can be evaluated even if the accompanying math in earlier sections is skipped.

The algorithm introduced in Section~\ref{sec:errors_indications} is essentially a robust generalization of the angular reconstitution method, which uses thousands of class averages simultaneously, with possibly a small number of raw-images averaged within each class.
We demonstrate our algorithm using two experimental data sets; one of them consisting of 3000 noisy class averages of as few as 3 raw projection-images per class. For both data sets, the resulting density maps have resolution of 20{\AA} or better, providing a much more accurate and reliable initialization for existing high-resolution refinement algorithms.
We also demonstrate how to detect a failure of the proposed algorithm, to avoid using wrong initial models in consecutive steps of the reconstruction process.
It is important to note that in its current formulation, the presented algorithm is applicable only to non-symmetric structures, similarly to the original angular reconstitution algorithm.

The contribution of this paper is thus a robust algorithm  for reference-free ab-initio reconstruction of non-symmetrical structures. The advantages of our algorithm are that it is not biased towards any initial model, and can be used as a black-box tool.
From the point of view of the refinement process, unbiased density maps with intermediate resolution are expected to assist refinement algorithms to converge to the global optimum rather than to a local one.
From the point of view of recent common lines approaches such as \protect{\cite{LYUMKIS2013417}}, the algorithm can be used as a robust replacement to the traditional angular reconstitution.
Moreover, the algorithm provides indicators to determine the reliability of the reconstructed initial model.

\section{Problem setup and roadmap}
\label{sec:intro_algo}

\ra{Under an ideal mathematical model, each image in a cryo-EM data set is given by a two-dimensional tomographic projection along the $z$-direction of a randomly rotated copy of the underlying molecule. Equivalently, we can model the process as a fixed molecule, with the microscope being randomly rotated around it. Specifically, the $i$-th image is generated by rotating  the microscope using a $3 \times 3$ rotation matrix $R_{i}$, followed by computing the integral of the (density function of the) molecule along the line from the electron source to the detector. We denote the resulting projection-image by $P_{R_i}$. The goal is then to recover the unknown molecule given only a finite set of its projection-images {$\left \{P_{R_{i}}\right \}_{i=1}^{N}$}, where the corresponding rotations $R_{i}$ are unknown. This may be achieved by first estimating the rotations $R_{i}$ from the images, followed by standard tomographic inversion algorithms~{\cite{Herman09}}.}

\ra{As is well known, reconstructions in SPR based on cryo-EM suffer from an inherent loss of handedness, namely, the handedness or the chirality of the reconstructed molecule cannot be deduced from common lines information. Equivalently, there exist two different molecules (related to each other by reflection) and two corresponding sets of rotations that result in the images {$P_{R_{1}},\ldots,P_{R_{N}}$}. We elaborate more on this point below.}


\ra{Algorithms for estimating the rotations $R_{1},\ldots,R_{N}$ using only the projection-images $P_{R_{1}},\ldots,P_{R_{N}}$ are often based on the well-known Fourier projection-slice theorem~{\cite{Natr2001a}}, which implies that any two projection-images share a common line in Fourier space. This is the well-known ``common line property". Specifically, given images $P_{R_i}$ and $P_{R_j}$, we denote the angle that their common line makes with the $x$-axis in $P_{R_i}$ by $\alpha_{ij}$. Similarly, we denote the angle it makes with the $x$-axis in $P_{R_j}$ by $\alpha_{ji}$. The direction vectors (of unit length) of this common line in the images $P_{R_i}$ and $P_{R_j}$ are given respectively by}
\begin{equation}\label{eq:cijcji}
c_{ij} = \left ( \cos \alpha_{ij}, \sin \alpha_{ij} \right
),\qquad c_{ji} = \left ( \cos \alpha_{ji}, \sin
\alpha_{ji}\right ).
\end{equation}
\ra{It was shown in~{\cite{sync2N}} how to use the common lines between pairs of images ($c_{ij}$ and $c_{ji}$ of}~\eqref{eq:cijcji}\ra{) to estimate for each $(i,j)$ either the rotation $R_{i}R_{j}^{-1}$ or the rotation $JR_{i}R_{j}^{-1}J$, where $J=\operatorname{diag}(1,1,-1)$ is a $3 \times 3$ diagonal matrix corresponding to a reflection with respect to the $xy$ plane. The rotations $R_{i}R_{j}^{-1}$ and $JR_{i}R_{j}^{-1}J$ are indistinguishable due to the handedness ambiguity. However, it was shown in~{\cite{sync3N}} how to choose consistently either the set $\left \{ R_{i} R_{j}^{-1} \right \}_{i,j=1}^{N}$ or the set $\left \{ J R_{i} R_{j}^{-1} J \right \}_{i,j=1}^{N}$ (rather than $R_{i}R_{j}^{-1}$ for certain pairs $(i,j)$ and $JR_{i}R_{j}^{-1}J$ for other pairs).}

Assume therefore, without loss of generality, that we have the set $\left \{ R_{i} R_{j}^{-1} \right \}_{i,j=1}^{N}$. The rotations $\{R_i\}_{i=1}^{N}$ can be derived from the relative rotations $\{ R_iR_j^{-1} \}_{i,j=1}^{N}$ as follows. Define the matrix $S$ of size $3N \times 3N$ whose $(i,j)$ block of size $3\times 3$ is given by $S_{ij}=R_{i}R_{j}^{-1}$. Also define
\begin{equation} \label{eq:H}
H = \left( R_1 \cdots R_N \right)^{T} \in \mathbb{R}^{3N \times 3}.
\end{equation}
By a direct calculation we get that
\begin{equation}\label{eq:sync1}
(SH)_{i} = \sum_{j=1}^{N} (R_iR_j^{-1})R_j = N R_{i} = N H_{i}.
\end{equation}
Hence, $N$ is an eigenvalue of $S$ with multiplicity 3 and the (columns of the) rotations $R_{1},\ldots,R_{N}$ are the corresponding eigenvectors. Note that~\eqref{eq:sync1} implies that $S=HH^T$. Moreover, from~\eqref{eq:sync1} and~\eqref{eq:H} it follows that the matrix $S$ is of rank 3, and so $N$ is its only eigenvalue which is different from 0.

In practice, we don't have the relative rotations $R_{i}R_{j}^{-1}$, but only their estimates derived from noisy input data (the noisy images and the common lines derived from them). Naturally, the estimates of $R_{i}R_{j}^{-1}$ for different pairs $(i,j)$ are not equally reliable. As described in~\cite{sync3N}, the \rda{influence}{contribution} of each estimate $R_{i}R_{j}^{-1}$ \rda{on}{to} the computation of $R_{1},\ldots,R_{N}$ can be controlled by replacing the block $S_{ij}=R_{i} R_{j}^{-1}$ with $\tilde{S}_{ij} = w_{ij} R_{i} R_{j}^{-1}$, where $w_{ij} \ge 0$ reflects our confidence in the accuracy of the estimate of $R_{i} R_{j}^{-1}$. Then, if we normalize $w_{ij}$ such that \rd{for every $i$} $\sum_j w_{ij} = N$ \ra{for every $i$}, we get that
\begin{equation}\label{eq:sync2}
 (\tilde{S}H)_{i} = \sum_{j=1}^N w_{ij}(R_iR_j^{-1})R_j = R_i \sum_{j=1}^N w_{ij} = N R_{i} = N H_{i},
 \end{equation}
that is, the rotations $R_{1},\ldots,R_{N}$ are also the eigenvectors of the modified matrix $\tilde{S}$ corresponding to eigenvalue $N$.

Equation~\eqref{eq:sync2} is beneficial for reducing the influence of inaccurate relative rotations $R_{i}R_{j}^{-1}$ on the computation of the rotations $R_{1},\ldots,R_{N}$.
Indeed, due to high noise in typical projection-images, many common lines estimates are inaccurate and result in erroneous relative rotations. This paper suggests a model to evaluate the reliability of every common line estimate, and a corresponding weighting scheme to be used in~\eqref{eq:sync2}.

The structure of this paper is as follows. Section~\ref{sec:errors_model} suggests an empirical model for \rd{the} common lines' detection errors, showing that every common line estimate is either \textit{indicative} (i.e. correct up to some deviation) or totally \textit{arbitrary} (see Section~\ref{sec:errors_model} for details).
Section~\ref{sec:errors_indications} uses Bayesian analysis to find the probability of any given common line to be indicative, based on its relations with other common lines. These probabilities are then used to derive a weighting scheme for setting the weights $w_{ij}$ in{~\eqref{eq:sync2}}. Section~\ref{sec:examples} demonstrates the advantages of the new weighting scheme in terms of reconstruction resolution.
Section~\ref{sec:summary} summarizes the main results and discusses some future possible extensions.

\section{Common lines errors model}
\label{sec:errors_model}

As explained in Section~\ref{sec:intro_algo}, the Fourier transforms of every two clean projection-images share a common line in Fourier space\rd{ (see \ref{trash}))}. However, when the images $P_{R_{i}}$ and $P_{R_{j}}$ are noisy, \rda{\ref{trash} does not hold}{the values along the common line between the two images are not identical}. Therefore, a common algorithm for detecting common lines in noisy projection-images is based on computing $L$ radial Fourier lines for each image, with $n$ samples per Fourier line (where $L$ and $n$ are parameters),  computing the correlation between all $L \times L$ pairs of Fourier lines of the two images, and picking the pair with the highest correlation.
As one can expect, in the presence of noise, the pair of Fourier lines with \ra{the} highest correlation is not necessarily in the direction (equal or close to) $c_{ij}$ and $c_{ji}$ of~{\eqref{eq:cijcji}}.

In this section, we suggest a model for the common lines' estimation errors under the detection algorithm described above, and validate the model using simulated projection-images of 5 different molecules.
The model is used in Section~\ref{sec:errors_indications} to derive a measure for the reliability of the estimated common lines, allowing to determine the weights of the relative rotations in~\eqref{eq:sync2} accordingly.

Let $\tilde{c}_{ij}$ and $\tilde{c}_{ji}$ be the estimated common line between the images $P_{R_{i}}$ and $P_{R_{j}}$, and let $c_{ij}$ and $c_{ji}$ be (as before) the true common line \rda{according to \ref{trash} and}{of~{\eqref{eq:cijcji}}}. The suggested errors model is as follows:
\begin{enumerate}
\item With probability $P$, the angle between $c_{ij}$ and $\tilde{c}_{ij}$, as well as the angle between $c_{ji}$ and $\tilde{c}_{ji}$, is distributed $N(0,\sigma^{2})$, in which case we say that the estimated common line $(\tilde{c}_{ij}, \tilde{c}_{ji})$ is \emph{indicative}.
\item With probability $1-P$, the angles of the vectors $\tilde{c}_{ij}$ and $\tilde{c}_{ji}$ are distributed uniformly over $[0,360^\circ)$, in which case we refer to the estimated common line $(\tilde{c}_{ij}, \tilde{c}_{ji})$  as \emph{arbitrary}.
\end{enumerate}

Note that the errors model distinguishes between two types of errors. One type is detection of random correlations in the noise of the images, which have no preferred directions, hence the estimate of the common line in such a case is totally random.
The other type is detection of correlations in the signal, which exist in the direction of the correct common line, as well as in lines which are adjacent to it, depending on the smoothness of the (noiseless signal in the) projection-image. Hence, in this case the deviation of the estimated common line from the correct direction is assumed to be normally distributed with standard deviation that depends on the noise.

According to the proposed model, the quality of common lines' estimates is defined by two parameters: the rate $P$ of the indicative common lines, and the typical error $\sigma$ (measured in degrees) of the indicative common lines.

It should be noted that the directions of the common lines are estimated only modulo $180^{\circ}$, since the correlation of any two Fourier lines in any two images is invariant to reversing the directions of both lines by $180^{\circ}$. Thus, in practice, we expect the errors' distribution to be the sum of a uniform distribution (corresponding to arbitrary common lines) and \textit{two} Gaussians (corresponding to indicative common lines) -- one around $0^{\circ}$ and one around $180^{\circ}$. As a result, the probability density function of the errors is expected to be
\begin{linenomath*}
\begin{equation}\label{eq:clerrpdf}
 f(x) = P \cdot \frac{1}{\sigma \sqrt{2\pi}} \Big[ e^{-\frac{x^2}{2\sigma ^2}} + e^{-\frac{(180^\circ-x)^2}{2\sigma ^2}} \Big] + (1-P) \cdot \frac{1}{180^\circ},
 \quad x \in [0,180^{\circ}],
\end{equation}
up to a negligible normalization error due to the missing tail of the normal distribution beyond $180^\circ$, which integrates to less than $2 \cdot 10^{-9}$ for $\sigma<30^\circ$.
$P$ and $\sigma$ are the parameters of the model that need to be estimated from the common lines data. In order to fit these parameters for particular common lines data, we compute\rd{d} the histogram (with 36 intervals of width $W_{interval}=5^\circ$) of the errors, and use\rd{d} the standard Matlab curve fitting toolbox to find the best fit of the form
$$ f(x)=\Big(N_{lines}\cdot P\Big) \cdot W_{interval} \cdot \frac{1}{\sigma \sqrt{2\pi}} \Big[ e^{-\frac{x^2}{2\sigma^2}} + e^{-\frac{(180^\circ-x)^2}{2\sigma^2}} \Big] +\Big( N_{lines} \cdot (1-P) \Big) \cdot W_{interval} \cdot \frac{1}{180^\circ}, $$
where $N_{lines}=2\binom{N}{2}$ (two lines for every pair of images).
\end{linenomath*}

\begin{figure}
	\begin{center}
		\subfloat[SNR=1/2.5]{\includegraphics[width=0.2\textwidth]{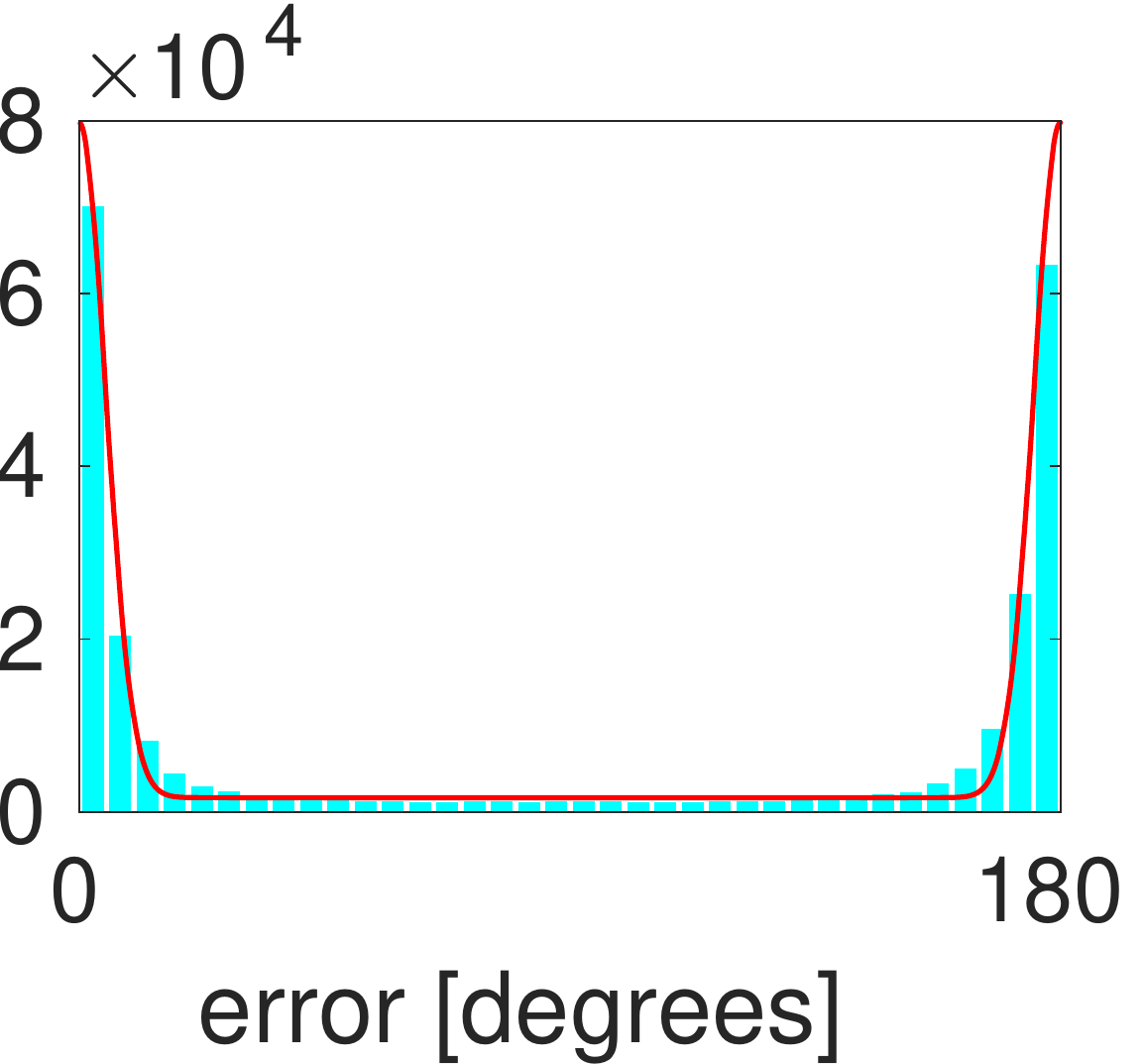}} \hfill
		\subfloat[SNR=1/5]{\includegraphics[width=0.2\textwidth]{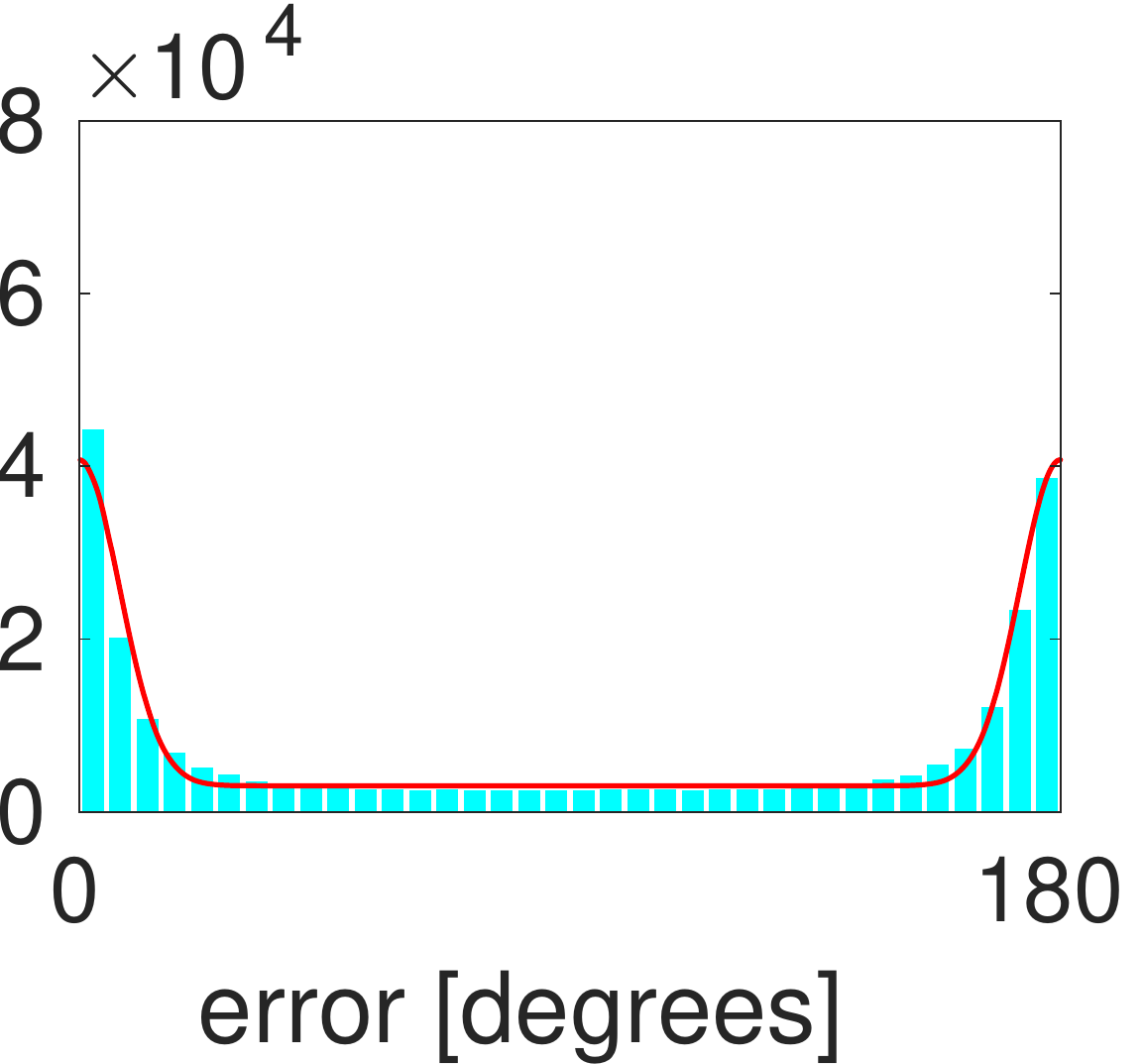}} \hfill
		\subfloat[SNR=1/7.5]{\includegraphics[width=0.2\textwidth]{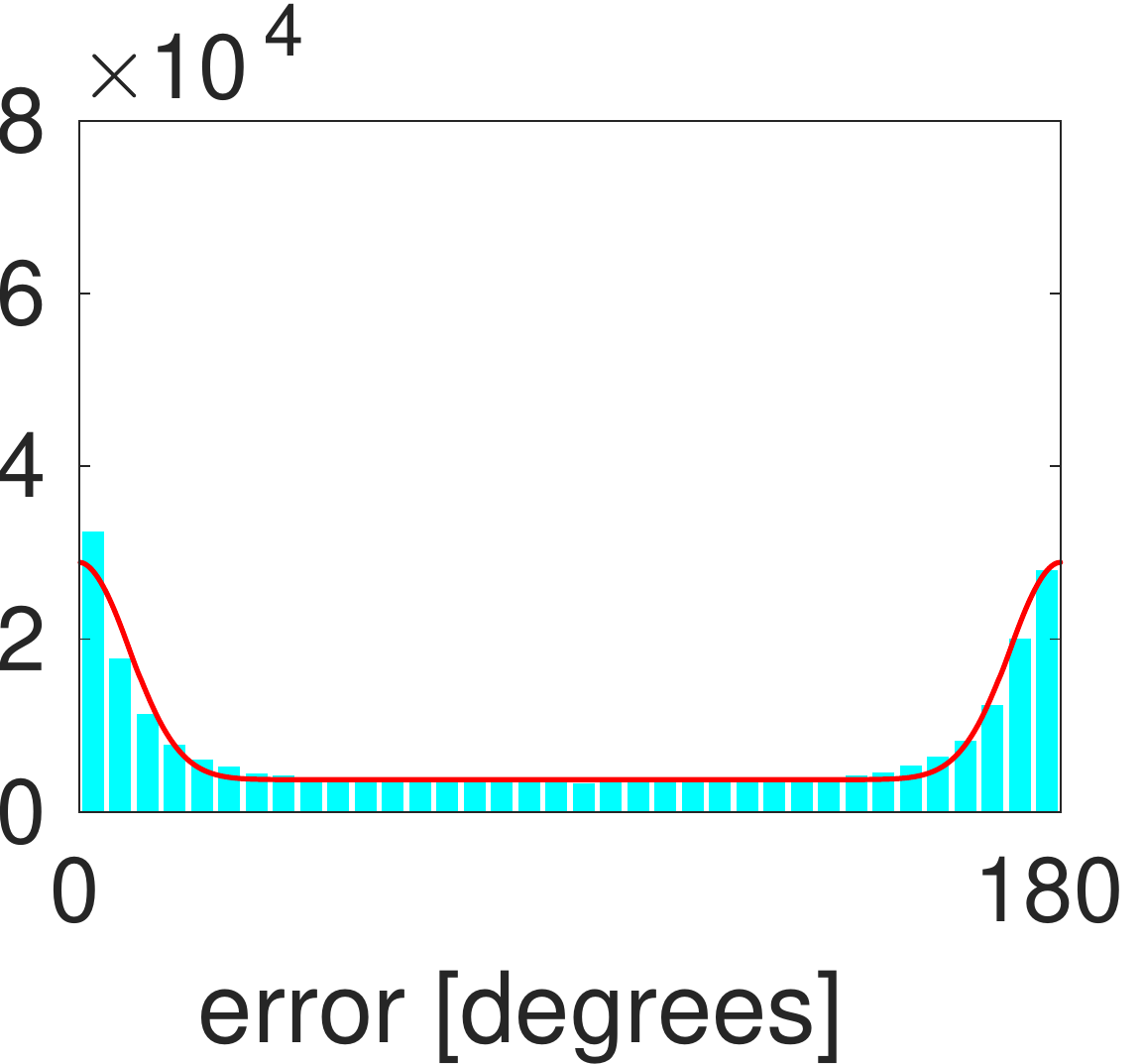}} \hfill
		\subfloat[SNR=1/10]{\includegraphics[width=0.2\textwidth]{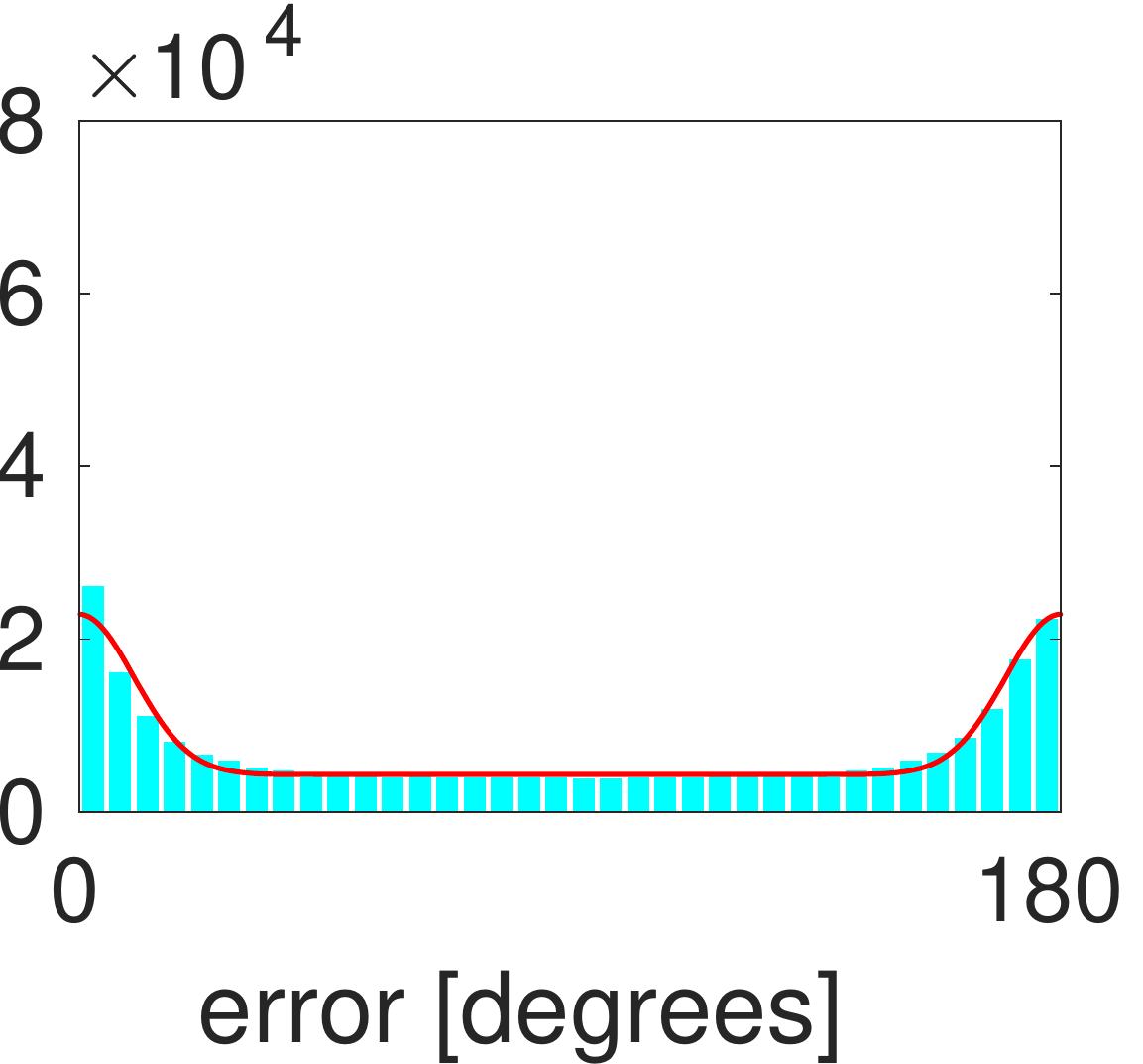}}\\
		\subfloat[SNR=1/12.5]{\includegraphics[width=0.2\textwidth]{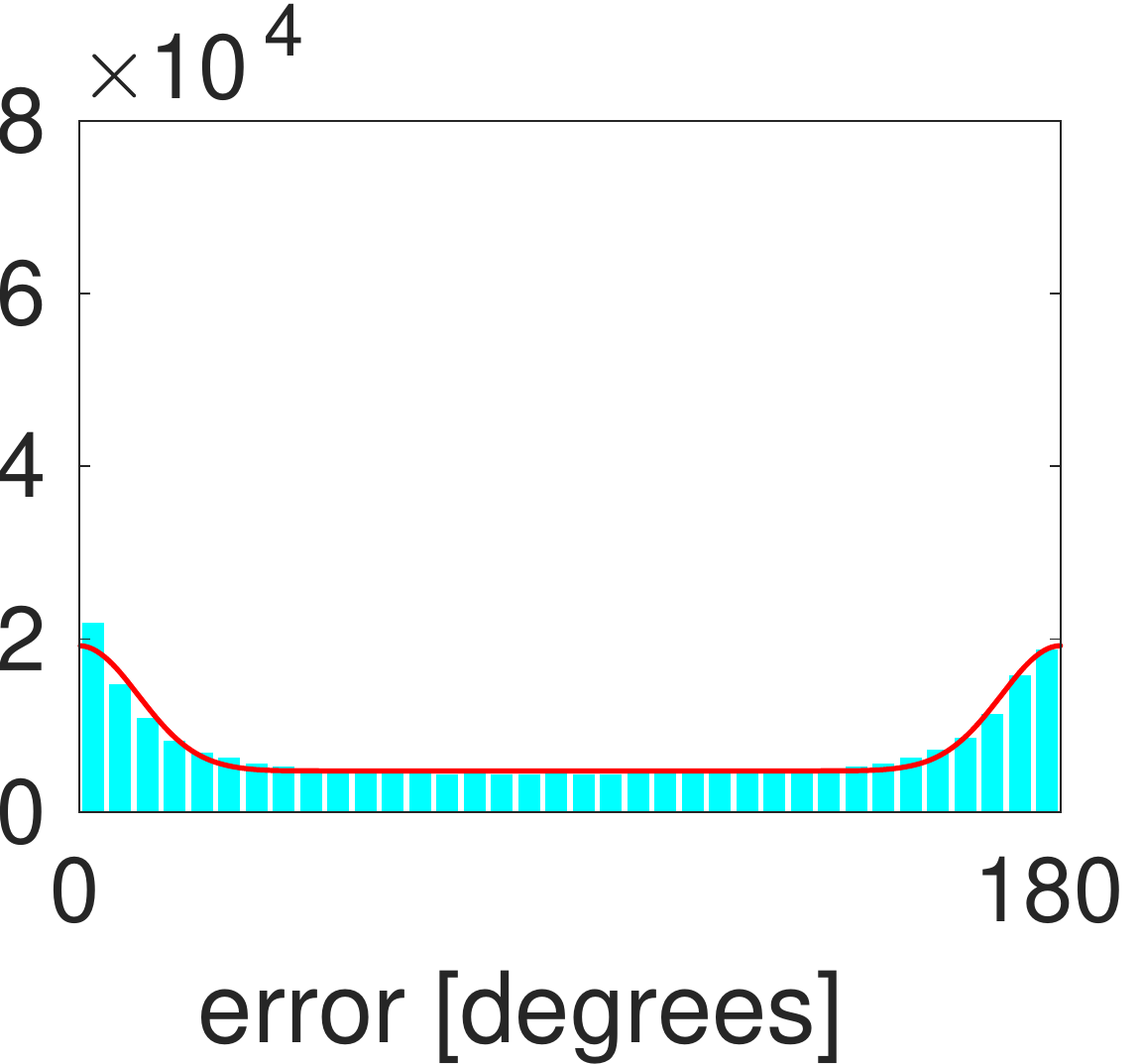}} \hfill
		\subfloat[SNR=1/15]{\includegraphics[width=0.2\textwidth]{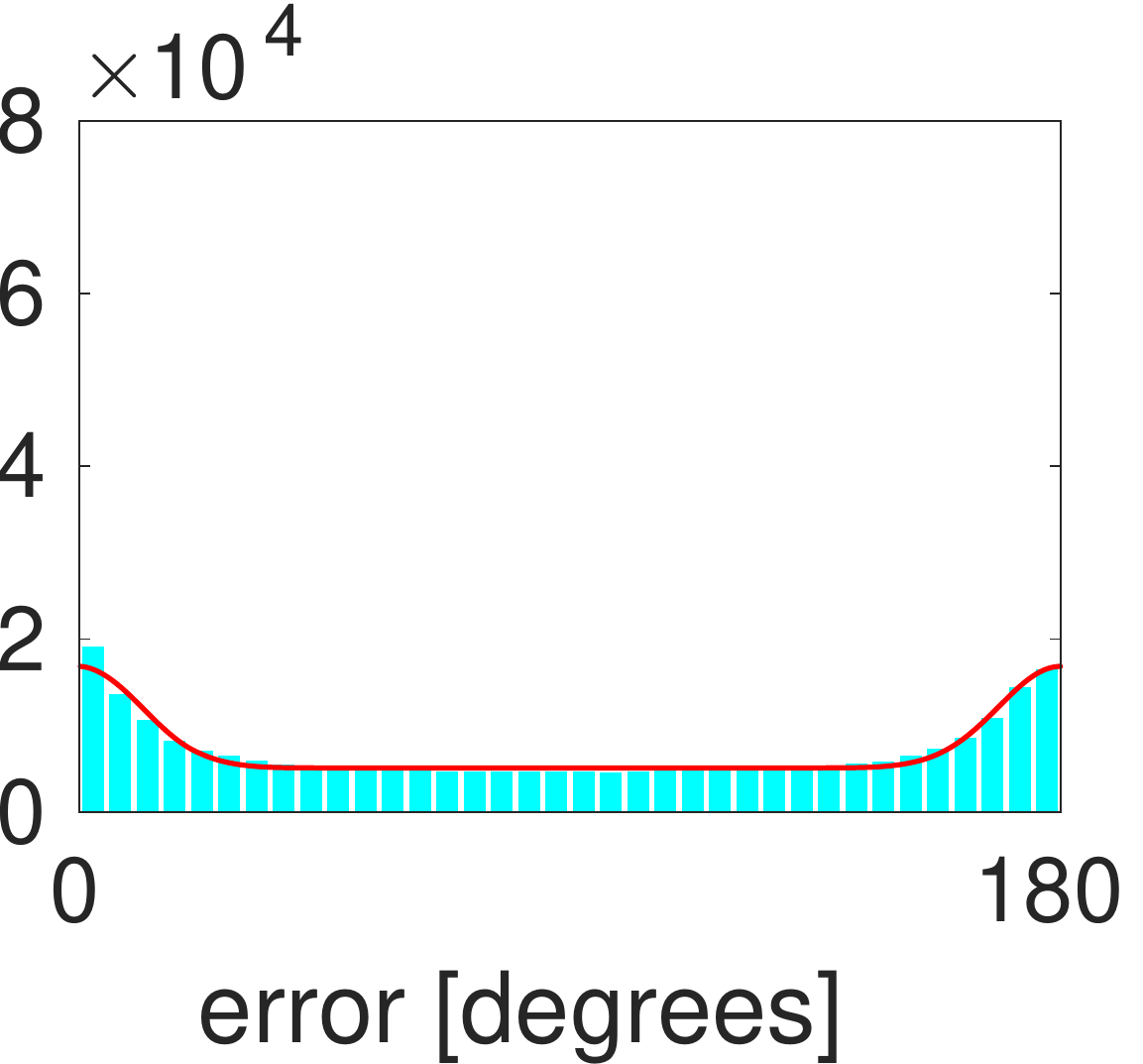}} \hfill
		\subfloat[SNR=1/17.5]{\includegraphics[width=0.2\textwidth]{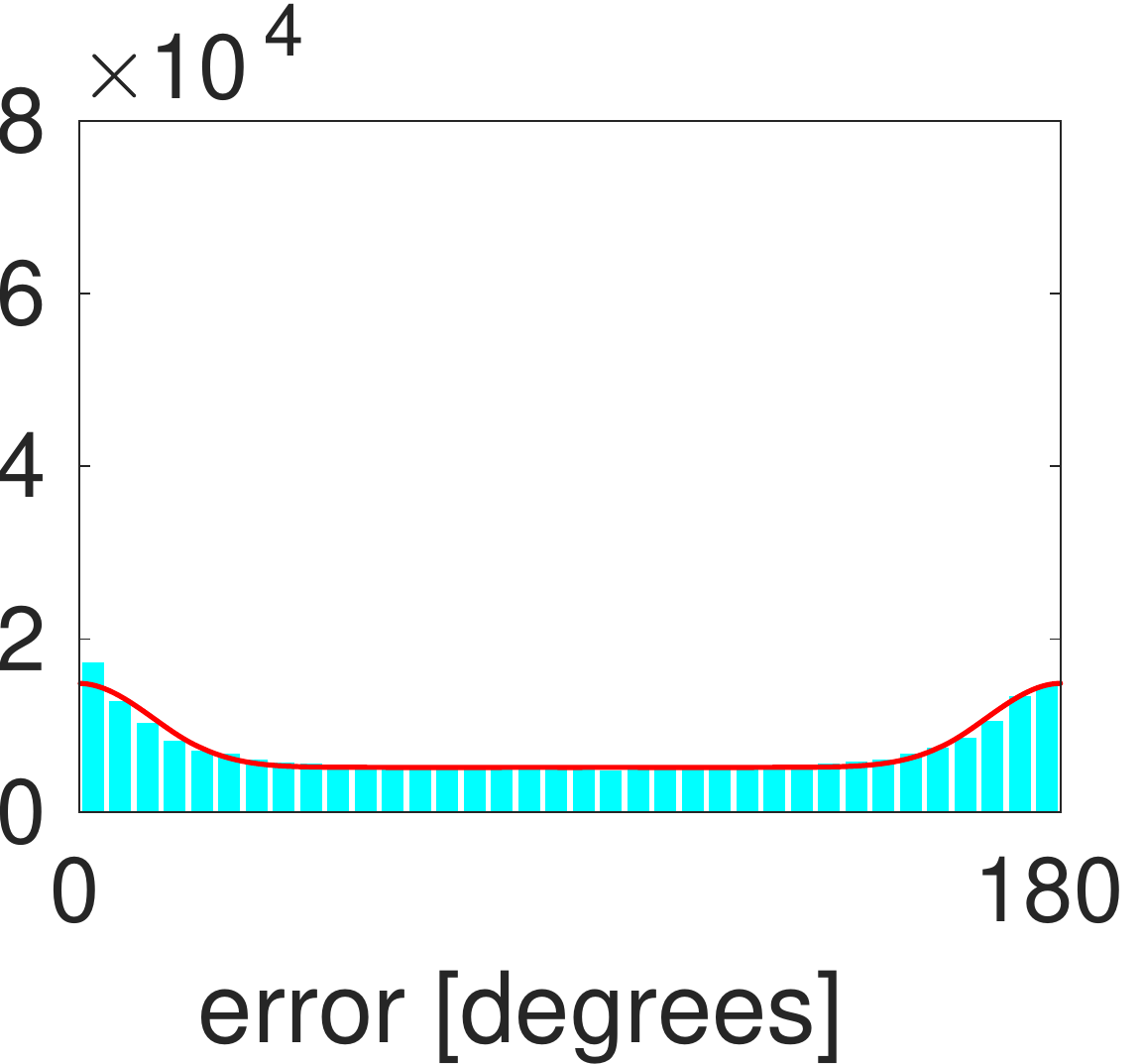}} \hfill
		\subfloat[SNR=1/20]{\includegraphics[width=0.2\textwidth]{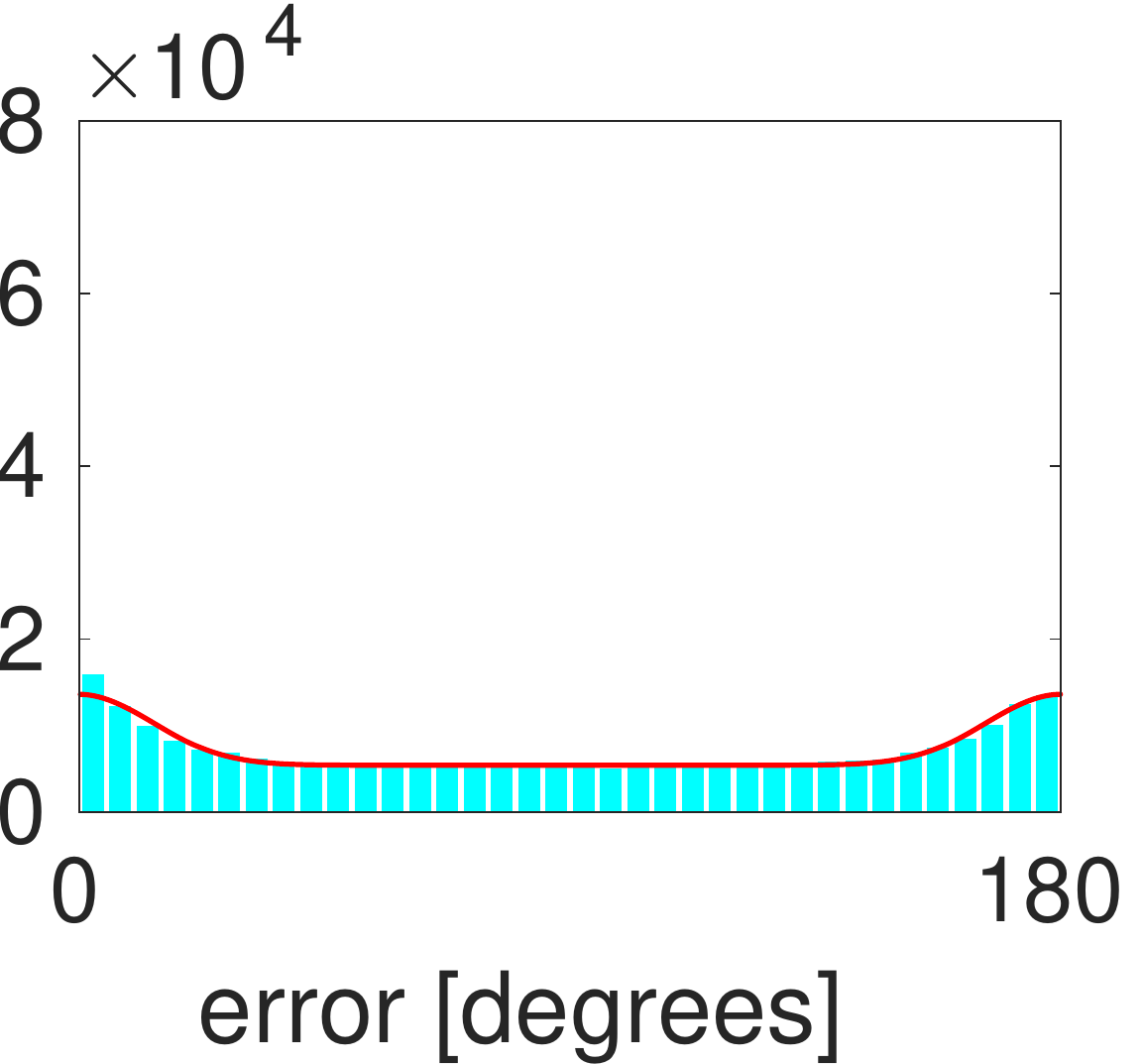}}

		\caption{Histograms of the errors of common lines' estimates and their fit according to the suggested errors' model, for common lines data generated using simulated projection-images of the 50S subunit of the E. Coli ribosome.}
		\label{model_fit_50S}
	\end{center}
\end{figure}

\begin{figure}
	\begin{center}
		\subfloat{\includegraphics[width=0.3\textwidth]{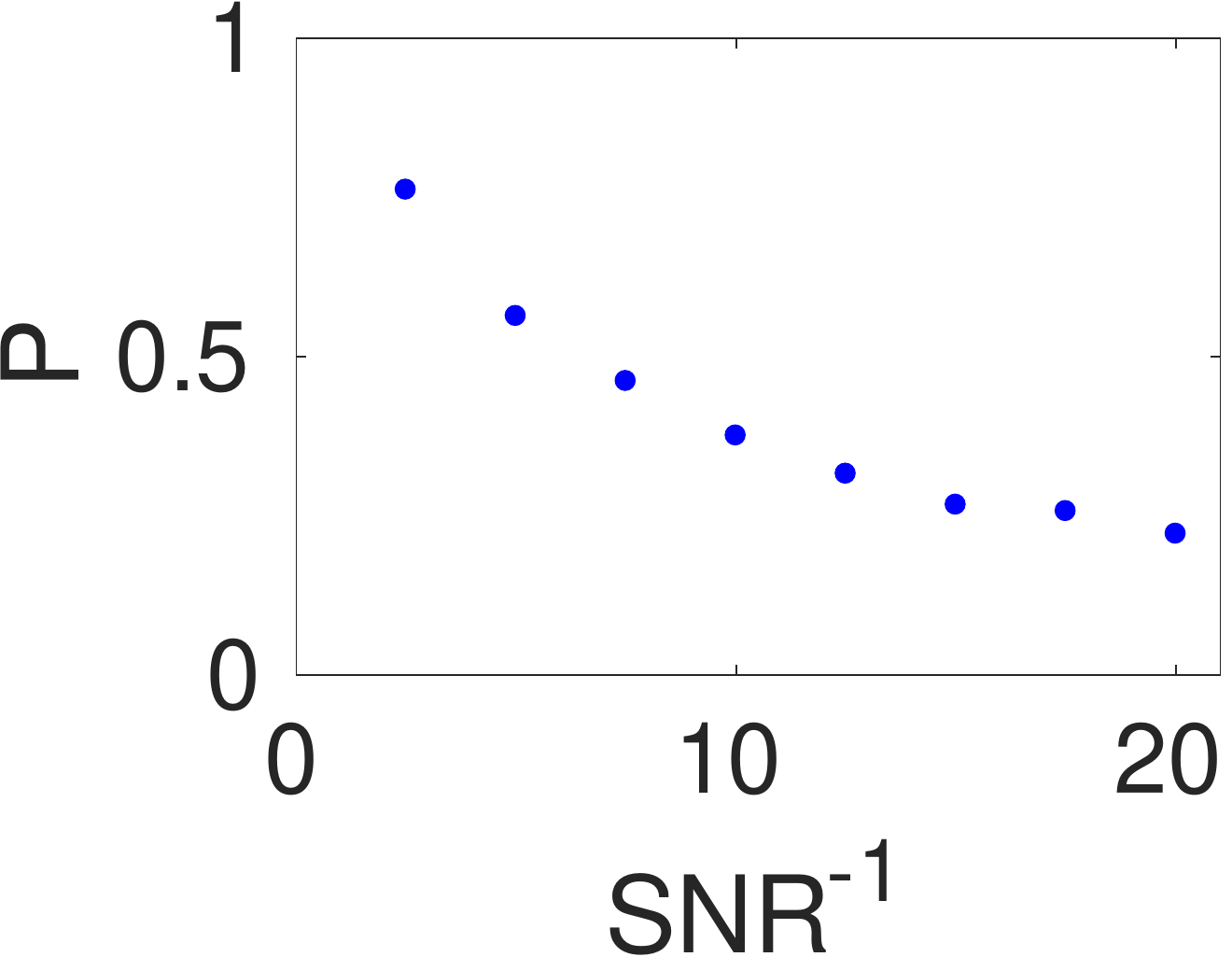}} \hspace{0.05\textwidth}
		\subfloat{\includegraphics[width=0.3\textwidth]{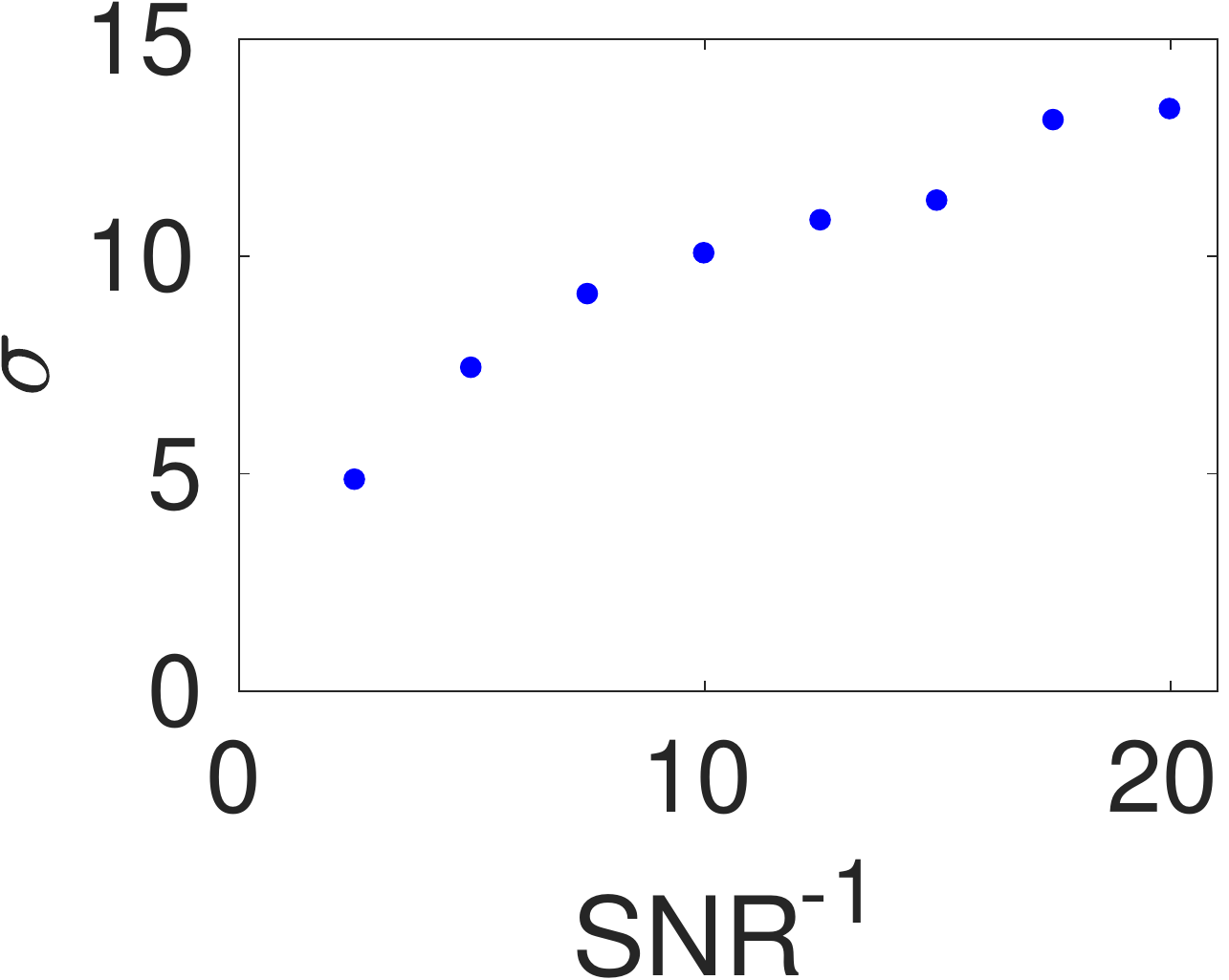}}
		\caption{Dependence of the model parameters on the SNR of the projection-images.}
		\label{param_50S}
	\end{center}
\end{figure}

To test the errors' model, we first used $N=500$ simulated (centered) projection-images of size $65 \times 65$ of the 50S subunit of the E. Coli ribosome, corrupted with Gaussian noise. For every pair of images, their common line was estimated according to the detection algorithm described above (with resolution of $L=360$ radial Fourier lines per image). Figure~\ref{model_fit_50S} presents the match of the model to the data based on the simulated projection-images, and Table~\ref{param_tab_50S} presents the fitted parameters $P$ and $\sigma$ as well as the R-squared value of the fit.
The parameters of the model $P$ and $\sigma$ clearly depend on the SNR (signal-to-noise ratio) of the projection-images, as shown in Figure~\ref{param_50S}.

\begin{table}
	\begin{center}
		\begin{tabular}{c|ccc}
			SNR  & $P$ & $\sigma$ & $R^2$
			\\ \hline &&& \\
			$1/2.5\phantom{0}$ & 76\% & 4.8 & 0.999 \\
			$1/5\phantom{.50}$ & 56\% & 7.4 & 0.998 \\
			$1/7.5\phantom{0}$ & 46\% & 9.1 & 0.997 \\
			$1/10\phantom{.5}$ & 38\% & 10.1 & 0.997 \\
			$1/12.5$ & 32\% & 10.8 & 0.997 \\
			$1/15\phantom{.5}$ & 27\% & 11.3 & 0.997 \\
			$1/17.5$ & 26\% & 13.1 & 0.997 \\
			$1/20\phantom{.5}$ & 22\% & 13.4 & 0.997 \\
		\end{tabular}
		\caption{Estimated model parameters for common lines data sets generated using simulated projection-images of the 50S subunit of the E. Coli ribosome for various SNR levels.}
		\label{param_tab_50S}
	\end{center}
\end{table}

\begin{figure}
	\begin{center}
		\subfloat[50s]{\includegraphics[width=0.18\textwidth]{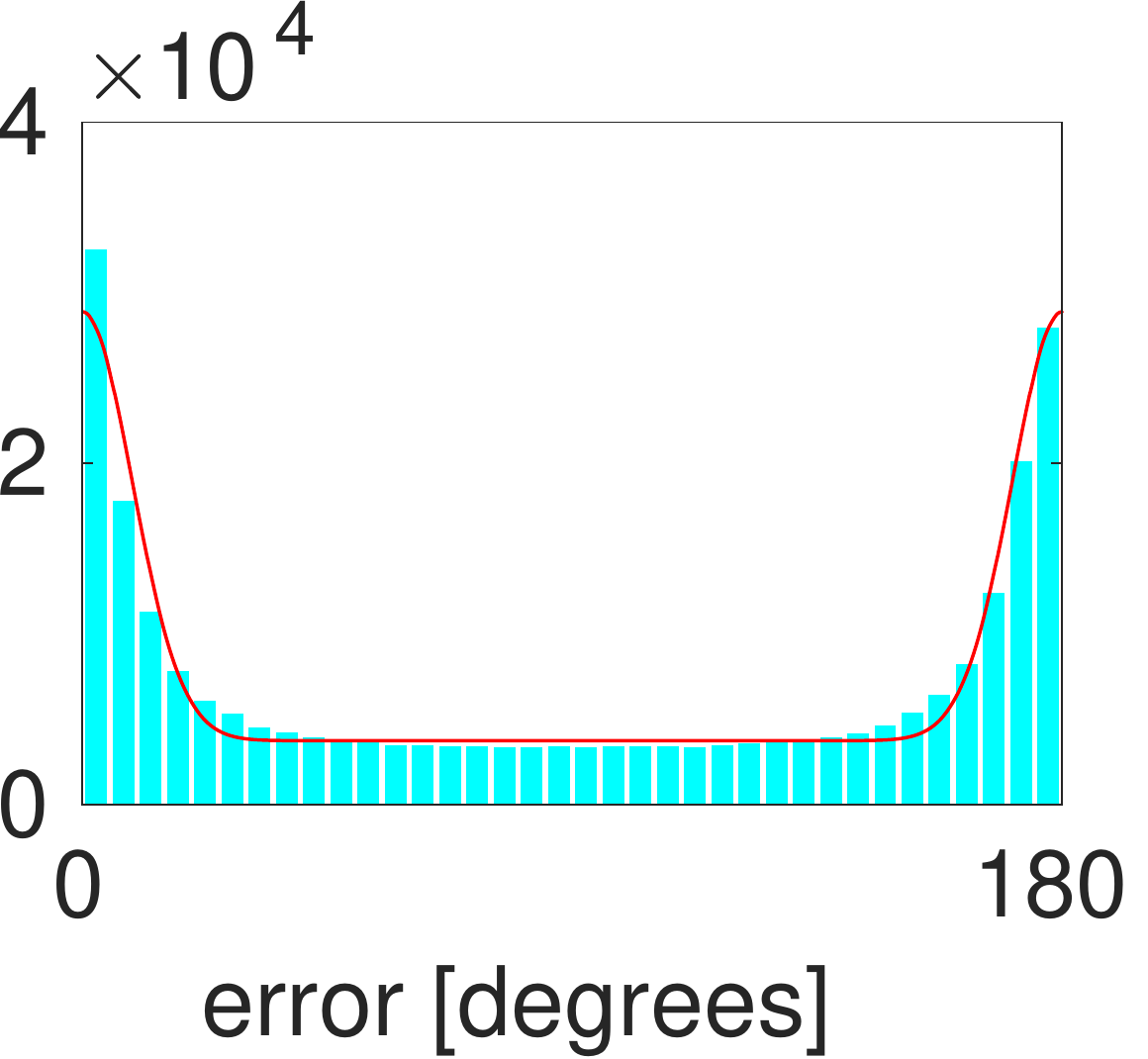}} \hfill
		\subfloat[\#2763]{\includegraphics[width=0.18\textwidth]{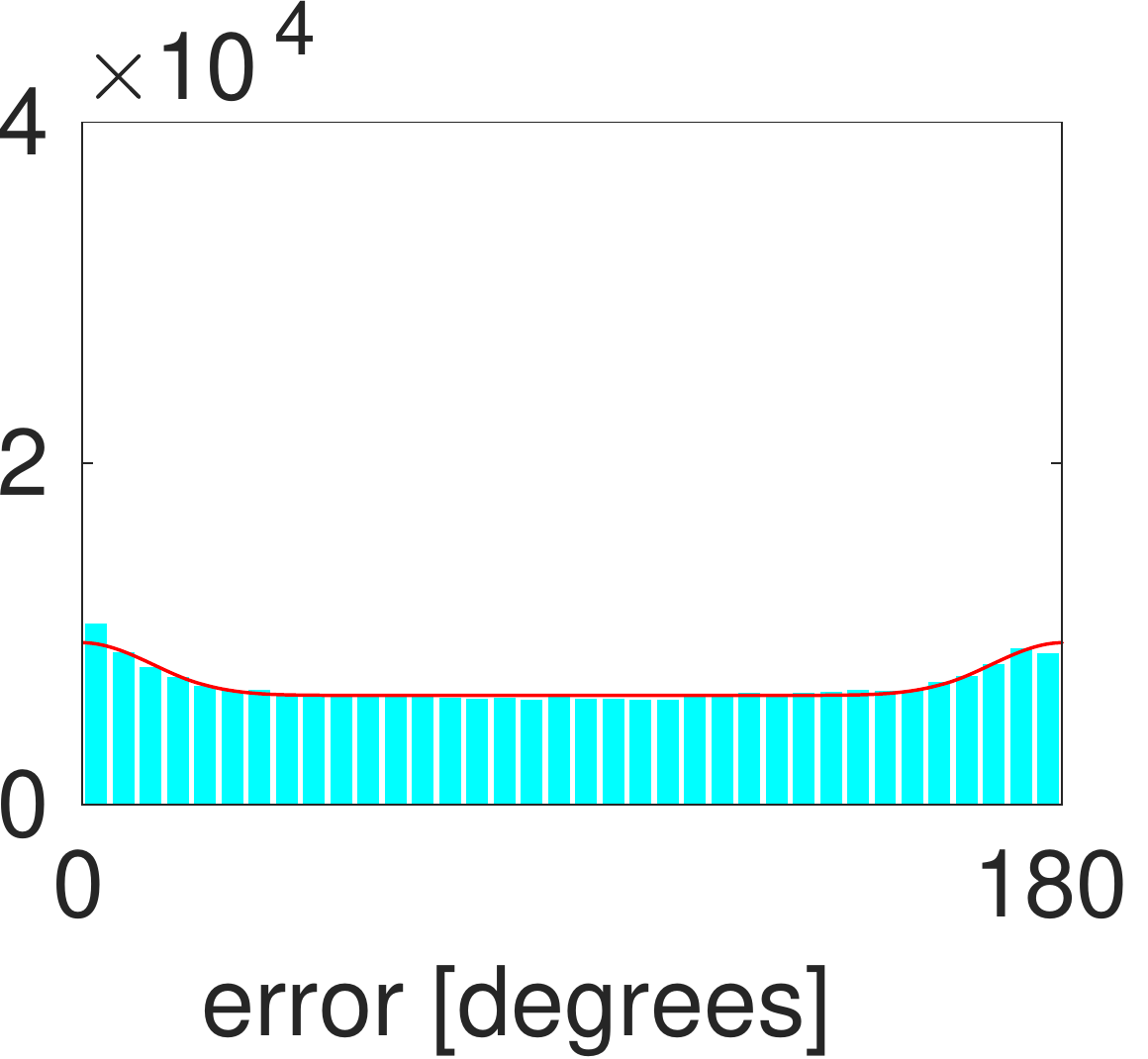}} \hfill
		\subfloat[\#2773]{\includegraphics[width=0.18\textwidth]{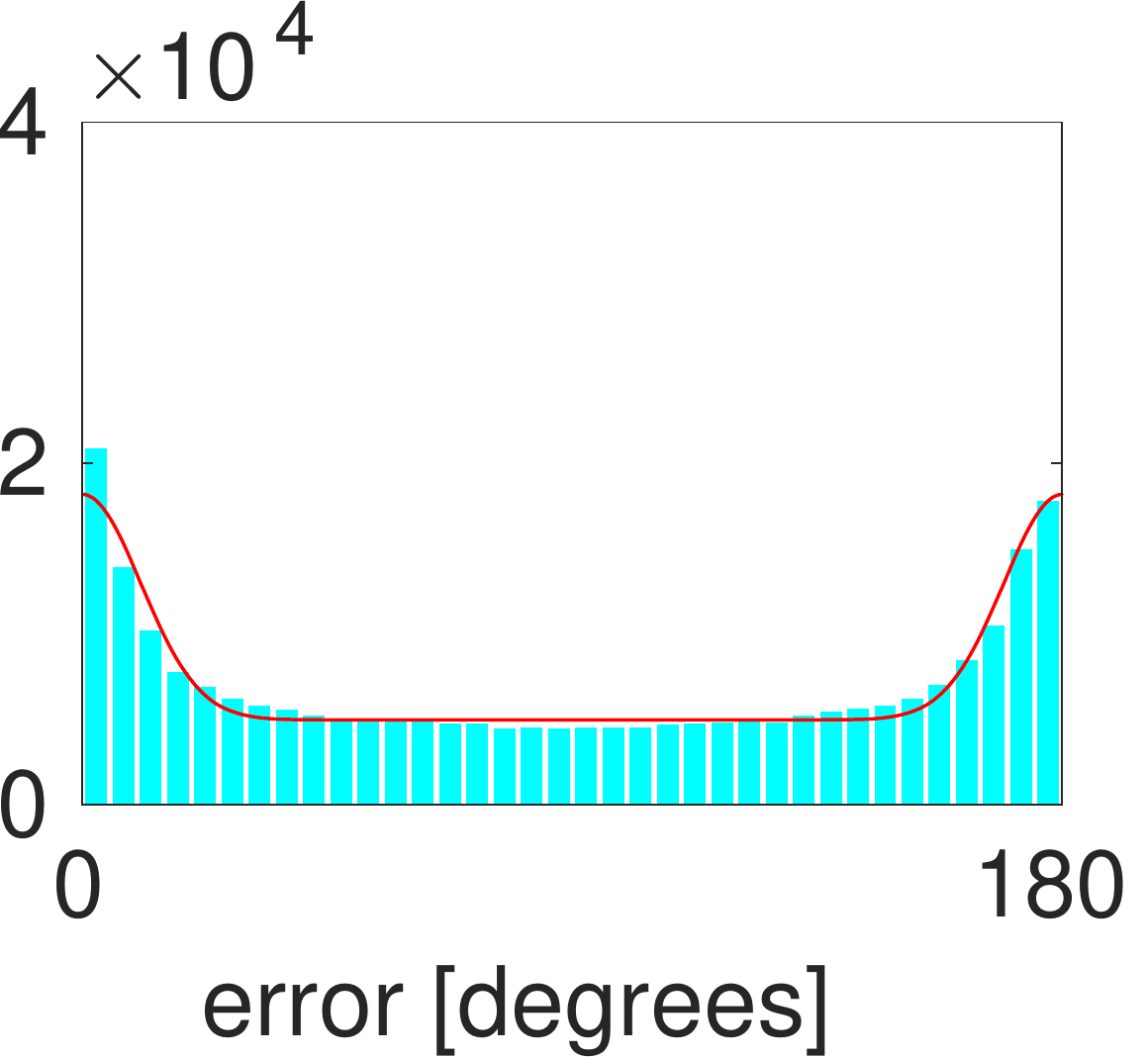}} \hfill
		\subfloat[\#2790]{\includegraphics[width=0.18\textwidth]{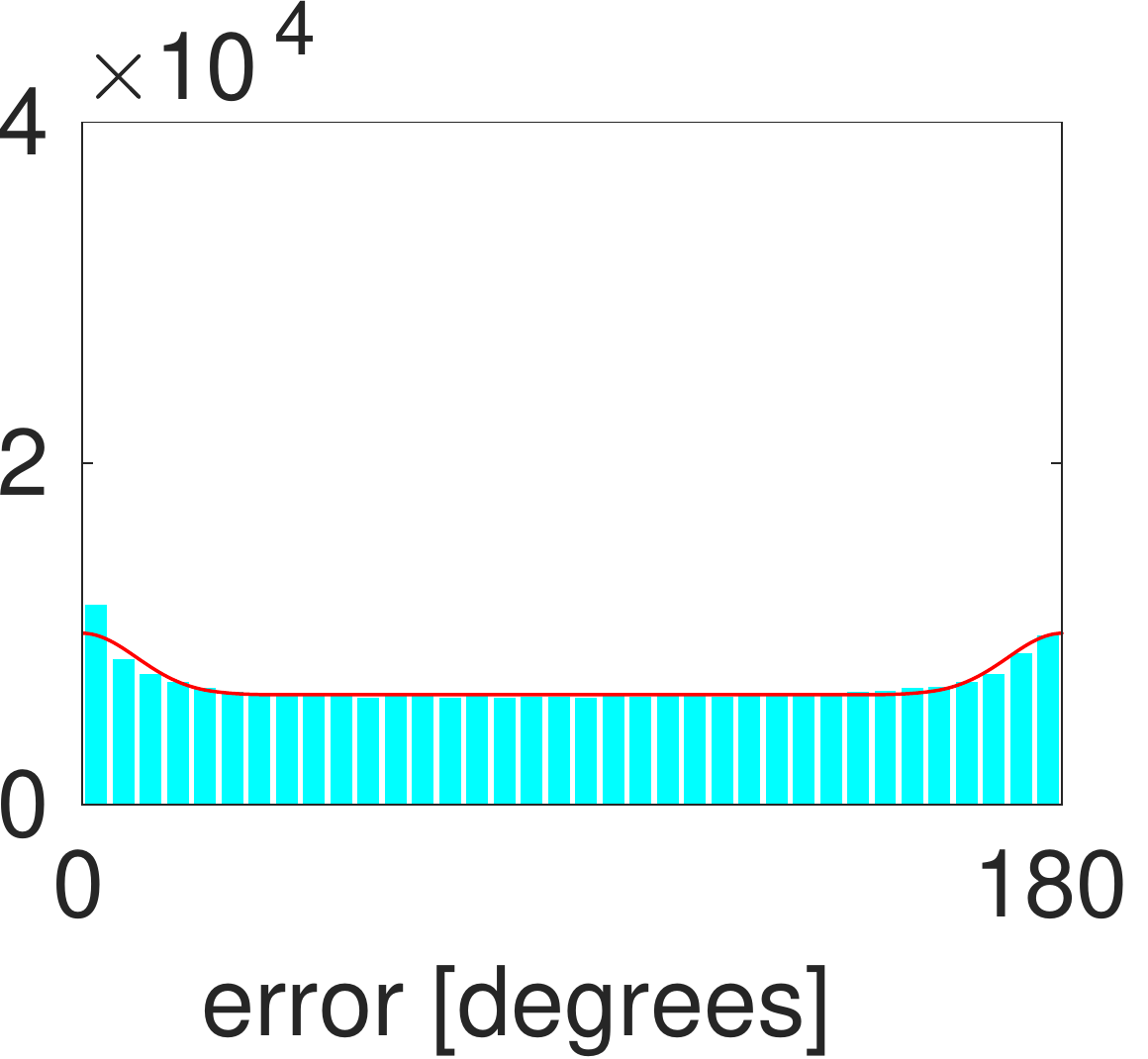}} \hfill
		\subfloat[\#6211]{\includegraphics[width=0.18\textwidth]{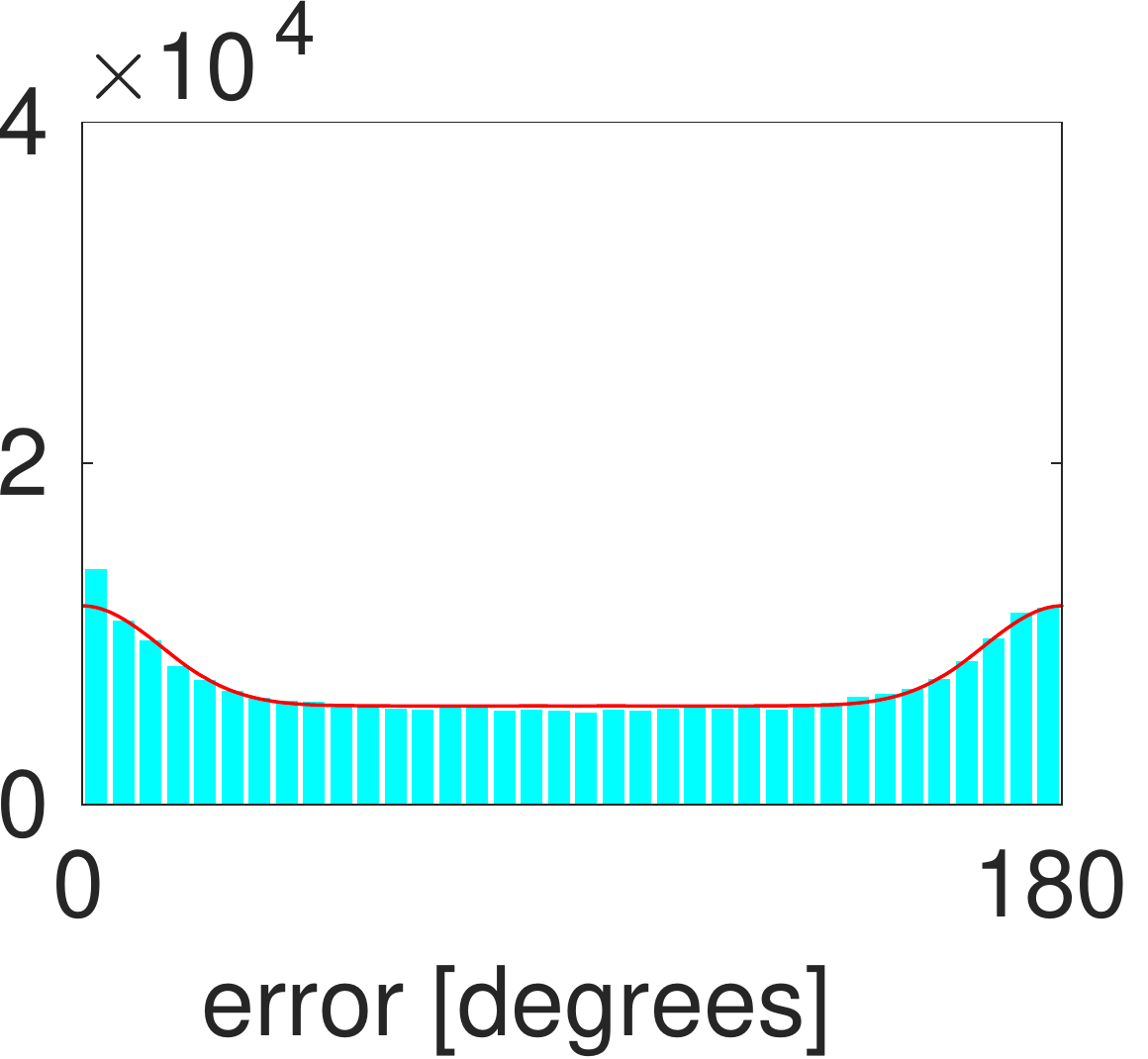}}

		\caption{Fits of the model to the empirical common lines errors' histograms, using simulated projection-images of various density maps at $\text{SNR}=1/7.5$.}
		\label{test_all}
	\end{center}
\end{figure}

Next, we tested the fit of the errors' model using simulated projection-images of four density maps taken from the Electron Microscopy Data Bank EMDB~\protect\cite{EMDB}: Partial yeast 48S preinitiation complex (EMD--2763), TnaC stalled E. Coli ribosome (EMD--2773), Molecular helix of eukaryotic polyribosomes (EMD--2790), \ra{and} Ribosomal protein S1 (EMD--6211). As before, for every volume and for several SNR levels, we simulated $N=500$ projection-images of size $65 \times 65$ pixels, estimated common lines between all pairs of images, and calculated the errors in the estimated common lines. For all volumes and all tested SNR levels, the resulting R-squared values of the fitted model satisfy $R^2>0.98$. Figure~\ref{test_all} demonstrates the fit of the model to the errors' histograms for all tested volumes for $\text{SNR}=1/7.5$.

\section{Common lines quality assessment}
\label{sec:errors_indications}

As explained in Section~\ref{sec:intro_algo}, we would like to distinguish between the indicative and the arbitrary common lines, in order to reduce the influence of the latter on the estimation of the rotations $R_{1},\ldots,R_{N}$. In this section we suggest a statistical model for evaluating the probability $P_{ij}$ of every common line pair $(c_{ij},c_{ji})$ (see~\eqref{eq:cijcji}) to be indicative, and justify the model empirically. Subsequently, in order to simplify the notation (and with \ra{a} slight abuse of notation), we denote by $c_{ij}$ the pair $(c_{ij},c_{ji})$ of~\protect\eqref{eq:cijcji}.

In a data set of $N$ projection-images, \rda{for every common line $c_{ij}$ there exist $N-2$ triplets of common lines}{every common line $c_{ij}$ participates in $N-2$ possible triplets} $\{(c_{ij},c_{jk},c_{ik})\}$ with $k \ne i,j$. In order to decide whether $c_{ij}$ is indicative or arbitrary, we start by proposing a score for every triplet $(c_{ij},c_{jk},c_{ik})$, which is shown to have a certain statistical distribution for triplets of indicative common lines, and a different distribution for arbitrary ones.
Once we find the two distributions, we can calculate the probability $P_{ij}$ that a certain common line is indicative, using Bayesian inference from the scores of all the triplets which include $c_{ij}$.

We start by briefly reviewing  some necessary background in Section~\ref{sec:j_sync}. Then, we introduce the statistical analysis of triplets' scores in Section~\ref{sec:triplets_scores}. Next, we derive an estimate for the probability $P_{ij}$ in Section~\ref{sec:triplets2pairs}, and finally we incorporate it into the reconstruction algorithm in Section~{\ref{sec:weighting_scheme}}.

\subsection{Handedness synchronization and triplets' scores}
\label{sec:j_sync}

According to~\cite{sync2N}, given projection-images $P_{R_{1}},\ldots,P_{R_{N}}$ corresponding to the (unknown) rotations $R_{1},\ldots,R_{N}$ (see Section~\ref{sec:intro_algo}), the common lines between the projection-images allow to determine for each pair $(i,j)$ either the rotation $R_{i}R_{j}^{-1}$ or the rotation $J R_{i} R_{j}^{-1} J$ (where $J=\operatorname{diag}(1,1,-1)$ \ra{is a reflection through the $xy$ plane}), not being able to distinguish between the two rotations.
This is a manifestation of the well-known handedness ambiguity in cryo-EM. Moreover, this ambiguity for a certain pair $(i,j)$ is independent of other pairs of indices. That is, it may be that for \rda{a}{the} pair $(i,j)$ we estimate $R_{i} R_{j}^{-1}$, and for another pair $(k,l)$ we estimate $J R_{k} R_{l}^{-1} J$.
\ra{However, to recover the volume underlying the images (or its reflected version, which is indistinguishable due to handedness), it is required to consistently estimate either all rotations $\left \{ R_{i} R_{j}^{-1} \right \}_{i,j=1}^{N}$ or all rotations $\left \{ J R_{i} R_{j}^{-1} J \right \}_{i,j=1}^{N}$.}
Such a procedure for consistent estimation was presented in~\cite{sync3N} under the name ``$J$-synchronization'', and is based on the identities
\begin{equation}
R_iR_j^{-1} \cdot R_jR_k^{-1} \cdot R_kR_i^{-1} - I = 0, \qquad JR_iR_j^{-1}J \cdot JR_jR_k^{-1}J \cdot JR_kR_i^{-1}J - I = 0.
\end{equation}

We denote by $R_{ij}$ the relative rotation estimated using common lines between the images $P_{R_i}$ and $P_{R_j}$ (using also some third image $P_{R_{k}}$ as required by the angular reconstitution). In the noiseless setting, it holds that $R_{ij} \in \left \{ R_{i}R_{j}^{-1}, \ JR_{i}R_{j}^{-1}J \right \}$. Given a triplet of indices $(i,j,k)$, in order to synchronize the relative rotations $R_{ij}$, $R_{jk}$, $R_{ki}$ (as required by~\cite{sync3N}), we define
\begin{equation}\label{eq:Cijk}
C_{ijk}(\mu_{ij},\mu_{jk},\mu_{ki}) = ||J^{\mu_{ij}}R_{ij}J^{\mu_{ij}} \cdot J^{\mu_{jk}}R_{jk}J^{\mu_{jk}} \cdot J^{\mu_{ki}}R_{ki}J^{\mu_{ki}} - I||_F,
\end{equation}
where $||\cdot||_F$ denotes the Frobenius norm, and exhaustively search over all possible triplets $(\mu_{ij},\mu_{jk},\mu_{ik}) \in \{0,1\}^3$ for the triplet which minimizes~\eqref{eq:Cijk}. The triplet $(\mu_{ij},\mu_{jk},\mu_{ik})$ corresponding to the minimum tells us how to $J$-conjugate the estimates $R_{ij}$, $R_{jk}$, $R_{ki}$ such that they equal either $R_{i}R_{j}^{-1}$, $R_{j}R_{k}^{-1}$, $R_{k}R_{i}^{-1}$, respectively, or $JR_{i}R_{j}^{-1}J$, $JR_{j}R_{k}^{-1}J$, $JR_{k}R_{i}^{-1}J$, respectively. In other words, this procedure ``$J$-synchronizes'' the triplet of rotations $R_{ij}$, $R_{jk}$, $R_{ki}$. The algorithm in~\cite{sync3N} shows how to take such triplets of rotations, where each triplet is $J$-synchronized, and to consistently construct either the set $\left \{ R_{i} R_{j}^{-1} \right \}_{i,j=1}^{N}$ or the set $\left \{ J R_{i} R_{j}^{-1} J \right \}_{i,j=1}^{N}$.

For noisy images, the value of $C_{ijk}$ in~\eqref{eq:Cijk} would not equal zero for any assignment $(\mu_{ij},\mu_{jk},\mu_{ik}) \in \{0,1\}^3$, simply because all estimates $R_{ij}$ are noisy and do not equal $R_{i}R_{j}^{-1}$ nor $J R_{i} R_{j}^{-1} J$. Still, lower values of $C_{ijk}$ are more likely to correspond to the correct assignment. Accordingly,~\cite{sync3N} defines the score of a triplet, denoted $s_{ijk}$, which expresses the certainty in the correctness of the $J$-synchronization, as the relative gap between the minimal value of $\{C_{ijk}(\mu_{ij},\mu_{jk},\mu_{ki})\ | \ (\mu_{ij},\mu_{jk},\mu_{ki}) \in \{0,1\}^3\}$, denoted by $C_{ijk}^{min}$, and the second lowest value $C_{ijk}^{alt}$, that is,
\begin{equation}\label{eq:score}
s_{ijk} = \frac{C_{ijk}^{alt}-C_{ijk}^{min}}{C_{ijk}^{alt}} = 1 - \frac{C_{ijk}^{min}}{C_{ijk}^{alt}}.
\end{equation}
Note that $s_{ijk}=1$ if and only if $C_{ijk}^{min}=0$ (i.e. there is a ''perfect'' $J$-conjugation), and $s_{ijk}=0$ if and only if $C_{ijk}^{min}=C_{ijk}^{alt}$ (i.e. there are two equivalent $J$-conjugations).

\subsection{Distribution of triplets' scores}
\label{sec:triplets_scores}

In this section, we model the distributions of the triplets' scores $s_{ijk}$ of~\eqref{eq:score}, separately for triplets of indicative common lines and for triples of arbitrary common lines. We will show that using this model along with the histogram of all the triplets' scores for given common lines data, one can find the total rate $P$ of indicative common lines of~\eqref{eq:clerrpdf}. This estimate of $P$ is used as the prior probability in Section~\ref{sec:triplets2pairs}, which applies Bayesian inference to evaluate the probability that a given common line is indicative.

\begin{figure}
\begin{center}
\subfloat[Arbitrary common lines]{\includegraphics[width=0.25\textwidth]{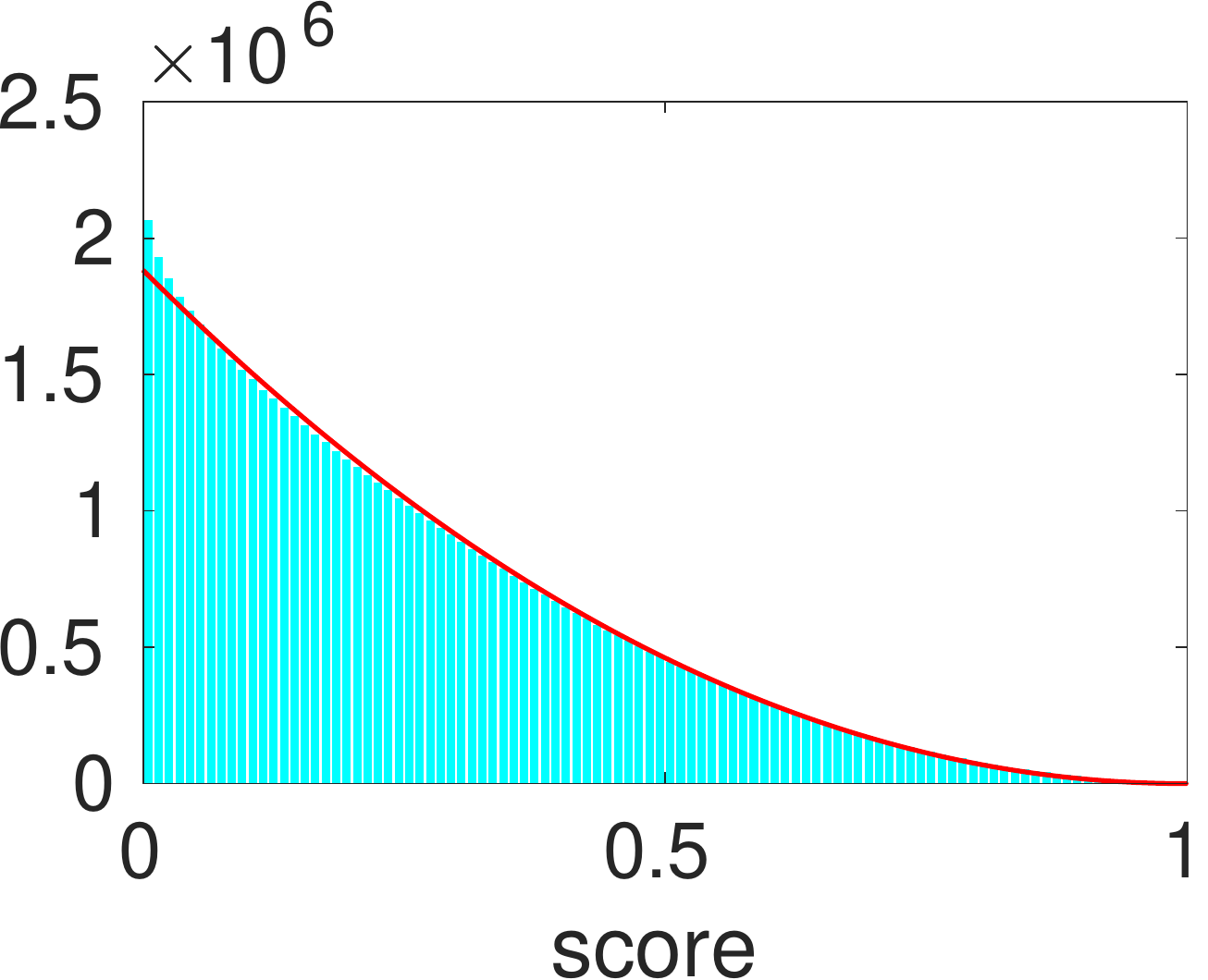}} \hfill
\subfloat[$\sigma=1^\circ$]{\includegraphics[width=0.25\textwidth]{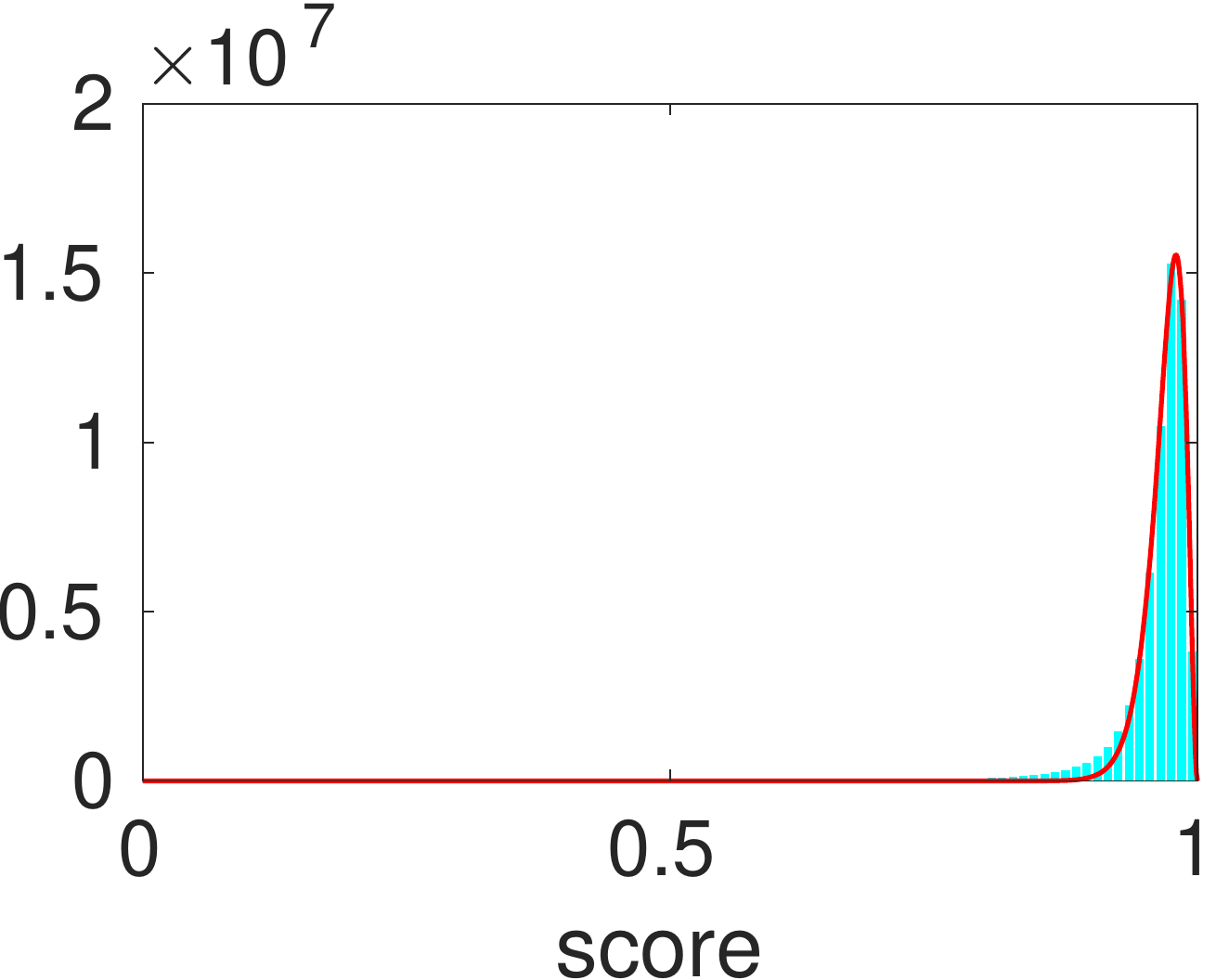}} \hfill
\subfloat[$\sigma=3^\circ$]{\includegraphics[width=0.25\textwidth]{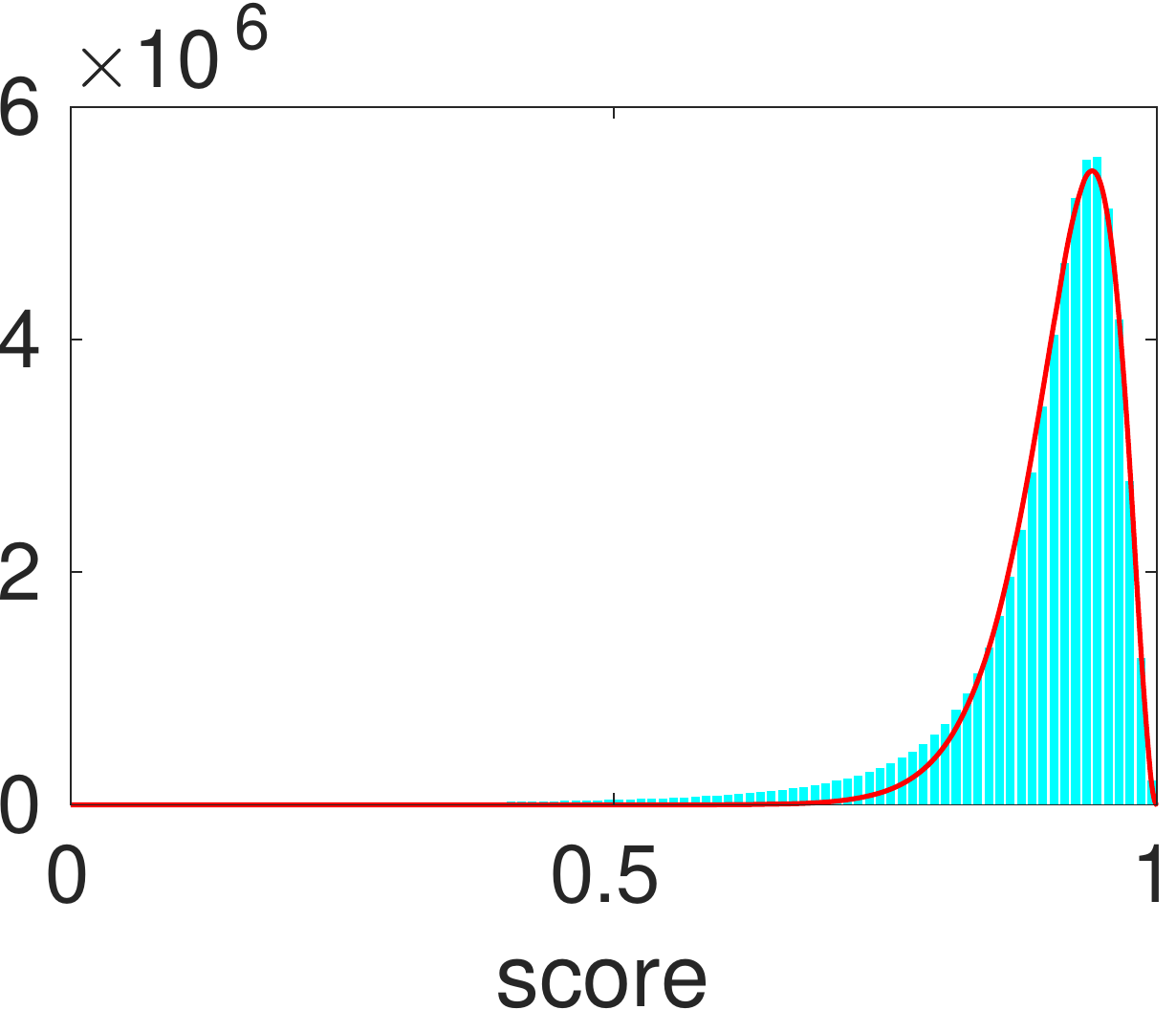}}\\
\subfloat[$\sigma=5^\circ$]{\includegraphics[width=0.25\textwidth]{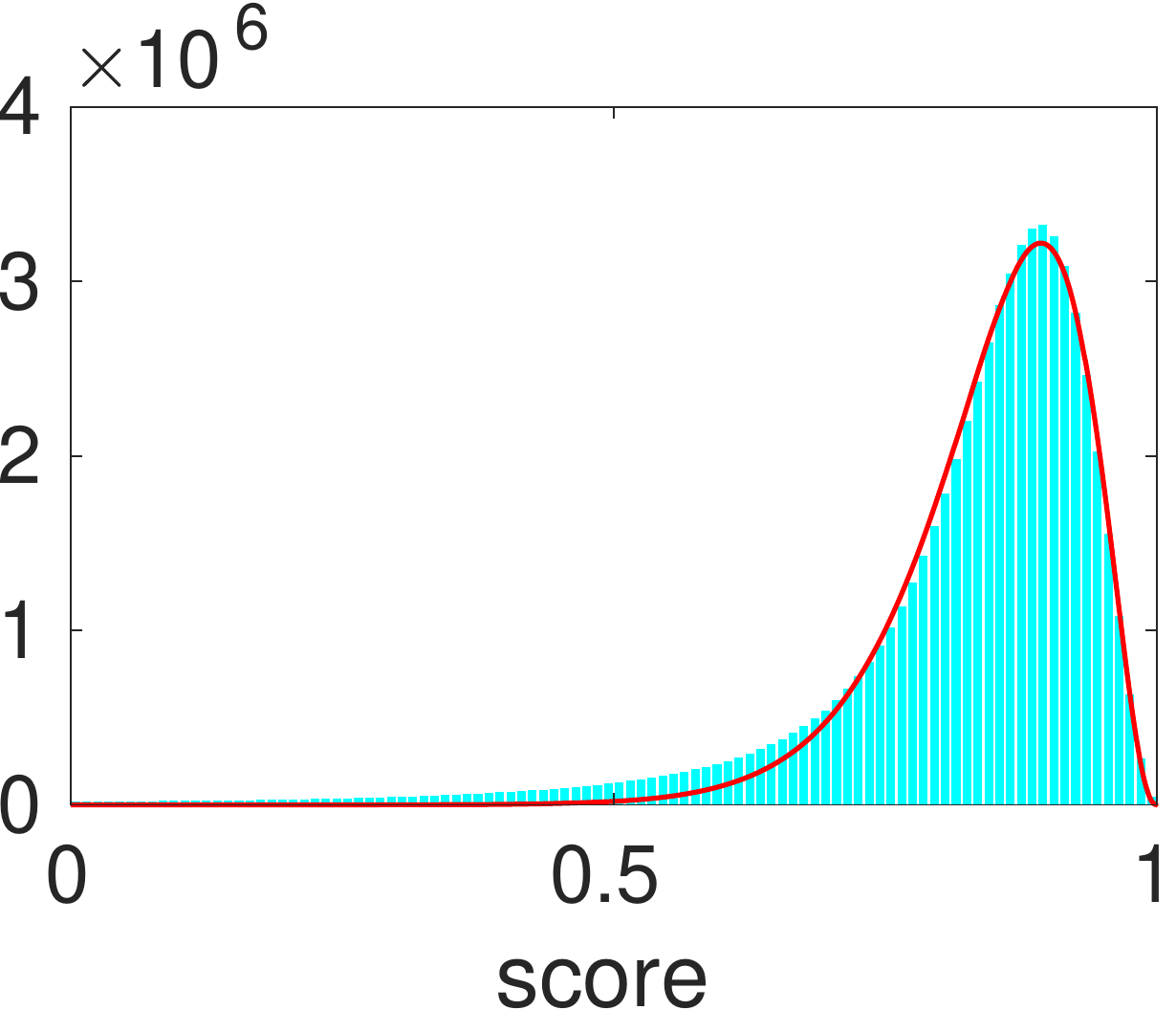}} \hfill
\subfloat[$\sigma=7^\circ$]{\includegraphics[width=0.25\textwidth]{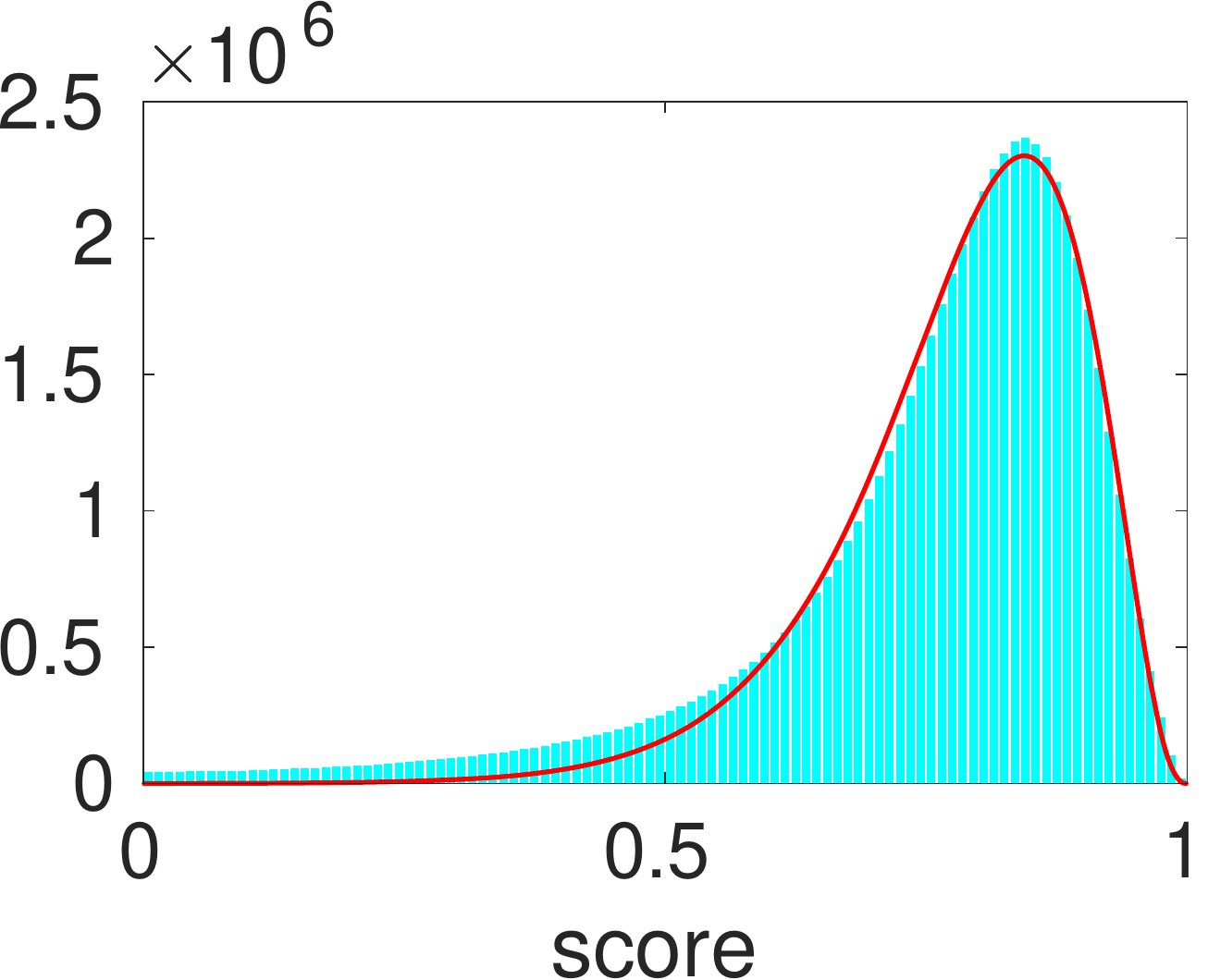}} \hfill
\subfloat[$\sigma=9^\circ$]{\includegraphics[width=0.25\textwidth]{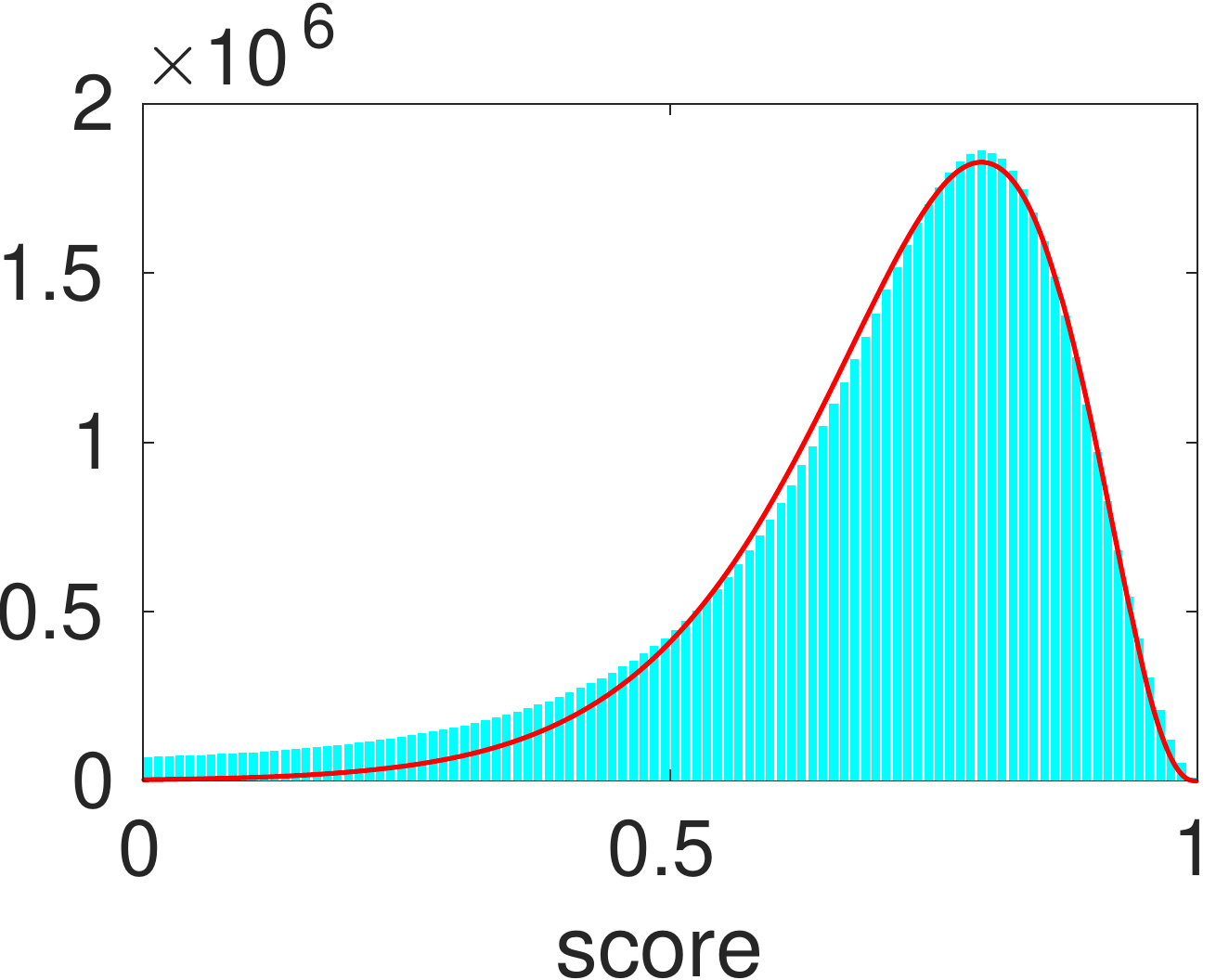}}\\
\subfloat[$\sigma=12^\circ$]{\includegraphics[width=0.25\textwidth]{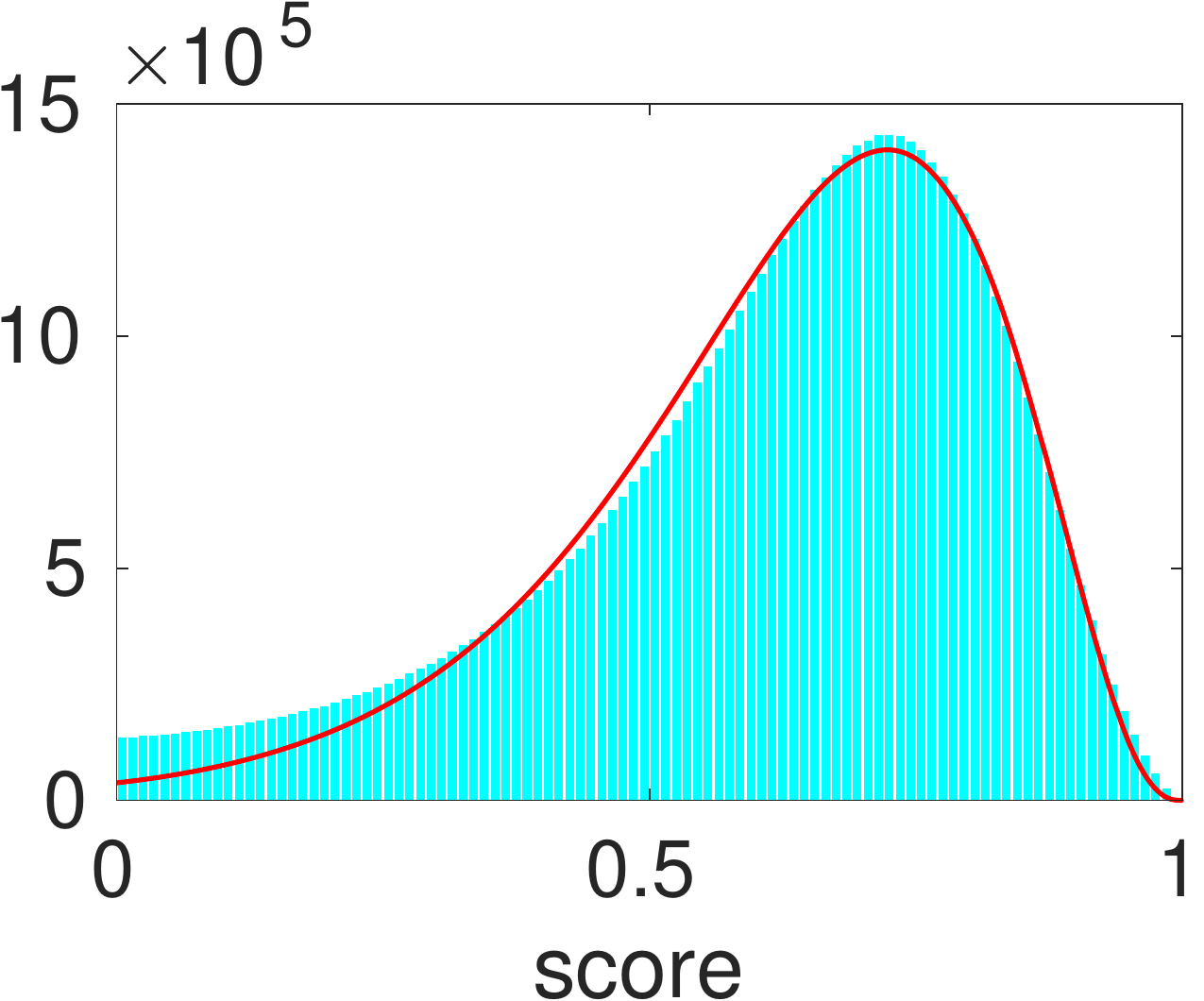}} \hfill
\subfloat[$\sigma=15^\circ$]{\includegraphics[width=0.25\textwidth]{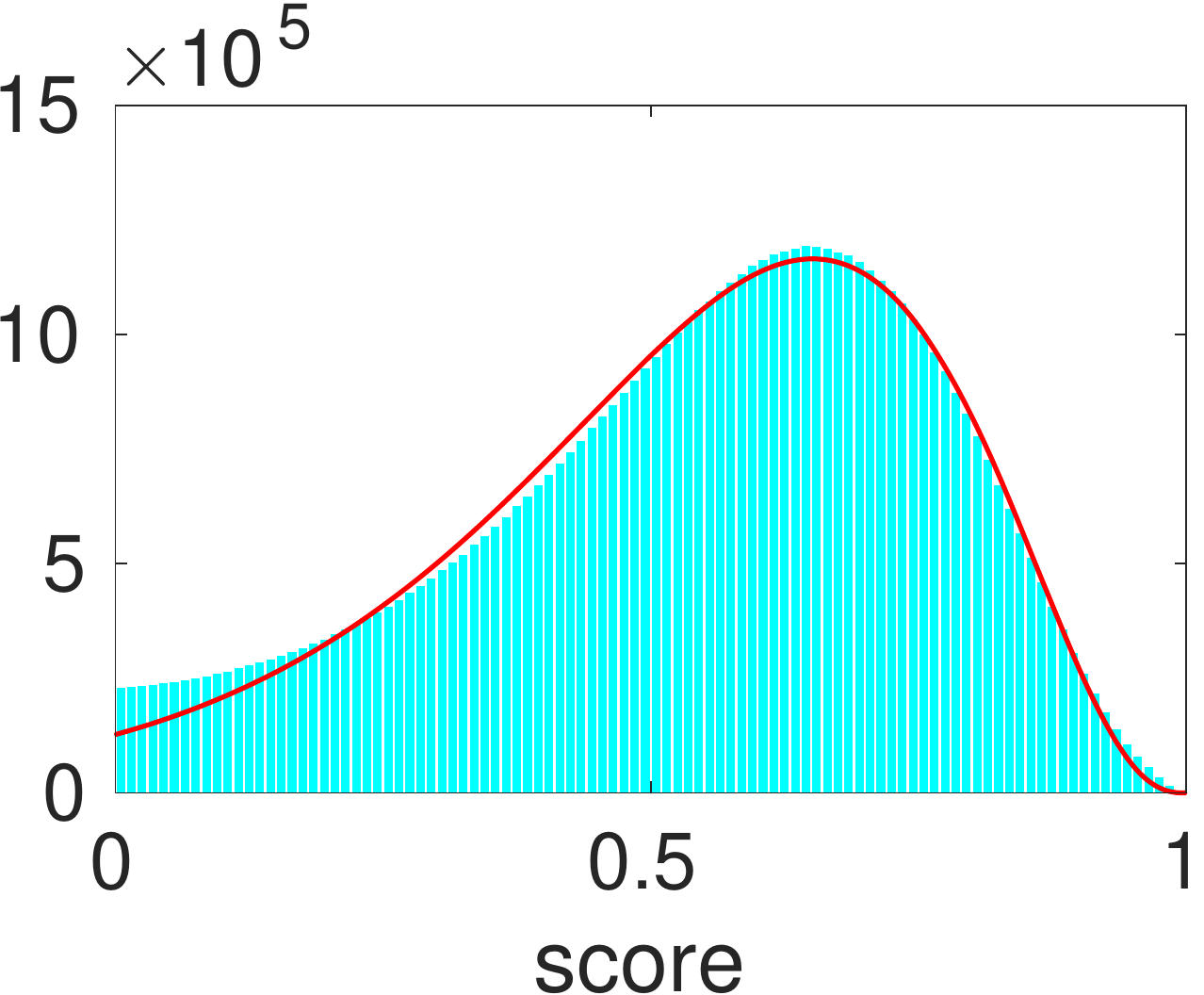}} \hfill
\subfloat[$\sigma=20^\circ$]{\includegraphics[width=0.25\textwidth]{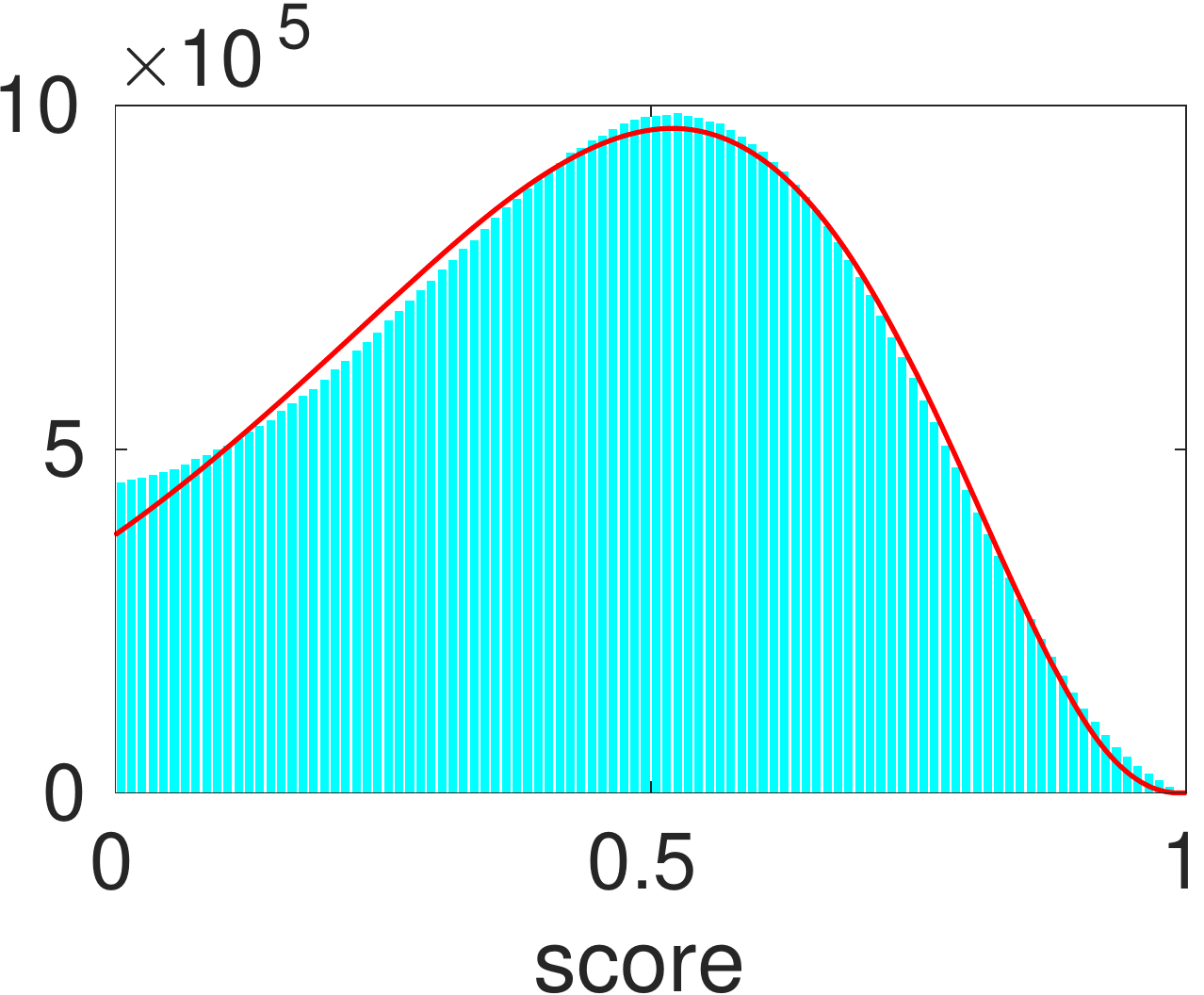}}

\caption{Histograms of triplets' scores for common lines simulated according to the errors model of Section~\ref{sec:errors_model}, for either arbitrary common lines, or indicative common lines with varying deviations $\sigma$.}
\label{scores_analysis}
\end{center}
\end{figure}

We start by analyzing the distribution of $s_{ijk}$ of~\eqref{eq:score} using common lines data simulated according to the model of Section~\ref{sec:errors_model}. We later show that this analysis agrees well also with experimental common lines data. For common lines simulated according to the model of Section~\ref{sec:errors_model}, we observed two types of triplets' scores histograms: one for triplets of three indicative common lines (\textit{indicative triplets}), and one for triplets where at least one common line is arbitrary (\textit{arbitrary triplets}). Figure~\ref{scores_analysis} shows several histograms corresponding to indicative triplets ($P=100\%$ in terms of Section~\ref{sec:errors_model}) with varying standard deviations $\sigma$ of the common lines, and one histogram corresponding to arbitrary triplets ($P=0$). Figure~\ref{scores_analysis} also suggests that the probability density functions of the triplets' scores can be approximated by
\begin{linenomath*}
\begin{align}
f(s_{ijk}=x | \text{ triplet } ijk \text{ is indicative}) &= B \cdot (1-x)^{\beta}\cdot e^{-\frac{\beta}{\hat{\sigma}} (1-x)}, \label{eq:fijk_ind} \\
f(s_{ijk}=x | \text{ triplet } ijk \text{ is arbitrary}) &= (\alpha+1)\cdot (1-x)^\alpha, \label{eq:fijk_arb}
\end{align}
\end{linenomath*}
where {$1-\hat{\sigma}$} (whose relation to the noise level {$\sigma$} will be clarified shortly) is the location of the maximum of~{\eqref{eq:fijk_ind}}, and {$B$} is a normalization constant. \rda{such that~{\eqref{eq:fijk_ind}} integrates to 1 (as required from a probability density function).}{Note that while {\eqref{eq:fijk_arb}} integrates to 1 (as required from a probability density function), analytic normalization of {\eqref{eq:fijk_ind}} requires computation of the complete gamma function. We avoid it by using the normalization constant~$B$ and computing it numerically when needed.}\footnote{\ra{While {\eqref{eq:fijk_ind}} and {\eqref{eq:fijk_arb}} bare some similarity to the beta and gamma distributions, this similarity was observed retrospectively and played no role in our modeling nor in our derivations.}}

\begin{figure}
	\begin{center}
		\includegraphics[width=0.3\textwidth]{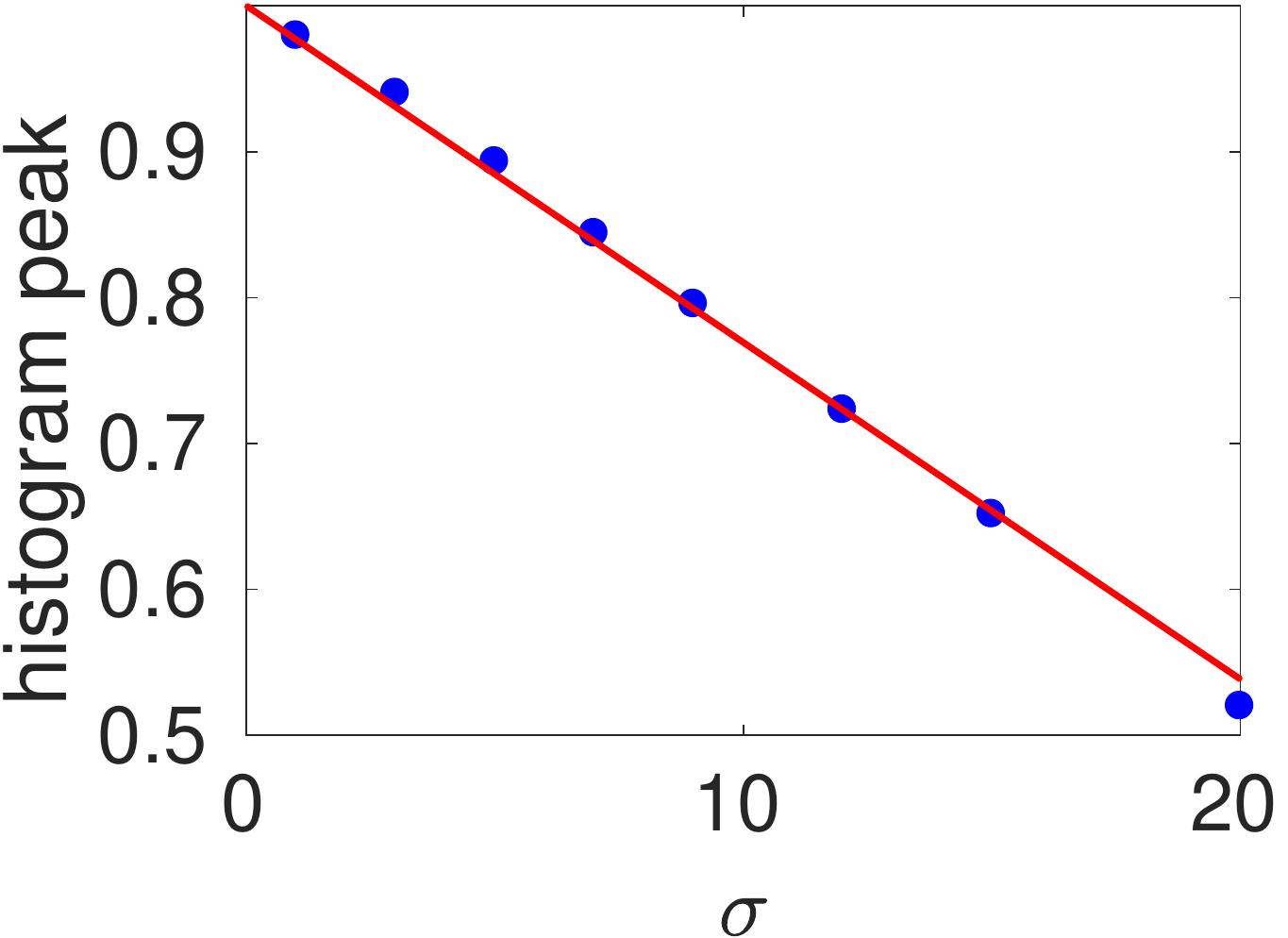}
		\caption{The peak $1-\hat{\sigma}$ of the triplets' scores histogram (for indicative triplets) vs. the deviation $\sigma$ in the estimated common lines.}
		\label{x0_vs_sigma}
	\end{center}
\end{figure}

By estimating the parameter in~\eqref{eq:fijk_arb}, we get $\alpha \approx 2.03$ with $R^2=0.99$. As~\eqref{eq:fijk_ind} depends on the noise level $\sigma$, Table~{\ref{scores_parameters}} presents the parameters (except for the normalization $B$) that were estimated for each value of $\sigma$. Note that $\sigma$ can be deduced from the location of the peak of~\eqref{eq:fijk_ind} (given by $1-\hat{\sigma}$). In particular, from Figure~\ref{x0_vs_sigma} we conclude the relation $\hat{\sigma}=a\cdot\sigma$ for $a=2.31\cdot 10^{-2}$, with the quality of the fit given by $R^2=0.99$.

\begin{table}
\begin{center}
\begin{tabular}{c|ccc}
$\sigma$ & $\beta$ & $1-\hat{\sigma}$ & $R^2$
\\ \hline \\
$1^{\circ}$ & 2.09 & 0.980 & 0.99 \\
$3^{\circ}$ & 2.19 & 0.940 & 0.99 \\
$5^{\circ}$ & 2.37 & 0.893 & 0.99 \\
$7^{\circ}$ & 2.55 & 0.844 & 0.99 \\
$9^{\circ}$ & 2.71 & 0.796 & 0.99 \\
$12^{\circ}$ & 2.73 & 0.723 & 0.99 \\
$15^{\circ}$ & 2.72 & 0.651 & 0.99 \\
$20^{\circ}$ & 2.71 & 0.520 & 0.99 \\
\end{tabular}
\caption{The parameters estimated by fitting the triplets' scores histograms of Figure~\ref{scores_analysis} to the model~\eqref{eq:fijk_ind}. The histograms were generated from indicative common lines simulated according to the model of Section~\ref{sec:errors_model}, with varying values of $\sigma$.}
\label{scores_parameters}
\end{center}
\end{table}

Generally, a histogram of triplets' scores consists of a mixture of both indicative and arbitrary triplets, since $P$ always satisfies $0 < P < 1$ (see also Figure~\ref{decomposition_test} below).
Such a mixed histogram is the sum of two ``sub-histograms'' -- one of the scores of the indicative triplets, and one of those of the arbitrary triplets.
The integral over each such sub-histogram is proportional to the fraction of triplets it contains, and hence, by decomposing a mixed histogram into its two sub-histograms, one can find the rate $P_{tri}$ of indicative triplets. Note that a triplet is indicative if and only if all its three common lines are indicative. Thus, by assuming independence between the three common lines, one has $P_{tri} = P^3$, from which the rate $P$ of indicative common lines can be extracted.

According to the last observations, the model of triplets' scores allows us to extract both parameters $P$ and $\sigma$ corresponding to a given set of common lines, provided we can reliably decompose a given scores' histogram into its two components. In order to verify that this is indeed feasible, we used Matlab's curve fitting toolbox to fit histograms of triplets' scores based on $\binom{500}{2}$ common lines simulated according to the model of Section~\protect\ref{sec:errors_model}, using $\sigma =7^\circ$ and various values of $P$.
Figure~\ref{decomposition_test} demonstrates the fits that were computed (by combining the models of~\protect\eqref{eq:fijk_ind} and~\protect\eqref{eq:fijk_arb}), and Table~\ref{decompositions} presents the estimated parameters $P$ and $\sigma$. It seems that the estimation of these parameters holds rather well for $P \gtrapprox 20\%$. For lower rates, however, the component in the histogram corresponding to the indicative triplets is apparently too small to be resolved and well-approximated. Note that for $P<20\%$ we have that $P_{tri} \approx P^3 < 1\%$, hence it is not surprising that the component corresponding to indicative triplets fails to be identified correctly.

\begin{figure}
\begin{center}
\subfloat[$P=15\%$]{\includegraphics[width=0.25\textwidth]{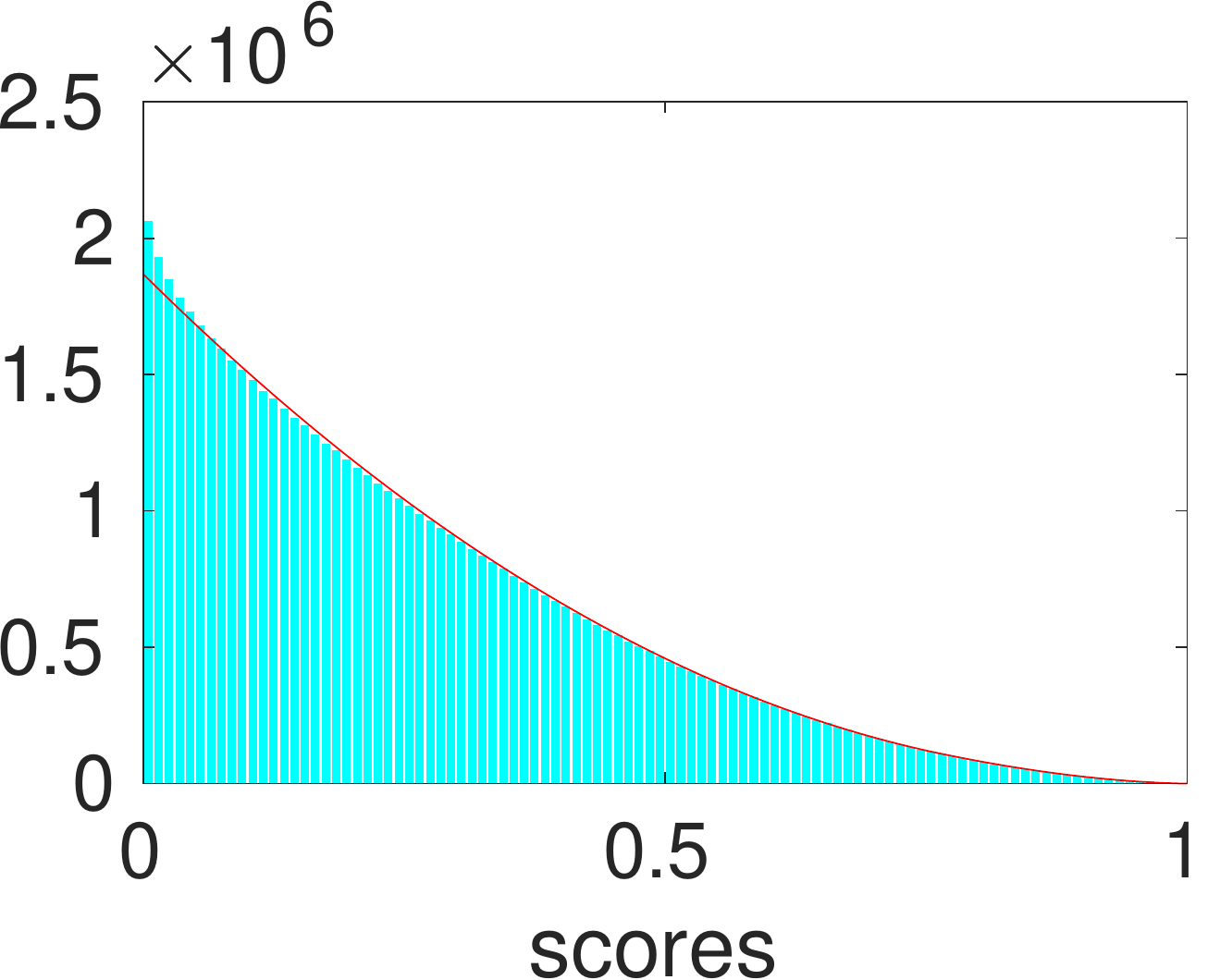}}
\hfill
\subfloat[$P=20\%$]{\includegraphics[width=0.25\textwidth]{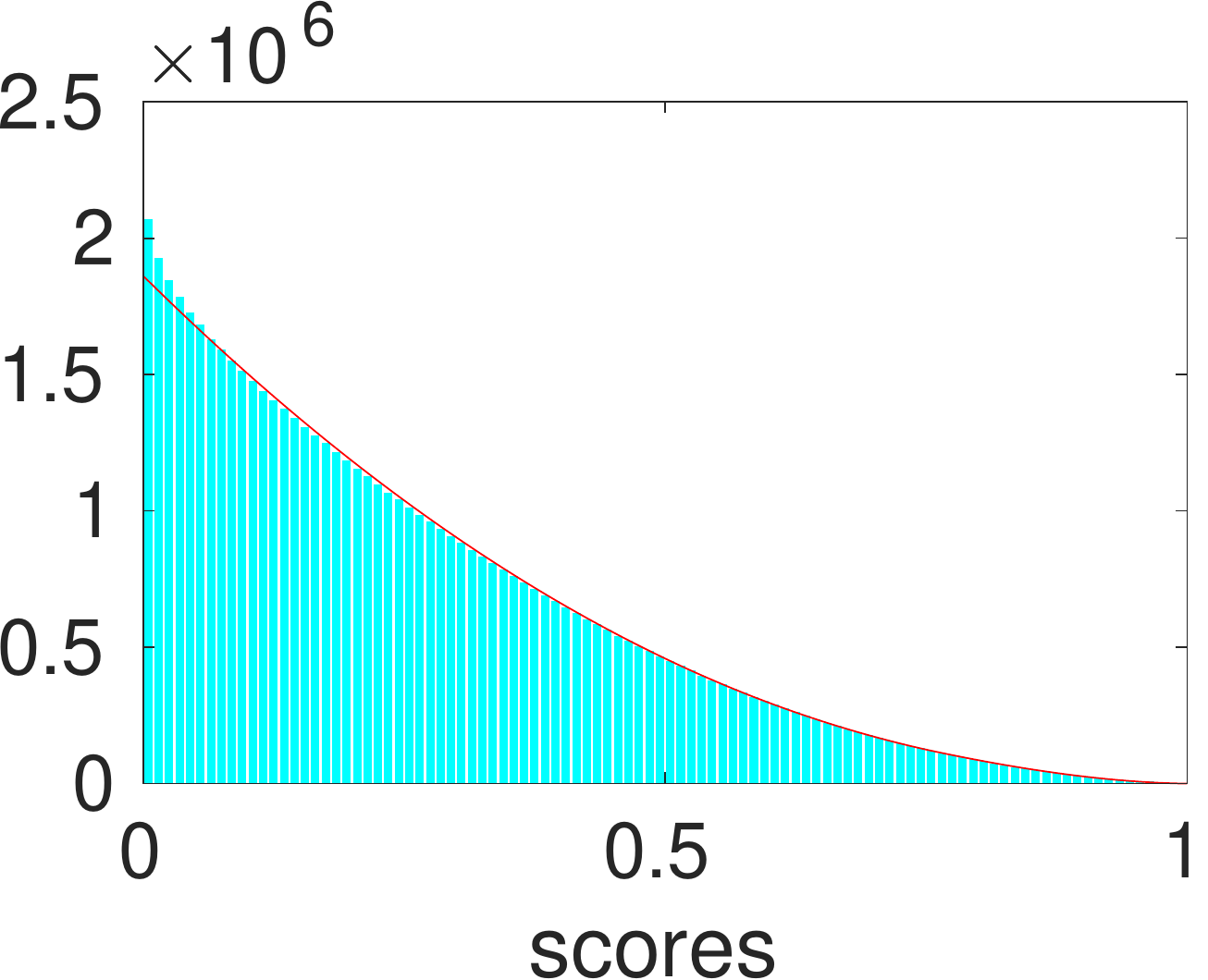}}
\hfill
\subfloat[$P=25\%$]{\includegraphics[width=0.25\textwidth]{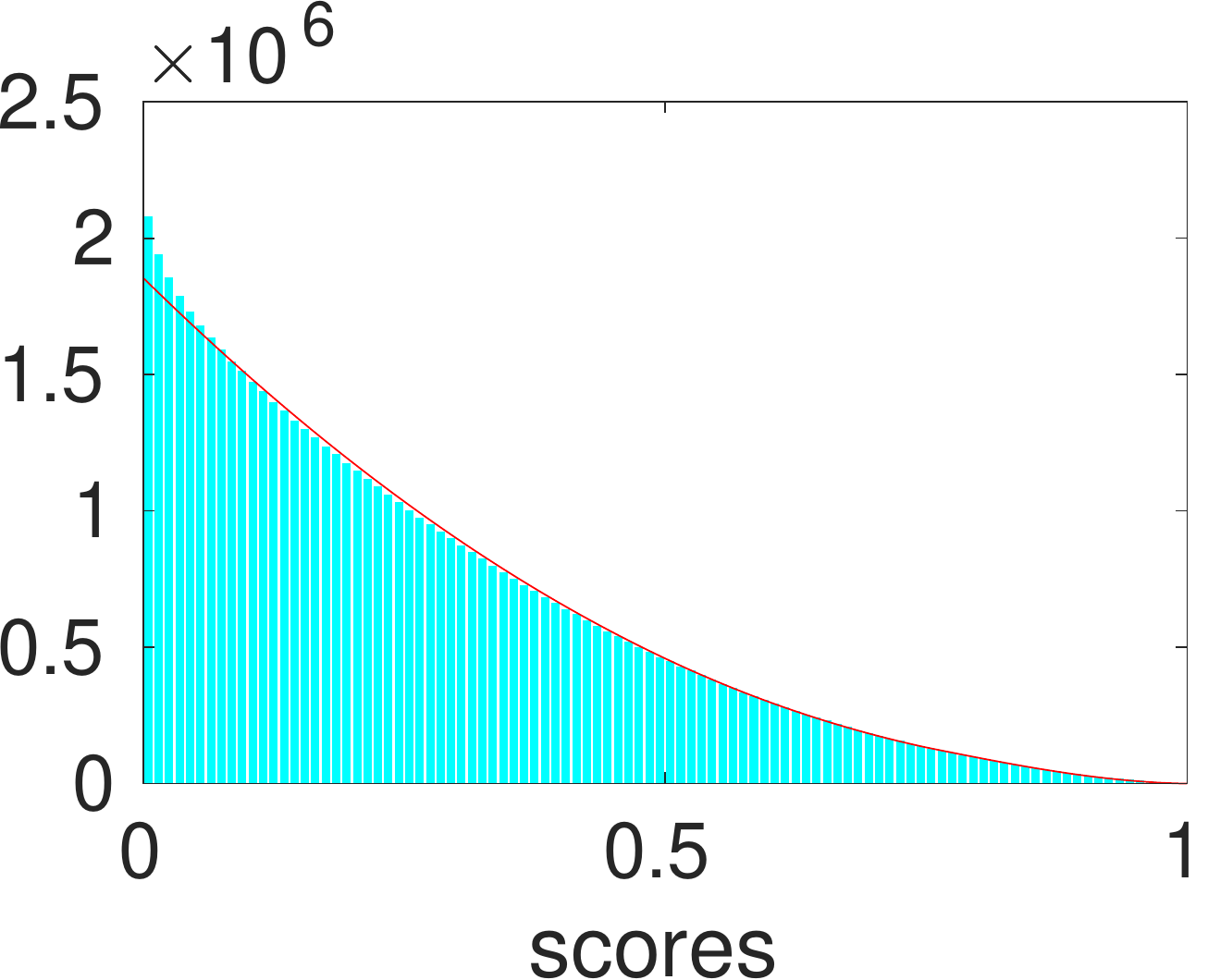}}\\
\subfloat[$P=30\%$]{\includegraphics[width=0.25\textwidth]{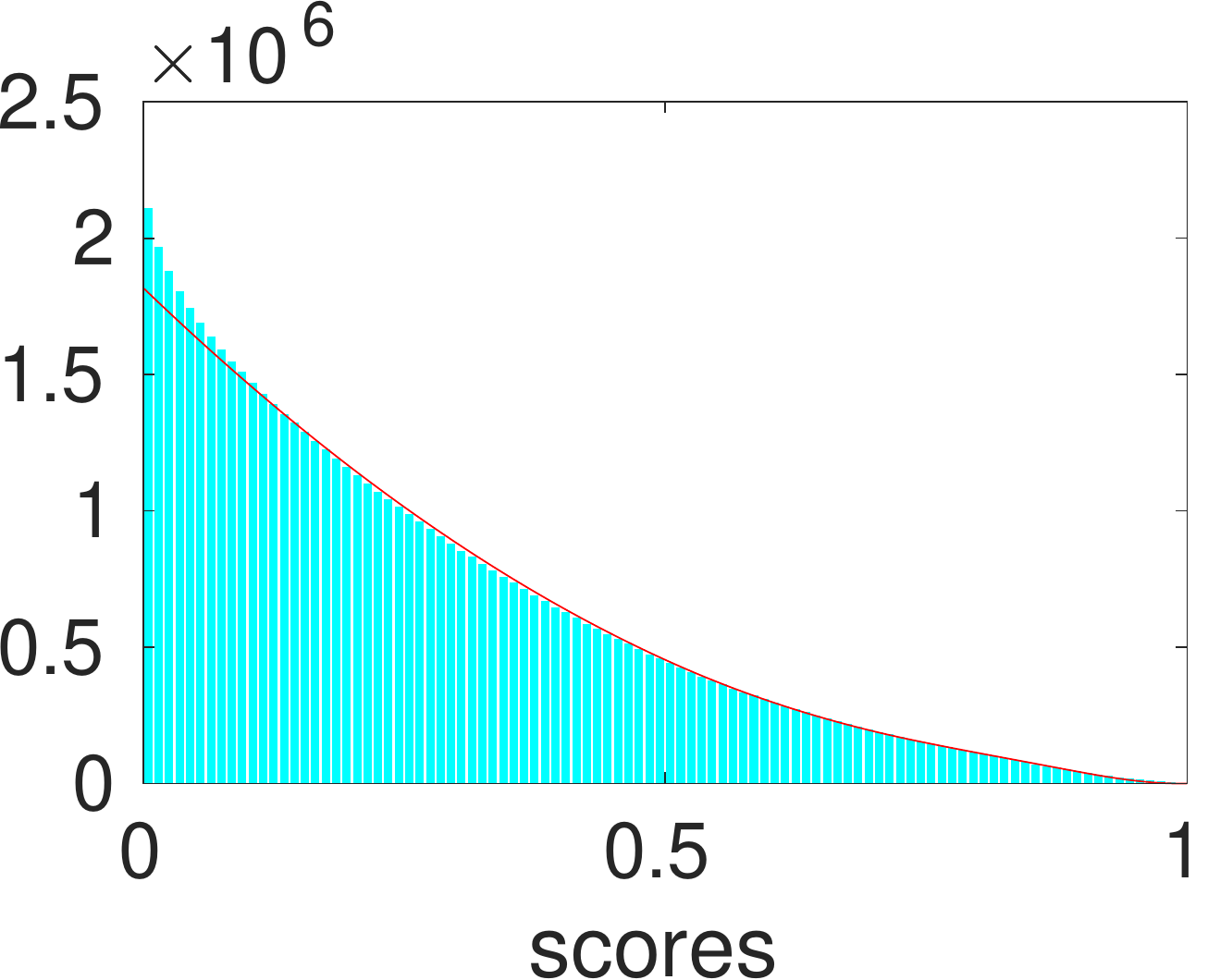}}
\hfill
\subfloat[$P=35\%$]{\includegraphics[width=0.25\textwidth]{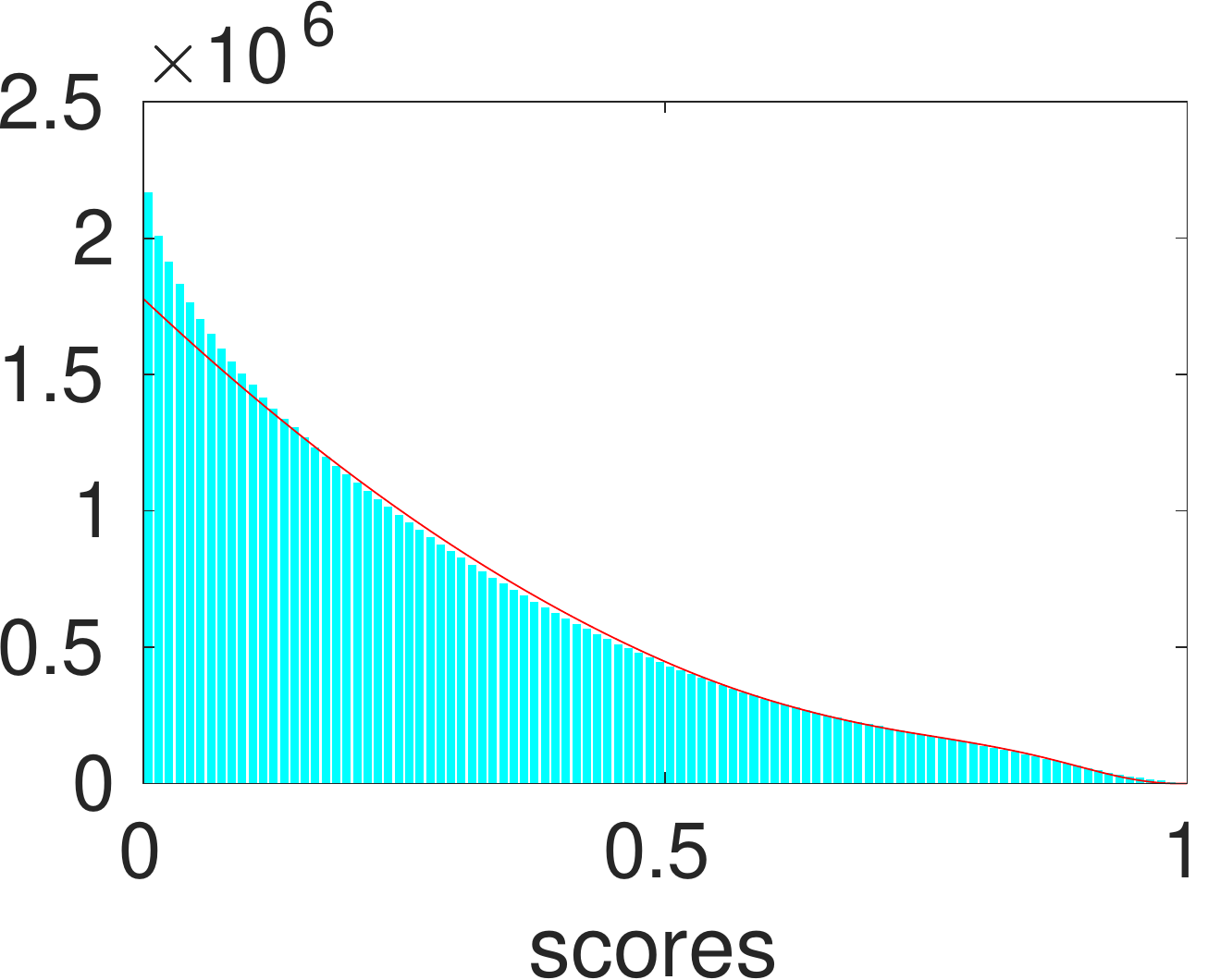}}
\hfill
\subfloat[$P=45\%$]{\includegraphics[width=0.25\textwidth]{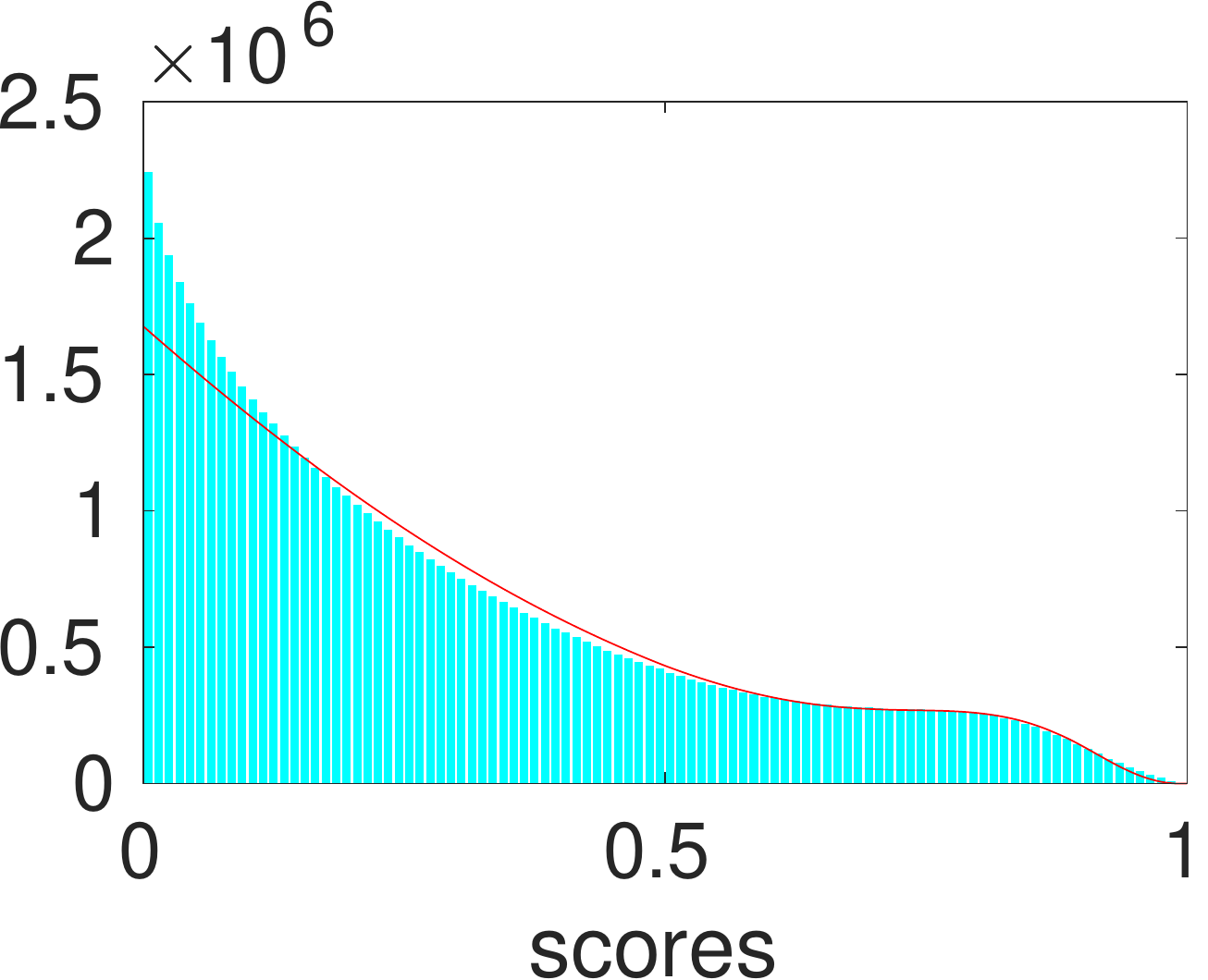}}\\
\subfloat[$P=55\%$]{\includegraphics[width=0.25\textwidth]{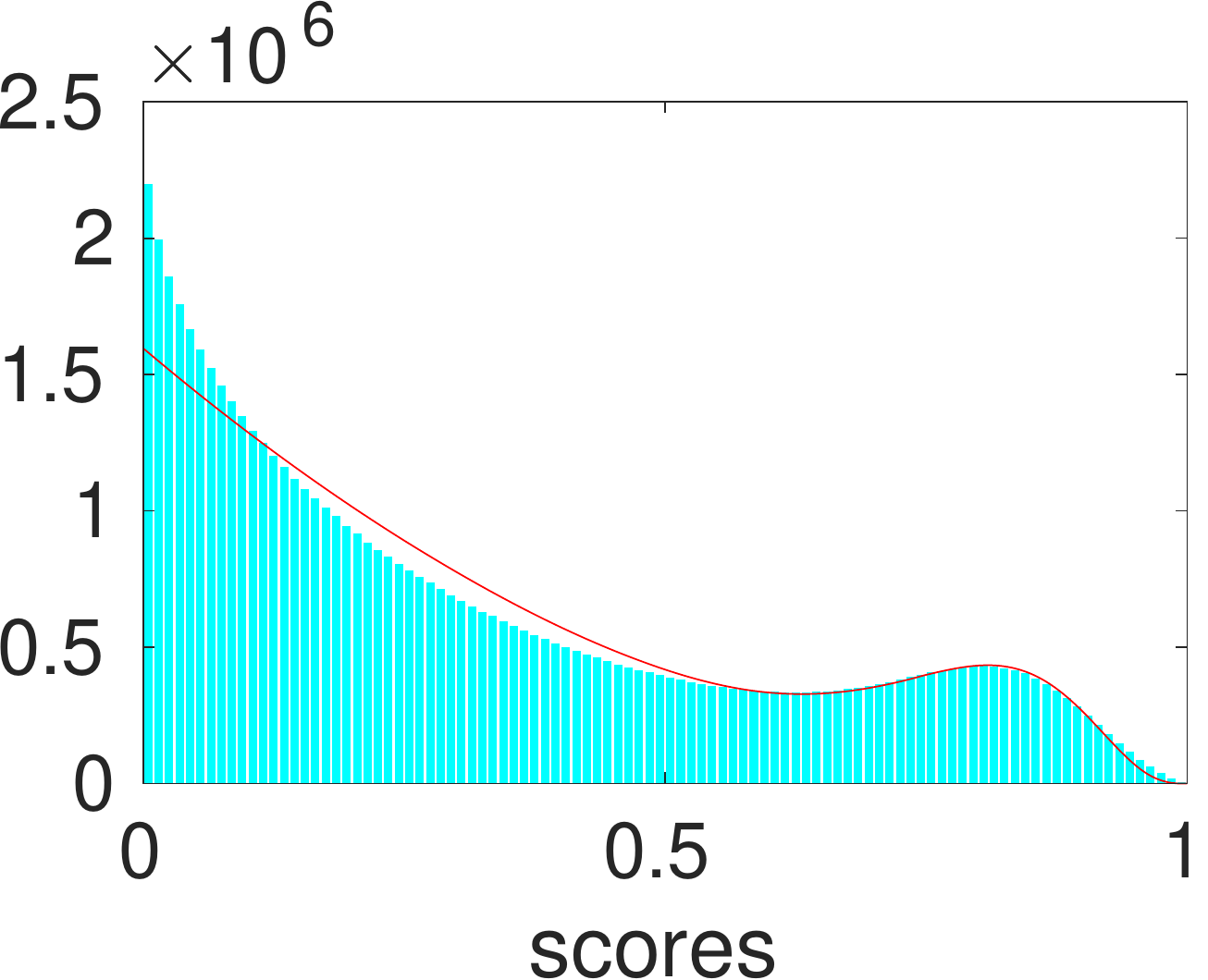}}
\hfill
\subfloat[$P=65\%$]{\includegraphics[width=0.25\textwidth]{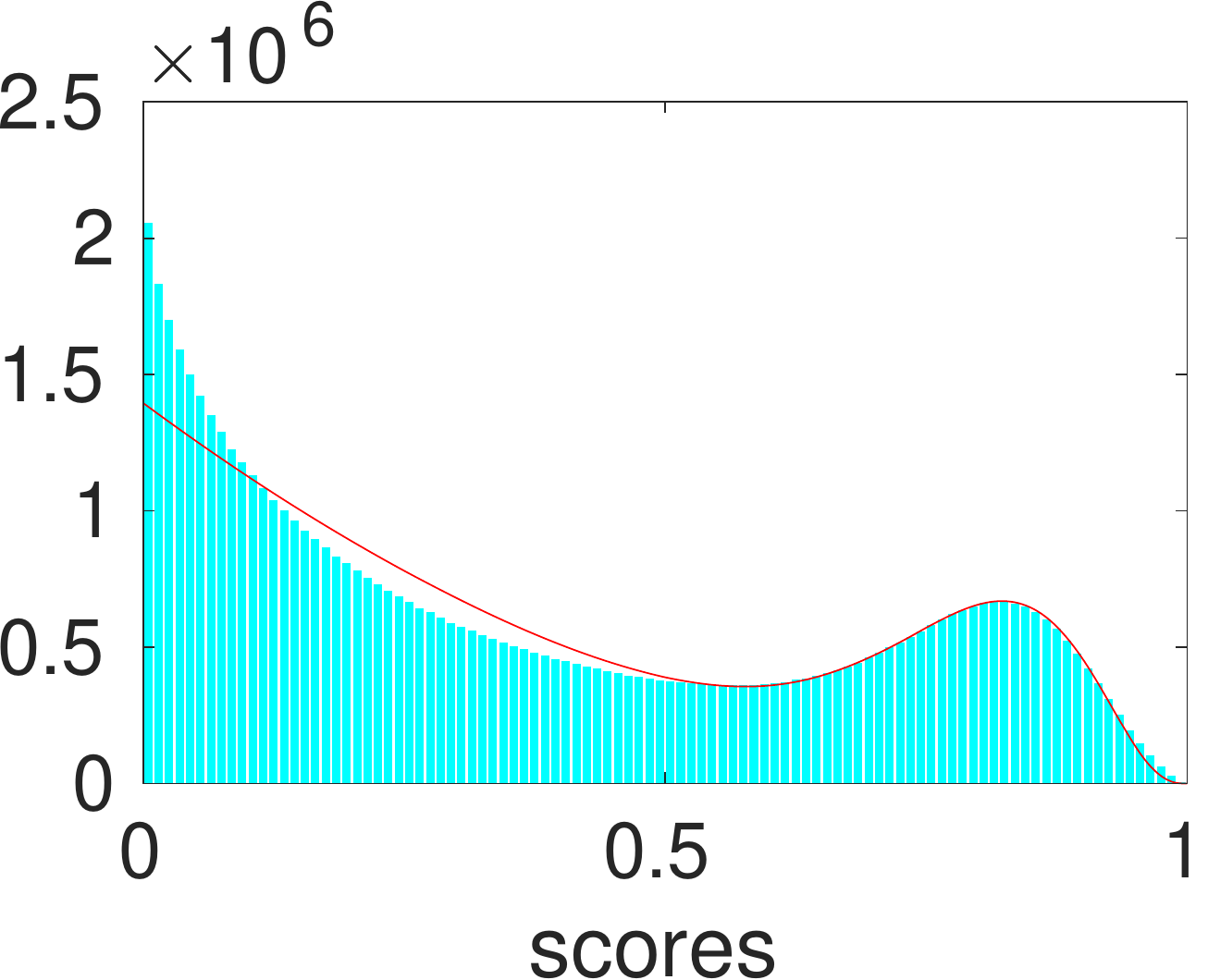}}
\hfill
\subfloat[$P=75\%$]{\includegraphics[width=0.25\textwidth]{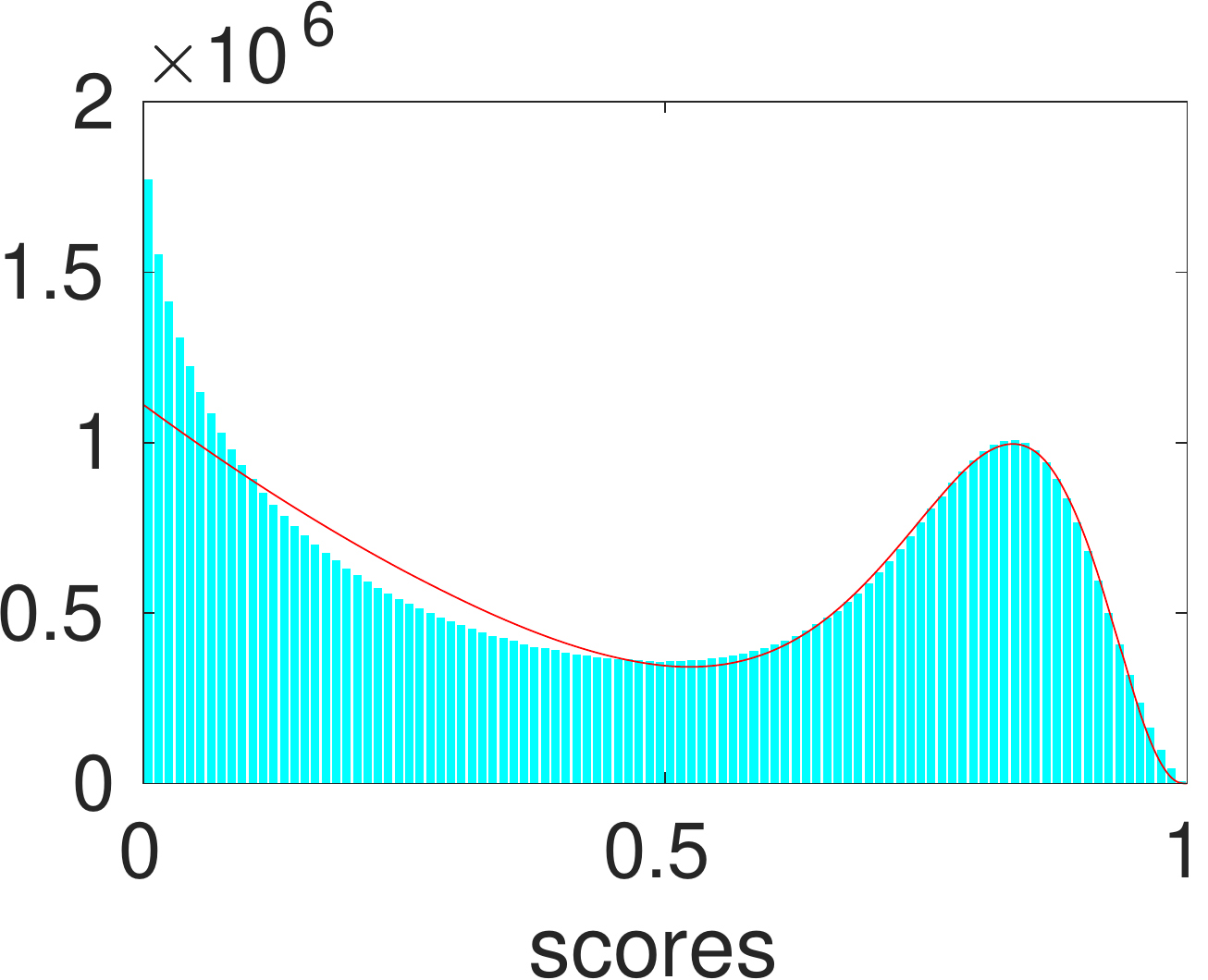}}

\caption{Triplets' scores histograms (based on simulated common lines with deviation $\sigma=7^\circ$ of the indicative lines) and their fits, based on decomposition into a combination of~\eqref{eq:fijk_ind} and~\eqref{eq:fijk_arb}, for varying rates $P$ of indicative common lines.}
\label{decomposition_test}
\end{center}
\end{figure}

\begin{table}
\begin{center}
\begin{tabular}{c|cc}
$P$  & Estimated $P$ & Estimated $\sigma$
\\ \hline
75\% & 74.2\% & 7.1 \\
65\% & 63.7\% & 7.3 \\
55\% & 53.3\% & 7.6 \\
45\% & 47.7\% & 8.0 \\
35\% & 38.0\% & 8.6 \\
30\% & 32.4\% & 8.9 \\
25\% & 24.4\% & 8.6 \\
20\% & 22.1\% & 8.2 \\
15\% & 19.2\% & 7.2 \\
\end{tabular}
\caption{Estimation of the parameters $P$ and $\sigma$ from the histograms in Figure~\ref{decomposition_test}, for simulated common lines generated using varying rates $P$ of indicative common lines, with deviation $\sigma=7^\circ$.}
\label{decompositions}
\end{center}
\end{table}

\begin{figure}
	\begin{center}
		\subfloat[$K = 6$: $P=46\%$, $\sigma=4.2^\circ$]{
			\includegraphics[width=0.3\textwidth]{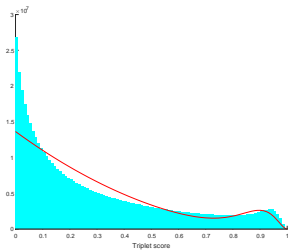}
		} \hspace{0.05\textwidth}
		\subfloat[$K = 5$: $P=39\%$, $\sigma=5.5^\circ$]{
			\includegraphics[width=0.3\textwidth]{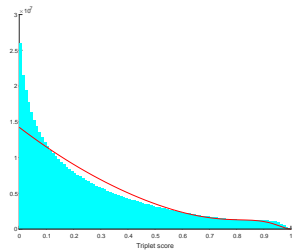}
		} \\
		\subfloat[$K = 4$: $P=31\%$, $\sigma=5.8^\circ$]{
			\includegraphics[width=0.3\textwidth]{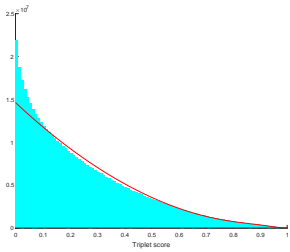}
		} \hspace{0.05\textwidth}
		\subfloat[$K = 3$: $P=26\%$, $\sigma=5.7^\circ$]{
			\includegraphics[width=0.3\textwidth]{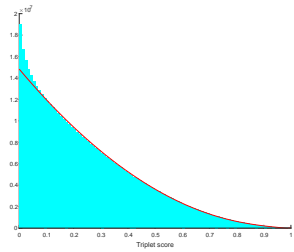}
		}
		
		\caption{Triplets' scores histograms and their fits, along with the estimated parameters $P$ and $\sigma$, for experimental class averages with a varying number $K$ of images per class.}
		\label{fig:real_scores}
	\end{center}
\end{figure}

Finally, Figure~\ref{fig:real_scores} demonstrates the histograms of triplets' scores for experimental class averages, as well as the fit of the model~\eqref{eq:fijk_ind}--\eqref{eq:fijk_arb} to the histograms.  Each histogram corresponds to $N=1000$ class averages of the Plasmodium falciparum 80S ribosome, generated as described in Section~{\ref{sec:examples}} below. In particular, each class average was generated by averaging $K$ properly aligned raw projection-images of size $179 \times 179$ pixels. The experiment was repeated for $3 \le K \le 6$, and the suggested model was found to match the triplets scores histograms with R-squared values of $0.81 \le R^2 \le 0.99$.
Note that the estimated $P$ in this case, which is a measure for the quality of the common lines data, consistently increases with $K$, in accordance with the objective of averaging to increase the SNR (signal-to-noise ratio) of the class averages. Similarly, the estimate for $\sigma$, which corresponds to the typical error in the indicative common lines, reduces with $K$.
Indeed, Section~\ref{sec:examples} below demonstrates higher reconstruction resolutions from class averages with larger $K$.

\subsection{From triplets' scores to common lines' reliabilities}
\label{sec:triplets2pairs}

In this subsection, we calculate the probability $P_{ij}$ of a common line $c_{ij}$ to be indicative, using the triplets' scores $\{s_{ijk}\}_{k \ne i,j}$ of~\eqref{eq:score} and the analysis of Section~\ref{sec:triplets_scores}.

According to~\eqref{eq:fijk_ind} and~\eqref{eq:fijk_arb}, we have the probability density functions of the triplets' scores for both indicative triplets and arbitrary ones, which we denote by $f_{ind\_tri}(s_{ijk})=f(s_{ijk}|\text{triplet } ijk \text{ is indicative})$ and $f_{arb\_tri}(s_{ijk})=f(s_{ijk}|\text{triplet } ijk \text{ is arbitrary})$, respectively. As mentioned above, both density functions integrate to 1 as required \--- \protect{\eqref{eq:fijk_ind}} numerically and {\eqref{eq:fijk_arb}} analytically. By Section~\ref{sec:triplets_scores}, we also have (under certain assumptions) the prior probability $P$ of a common line to be indicative. Hence, by assuming independence between the triplets' scores $\{s_{ijk}\}_{k \ne i,j}$ corresponding to an indicative common line $c_{ij}$, we have that their joint probability density function, denoted $f_{ind\_cl}$, is given by
\begin{equation}
\label{eq:ind_cl}
\begin{aligned}
f_{ind\_cl}(\{s_{ijk}\}_k) &= f(\{s_{ijk}\}_k|c_{ij} \text{ is indicative}) = \prod_k f(s_{ijk}|c_{ij} \text{ is indicative}) \\
&= \prod_k \bigl[ \mathbf{P}\big( \text{triplet } ijk \text{ is indicative } | \text{ } c_{ij} \text{ is indicative} \big) \cdot f_{ind\_tri}(s_{ijk}) \\ & \qquad\qquad + \mathbf{P}\big( \text{triplet } ijk \text{ is arbitrary } | \text{ } c_{ij} \text{ is indicative} \big) \cdot f_{arb\_tri}(s_{ijk}) \bigr] \\
&= \prod_k \bigl[ P^2 \cdot f_{ind\_tri}(s_{ijk}) + (1-P^2) \cdot f_{arb\_tri}(s_{ijk}) \bigr].
\end{aligned}
\end{equation}
In a similar fashion, the probability density function of triplets' scores corresponding to an arbitrary common line $c_{ij}$ (in which case the triplets are necessarily arbitrary, see Section~\ref{sec:triplets_scores}), denoted $f_{arb\_cl}$, is given by
\begin{align}
\label{eq:arb_cl}
f_{arb\_cl}(\{s_{ijk}\}_k) = f(\{s_{ijk}\}_k|c_{ij} \text{ is arbitrary}) = \prod_k f_{arb\_tri}(s_{ijk}).
\end{align}

According to~\eqref{eq:ind_cl} and~\eqref{eq:arb_cl}, the probability of a common line $c_{ij}$ to be indicative, assuming it corresponds to triplets with scores $\{s_{ijk}\}_k$, is given according to Bayes theorem by
\begin{equation}\label{eq:Pij}
P_{ij} = \mathbf{P} \left( c_{ij} \text{ is indicative} | \{s_{ijk}\}_k \right) = \frac{P \cdot f_{ind\_cl}(\{s_{ijk}\}_k)}{P \cdot f_{ind\_cl}(\{s_{ijk}\}_k) + (1-P) \cdot f_{arb\_cl}(\{s_{ijk}\}_k)}.
\end{equation}

\subsection{Weighting scheme}
\label{sec:weighting_scheme}

According to Section~\ref{sec:intro_algo}, the rotations $R_{1},\ldots,R_{N}$ \rd{of \ref{trash} }(\rda{corresponding to}{defining} the viewing directions of the projection-images) can be extracted from the eigenvectors of the matrix $S$ of~\eqref{eq:sync1}, which consists of the blocks $\{R_iR_j^{-1} \}_{i,j=1}^N$. Section~\ref{sec:intro_algo} also shows that the rotations $R_{1},\ldots,R_{N}$ can be extracted from the eigenvectors of any matrix that is obtained from $S$ by multiplying its blocks  by arbitrary non-negative weights, as long as the weights are correctly normalized.

In practice, we only have a noisy estimate $S^{est}$ of $S$, where each $3\times 3$ block $R_{ij}$ of $S^{est}$ is an estimate of $R_iR_j^{-1}$. To compute the estimated block $R_{ij}$, we \rda{use the fact that it can be determined by any triplet of common lines $(c_{ij},c_{jk},c_{ki})$, and apply}{use} the voting algorithm~\cite{voting}, which takes into account all \rda{possible $k$'s}{triplets $(c_{ij},c_{jk},c_{ki})$} ($k\ne i,j$) when computing $R_{ij}$. Thus, a necessary condition to estimate $R_{ij}$ is that the common line $c_{ij}$ is approximately correct, that is, indicative in the sense of \rda{Section~{\ref{sec:intro_algo}}}{Section~{\ref{sec:errors_model}}}. Accordingly, We would like to reduce the weights of blocks in $S^{est}$ corresponding to non-indicative common lines, so that they do not affect the estimation of the rotations. According to Section~\ref{sec:triplets2pairs}\ra{,} we only know the probability $P_{ij}$ of a common line to be indicative (rather than arbitrary), and thus, we would like the weight $w_{ij}$ of $R_{ij}$ in~\eqref{eq:sync2} to increase with $P_{ij}$. The simplest corresponding choice of weights that complies with the required normalization is
\begin{equation}\label{eq:wij}
w_{ij} = N \cdot \frac{P_{ij}}{\sum_{k\ne i} P_{ik}},\qquad 1 \le i,j \le N.
\end{equation}

We thus propose to incorporate the weighting scheme of~\eqref{eq:wij} into the reconstruction algorithm of~\cite{sync3N}. The benefits of the resulting algorithm are demonstrated in Section~\ref{sec:examples}.

\section{Numerical examples}
\label{sec:examples}

As explained in Section~\ref{sec:background} and Section~\ref{sec:errors_indications}, we suggest in this paper a reference-free ab-initio reconstruction algorithm, which is based on the algorithm of~\cite{sync3N} along with the weighting scheme suggested in~\eqref{eq:wij}.
In this section we demonstrate the algorithm on several sets of class averages, generated from two different data sets of raw projection-images. In Section~\ref{sec:80S} \rda{We}{we} apply the algorithm on class averages of the Plasmodium falciparum 80S ribosome generated from the EMPIAR-10028 data set~\cite{EMPIAR10028}. In Section~\ref{sec:yeast} we apply the algorithm on the EMPIAR-10073 data set~\cite{EMPIAR10073}. In Section~\ref{sec:noise robustness} we demonstrate the robustness of the algorithm to noise by applying it on very noisy class averages.\rd{ In Section~\ref{sec:bad images} we demonstrate the robustness of the algorithm to ``bad'' class averages that do not correspond to any view of the molecule.} In Section~\ref{sec:reliability} we demonstrate that the algorithm provides measures that allow to assess the reliability of the resulting ab-initio model. Finally, in Section~\ref{sec:running time} we provide some details on the implementation and running time of the algorithm.

\subsection{Plasmodium falciparum 80S dataset}
\label{sec:80S}
\begin{figure}
	\begin{center}
		\subfloat{
			\includegraphics[width=0.23\textwidth]{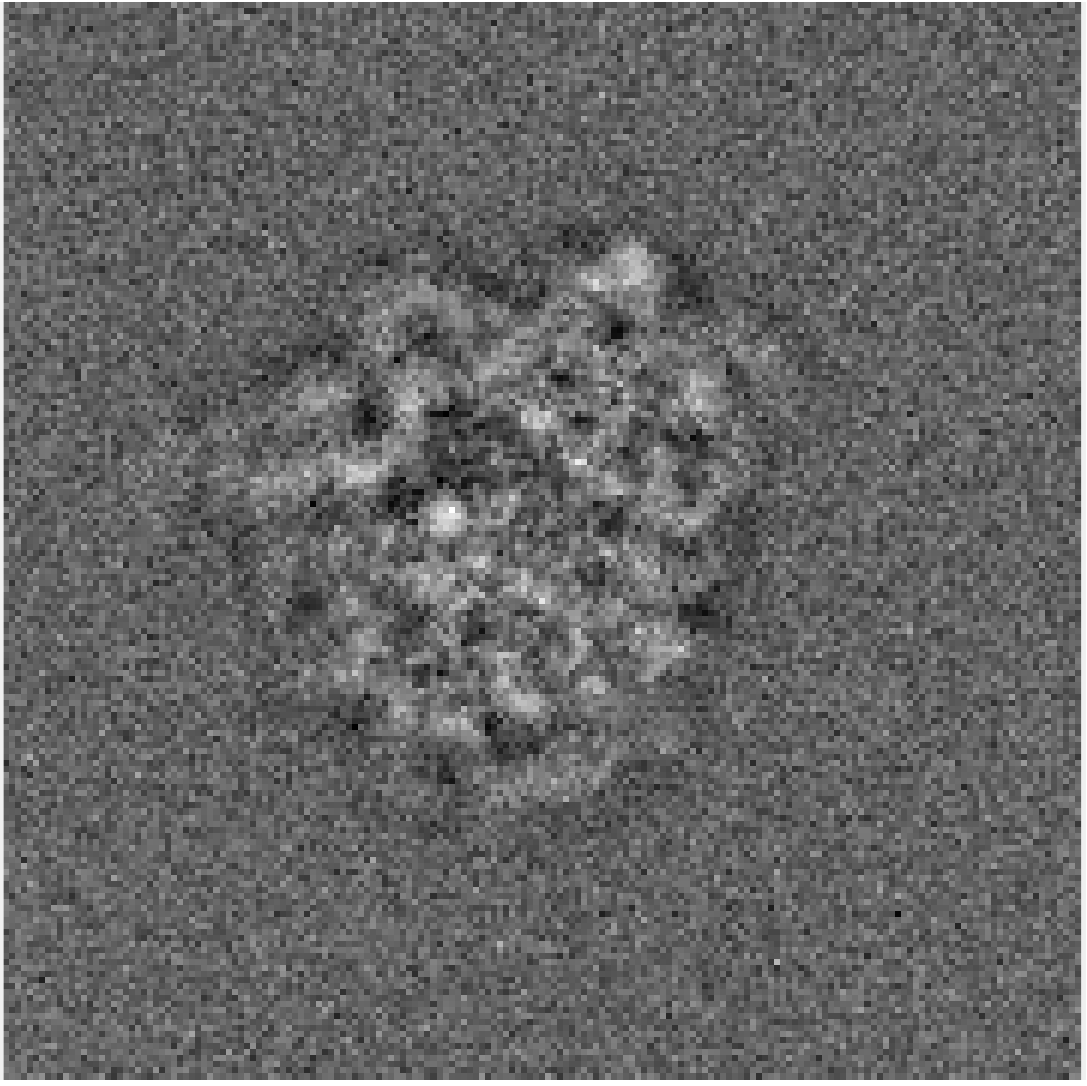}
		} 
		\subfloat{
			\includegraphics[width=0.23\textwidth]{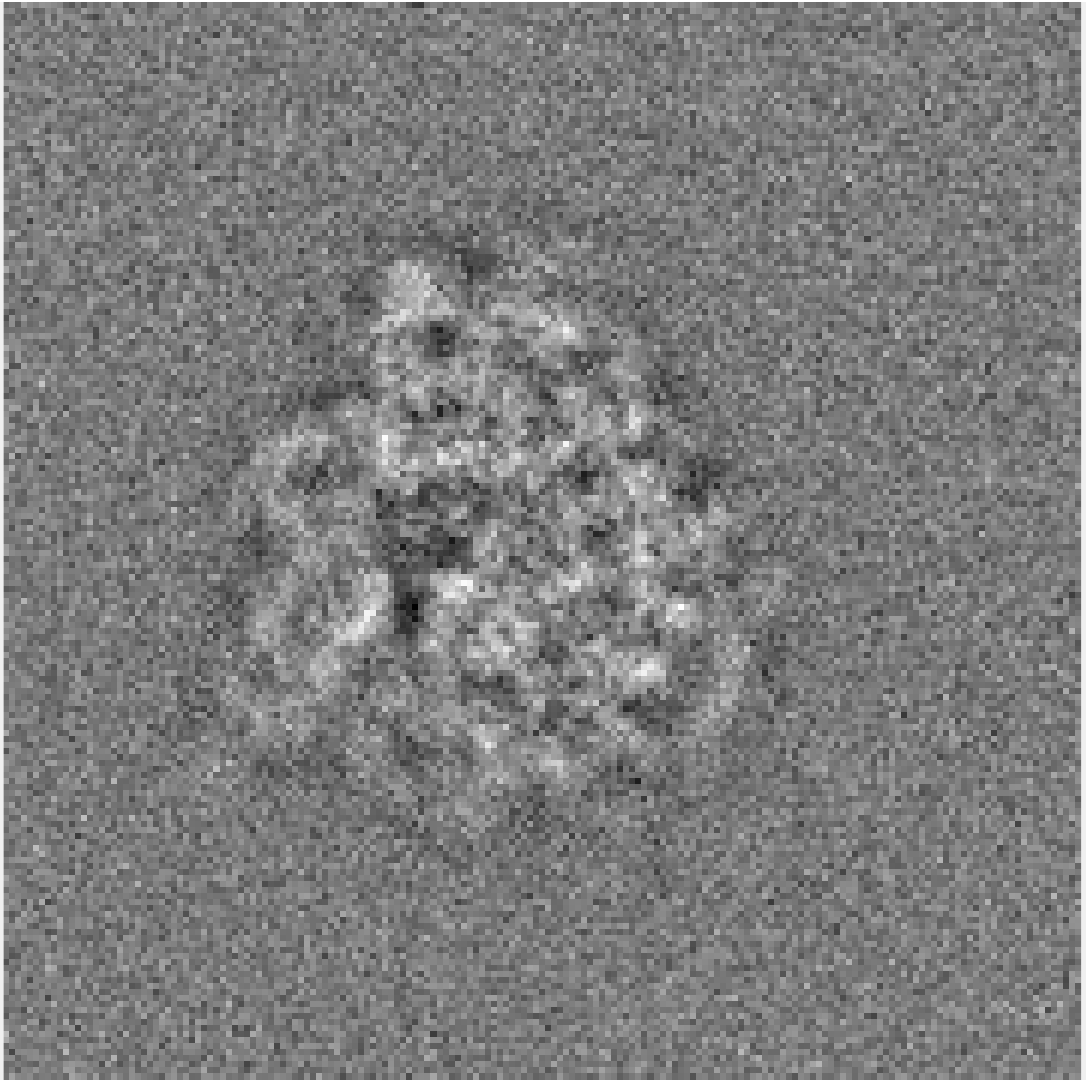}
		} 
		\subfloat{
			\includegraphics[width=0.23\textwidth]{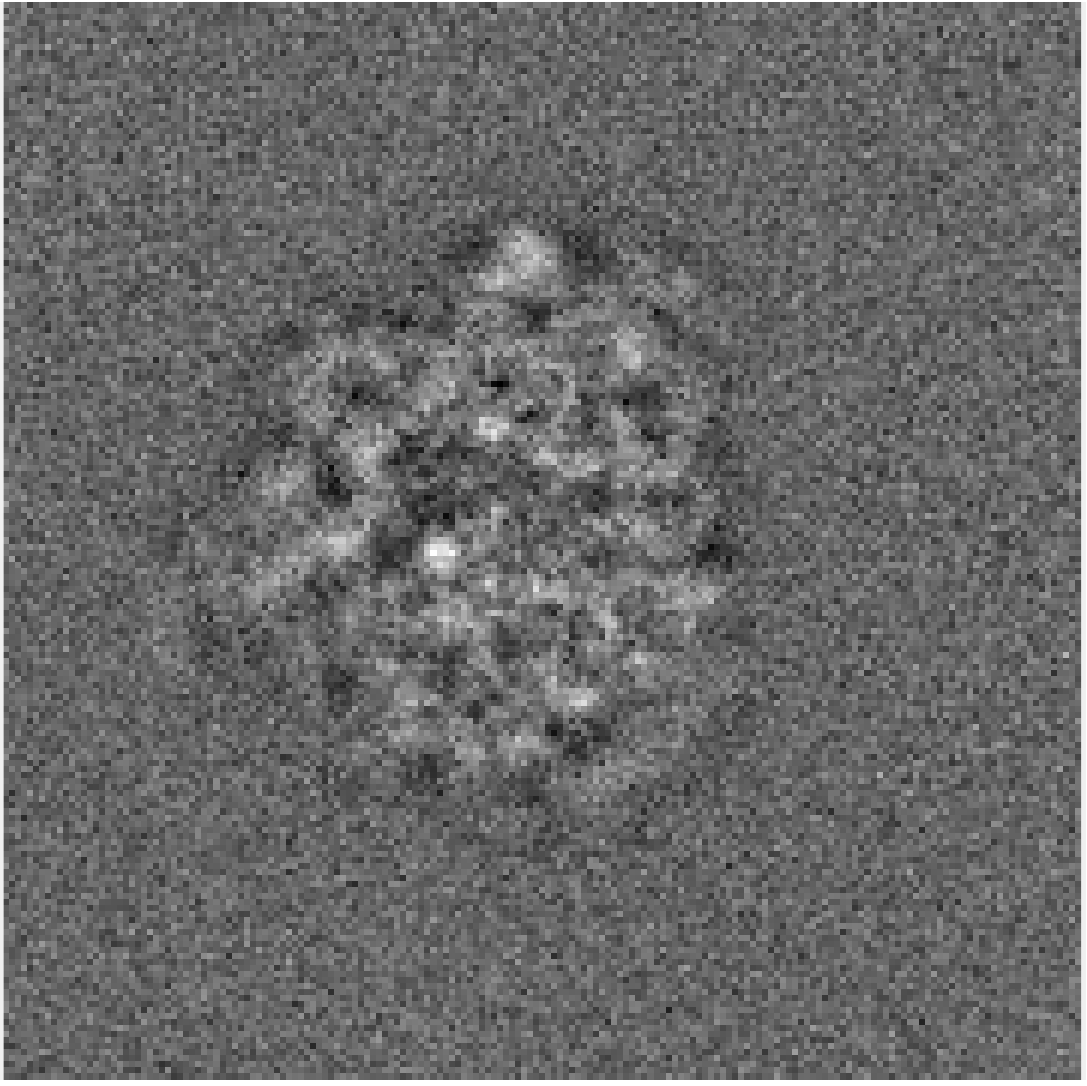}
		} 
		\subfloat{
			\includegraphics[width=0.23\textwidth]{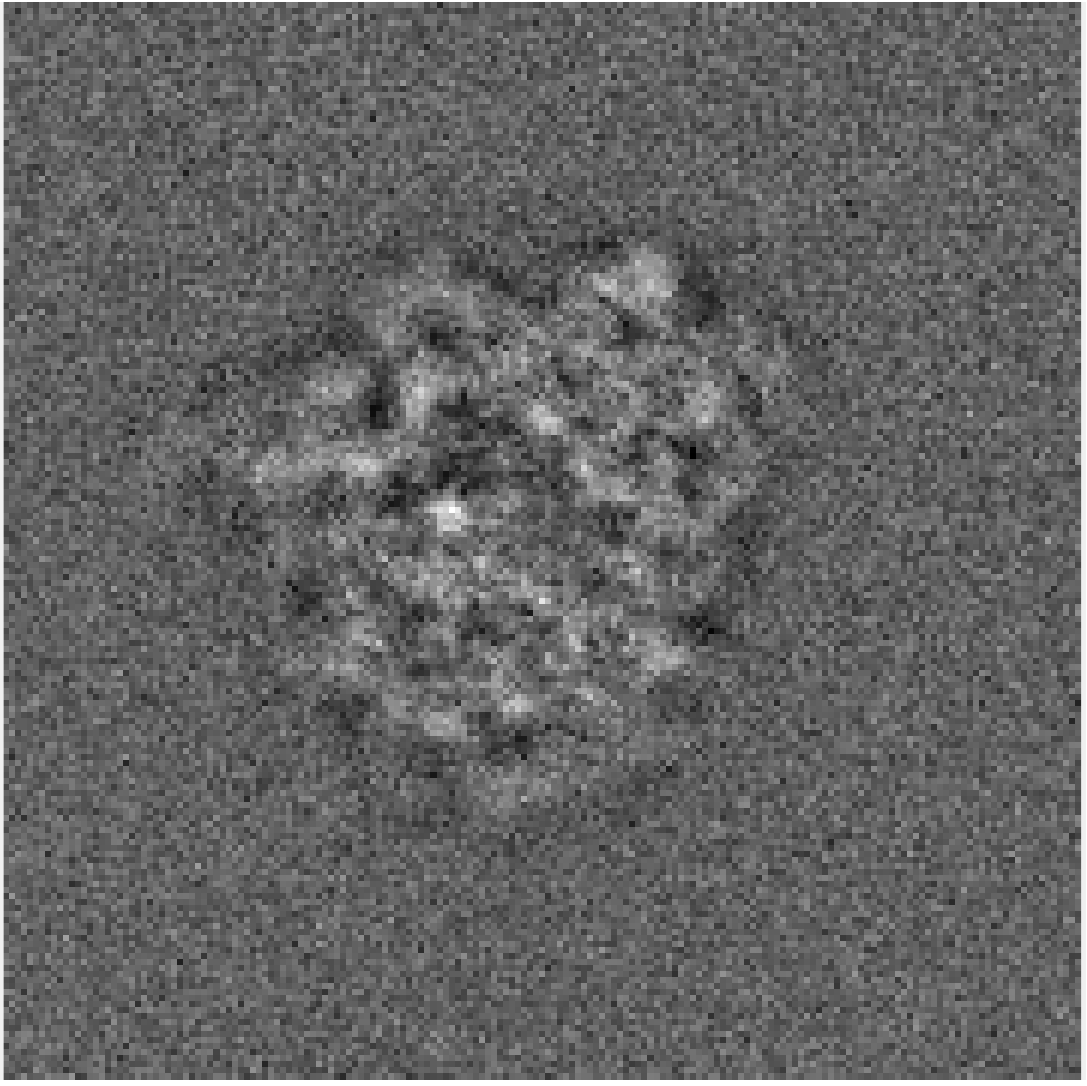}
		}\\
		
		\subfloat{
			\includegraphics[width=0.23\textwidth]{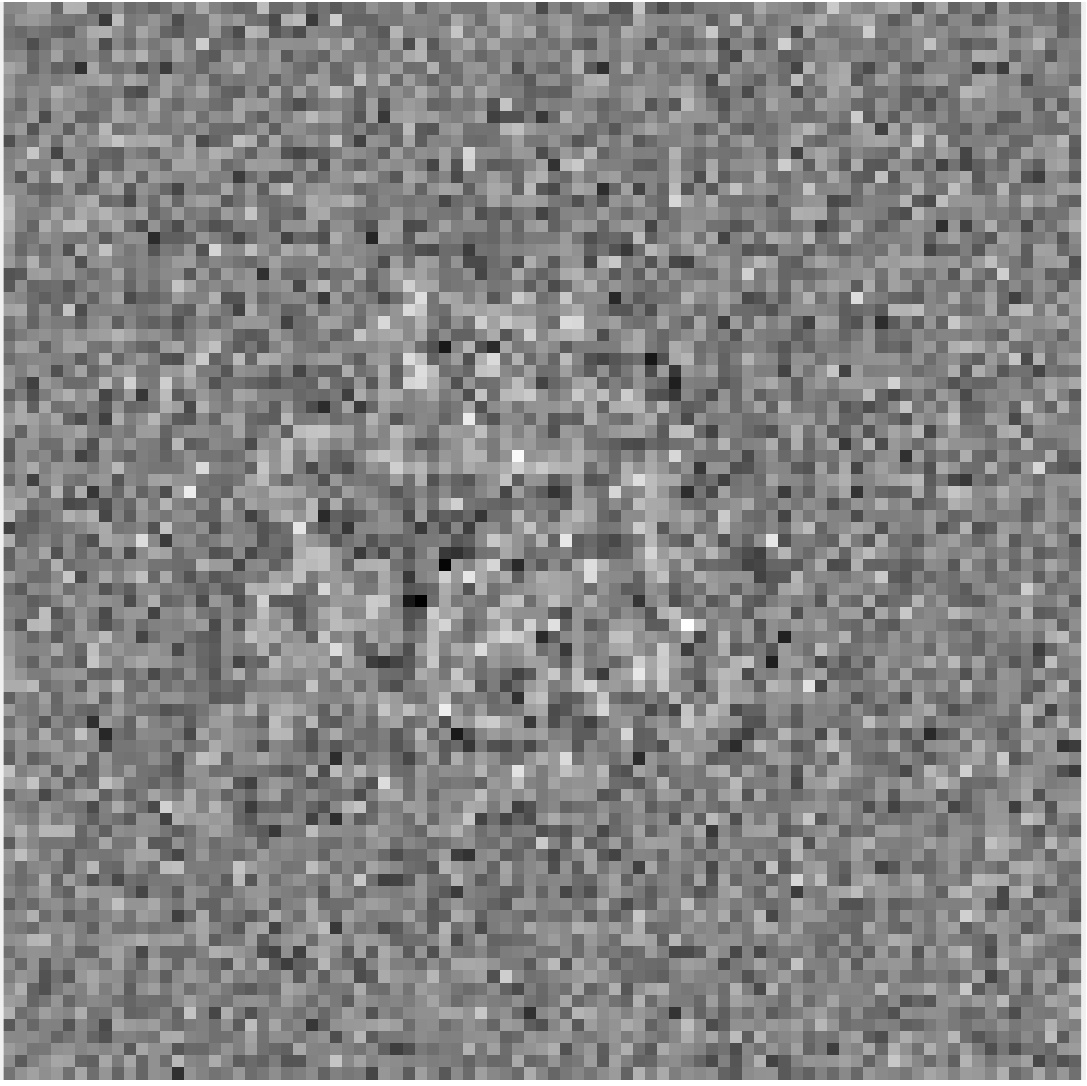}
		} 
		\subfloat{
			\includegraphics[width=0.23\textwidth]{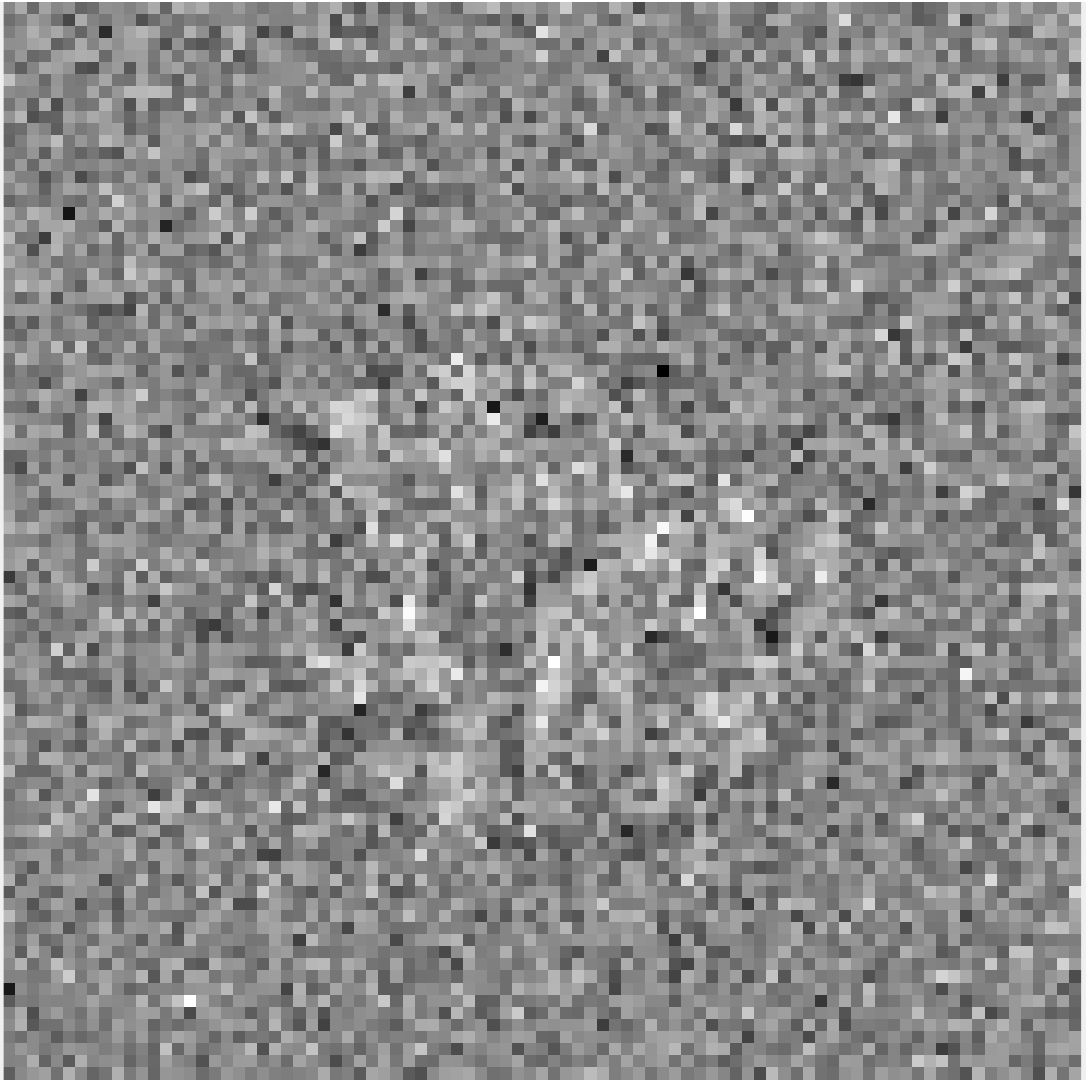}
		} 
		\subfloat{
			\includegraphics[width=0.23\textwidth]{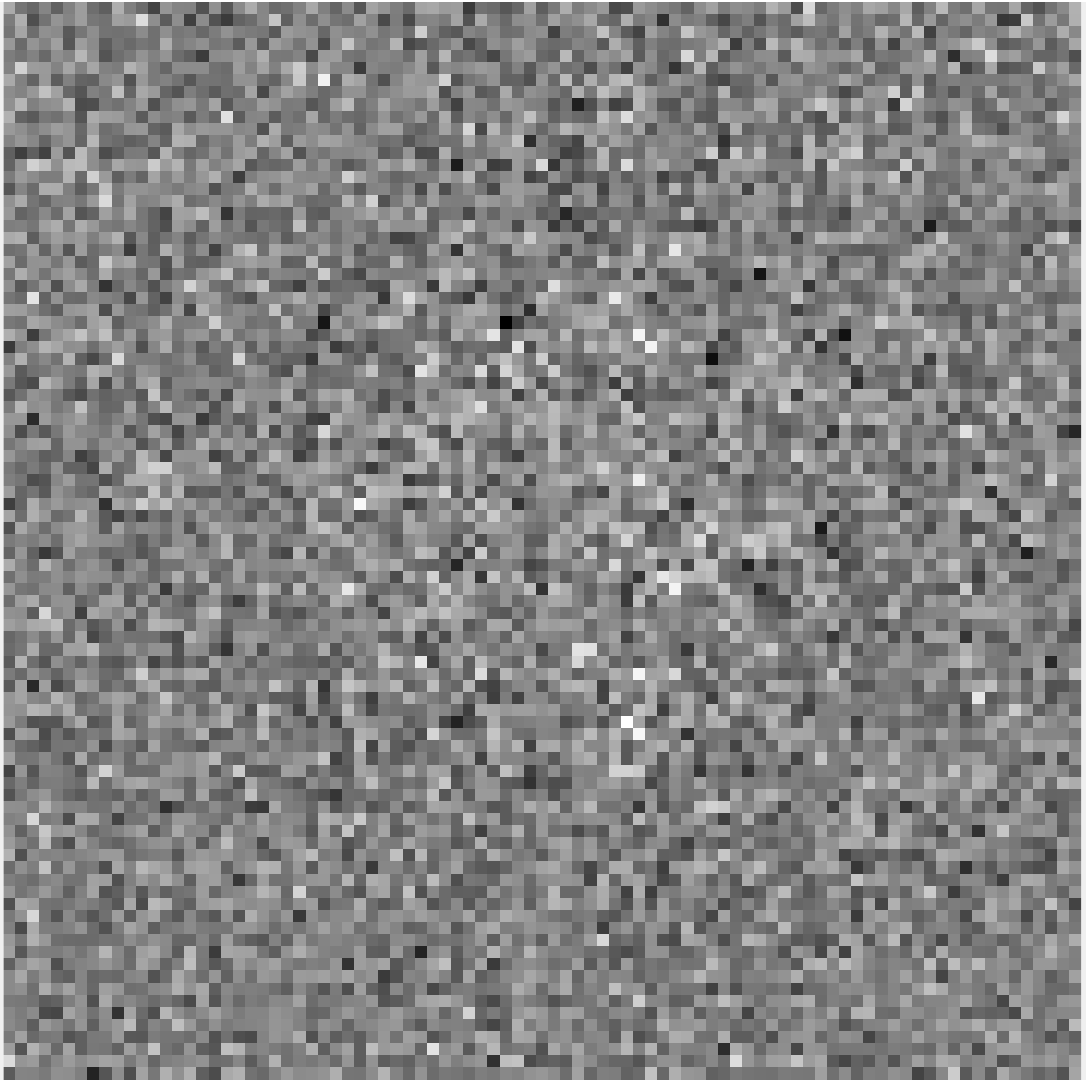}
		} 
		\subfloat{
			\includegraphics[width=0.23\textwidth]{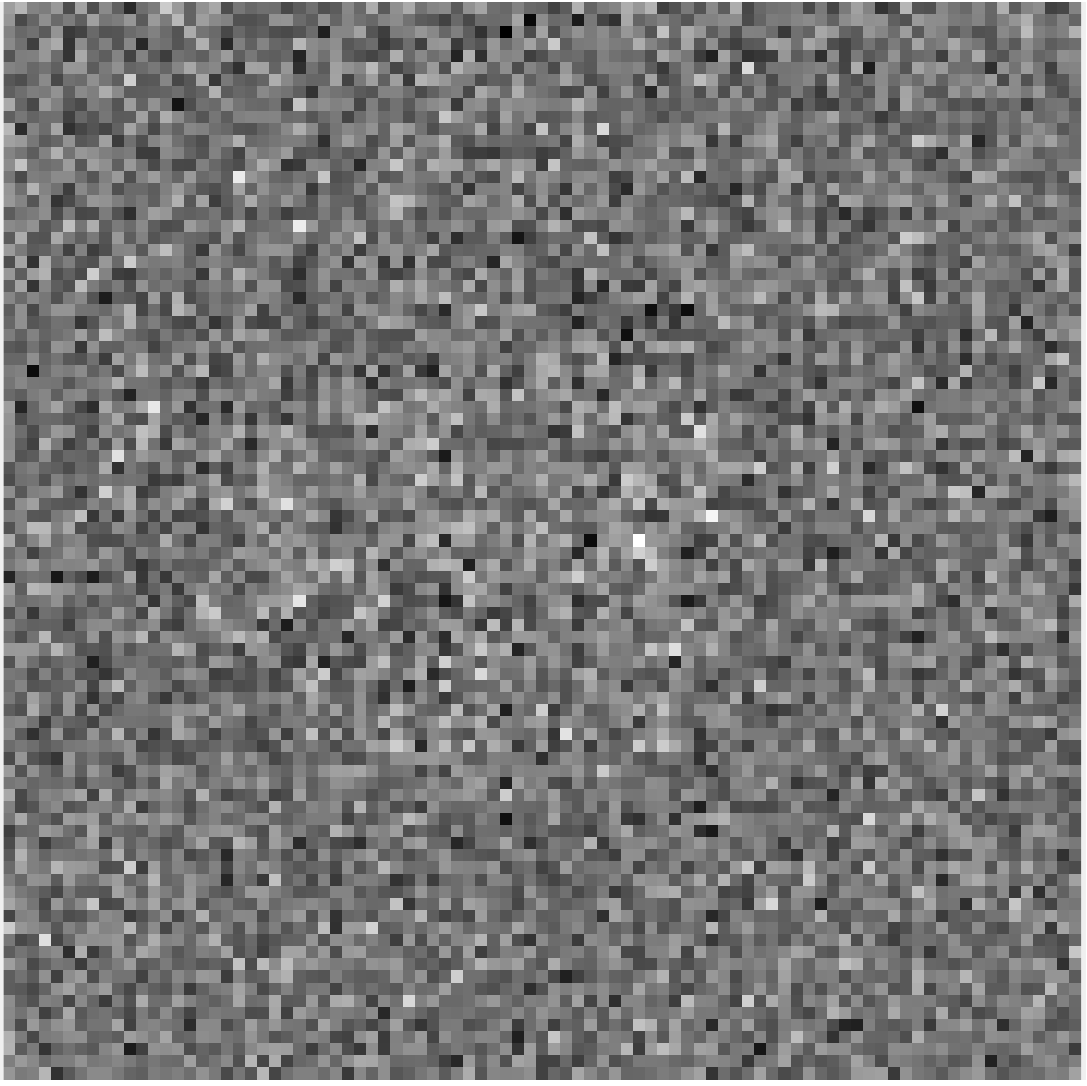}
		}
		\caption{A sample of class averages of the 80S data set with $K$ raw projection-images per class. Top: averages of size  $179 \times 179$ with $K=50$; Bottom: averages of size $89 \times 89$ with $K=3$.}
		\label{fig:class_averages 80S}
	\end{center}
\end{figure}

First, we tested the algorithm on class averages of the Plasmodium falciparum 80S ribosome, generated from the particle images provided in the EMPIAR-10028 data set~\cite{EMPIAR10028} from the EMPIAR archive~\cite{EMPIAR}. The data set consists of 105,247 raw particle images, each of size $360\times 360$ pixels, with pixel size of 1.34\AA. The class averages were generated using the ASPIRE software package~\cite{aspire} as follows. First, all images were phase-flipped, downsampled to size $179\times 179$ pixels, and normalized to background mean~0 and background variance 1. We next used the class averaging routine~\cite{avgVDM} implemented in ASPIRE~\cite{aspire} to generate sets of class averages. Each set of class averages was generated by averaging each raw image with its $K-1$ properly-aligned most-similar images, where in this case we used $K=50$. Note that averaging an image and its $K-1$ most similar images results in class averages that consist of a total of $K$ images averaged together. Also, note that unlike other 2D classification algorithms, there is no clustering process involved, but rather the classes are not mutually disjoint, and the number of resulting class averages is equal to the number of processed images, independently of the size of the classes. Next, we sorted each set of class averages according to the contrast of the averages, where the contrast of an image is simply the standard deviation of its pixel values. The input to the subsequent reconstruction procedure was the top (highest contrast) $N=3000$ class averages. A sample of the class averages is displayed \rda{in}{at} the top row of Figure~\protect\ref{fig:class_averages 80S}.

\begin{figure}
	\begin{center}
		\includegraphics[width=0.5\textwidth]{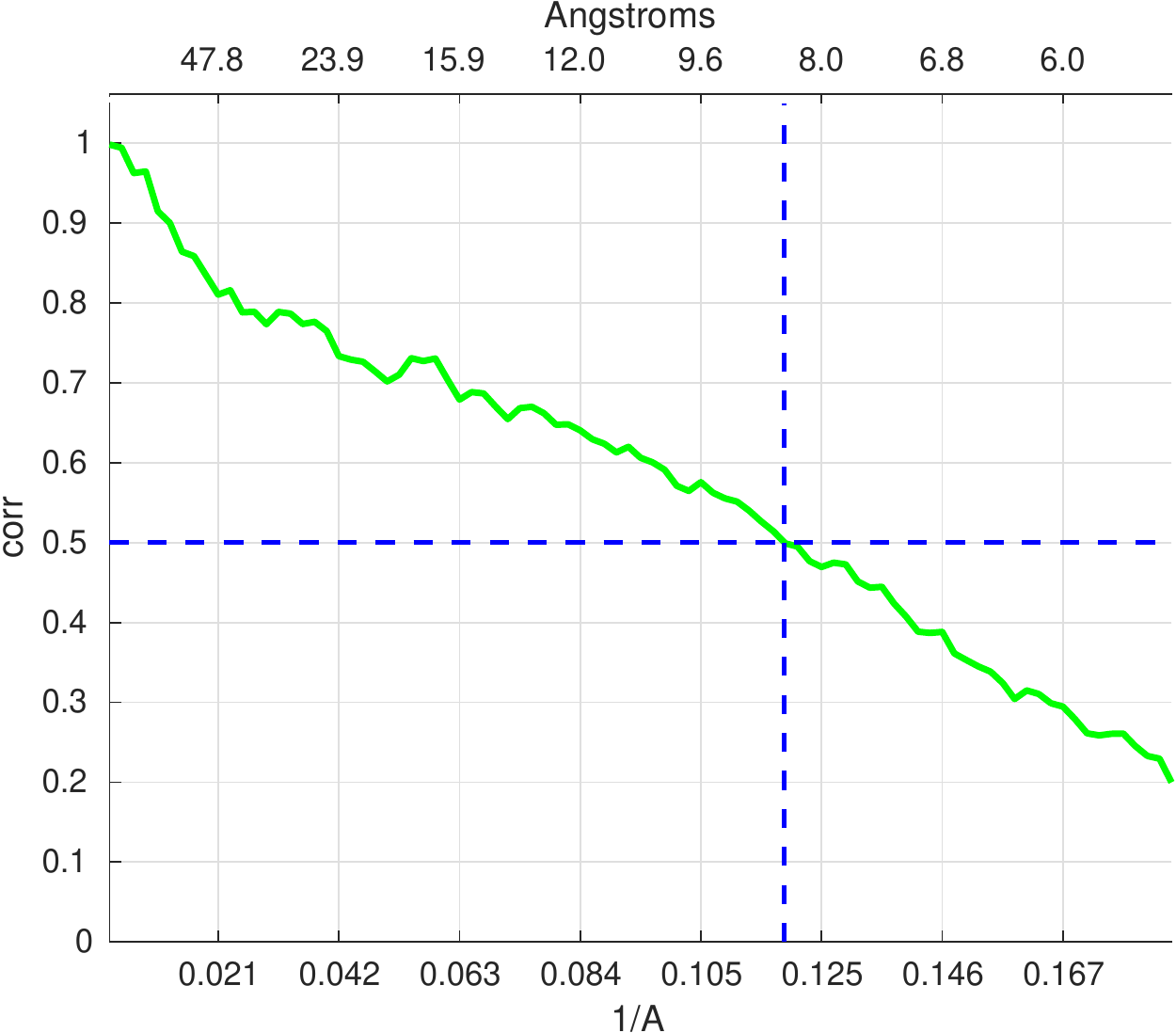}
		\caption{Fourier shell correlation curve for the reconstruction of the 80S subunit from class averages of size $179 \times 179$ with $K=50$ images per class, against the reference density map EMD-2660 of EMDB~{\cite{EMDB}}.}
		\label{fig:fsc_80s_nn49}
	\end{center}
\end{figure}

\begin{figure}
	\begin{center}
		\subfloat[Reference]{
            \label{fig:refmap}
			\includegraphics[width=0.2\textwidth]{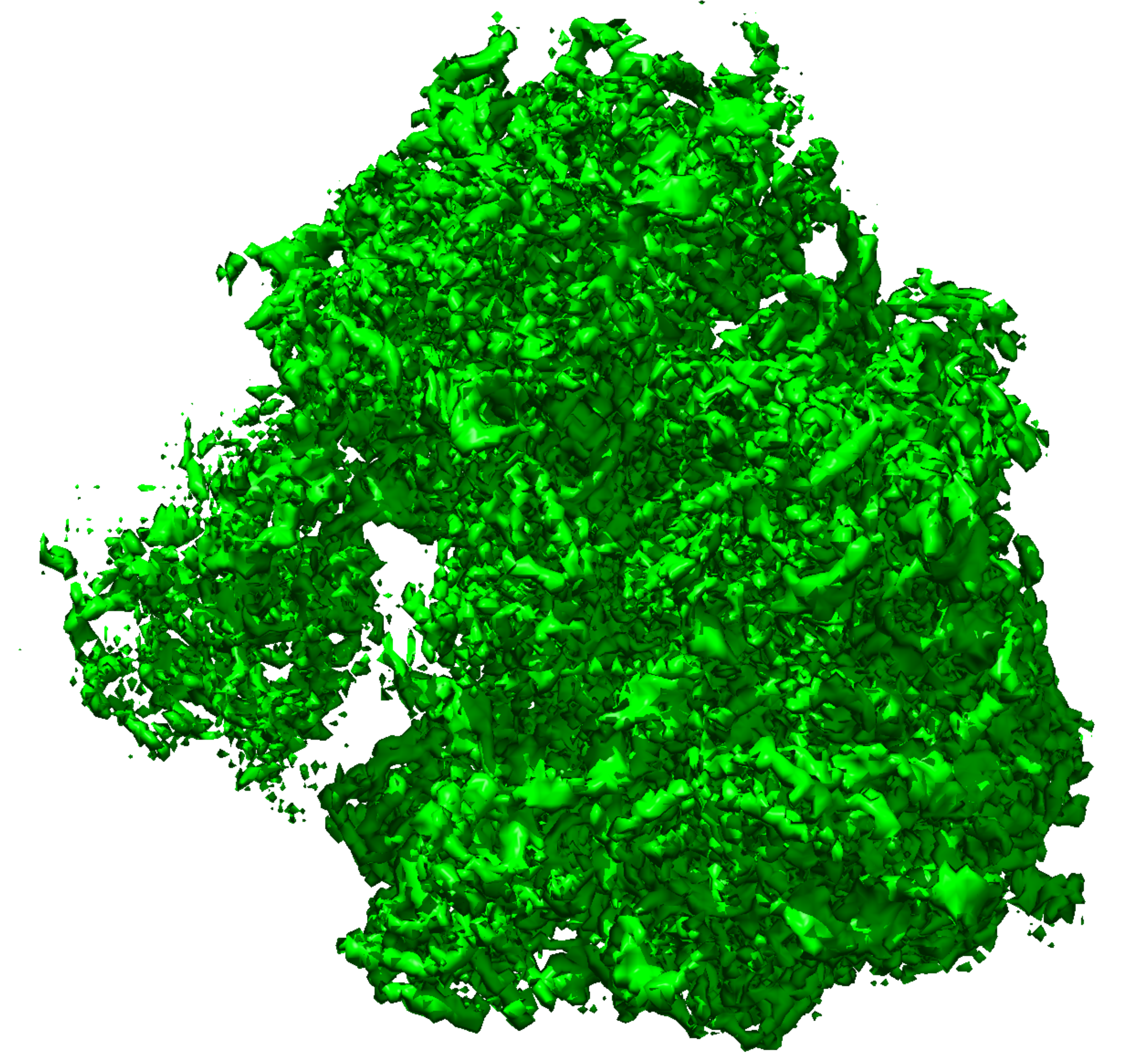}
		} \hspace{0.05cm}
		\subfloat[$K=50$]{
			\label{fig:vol_K50}
			\includegraphics[width=0.2\textwidth]{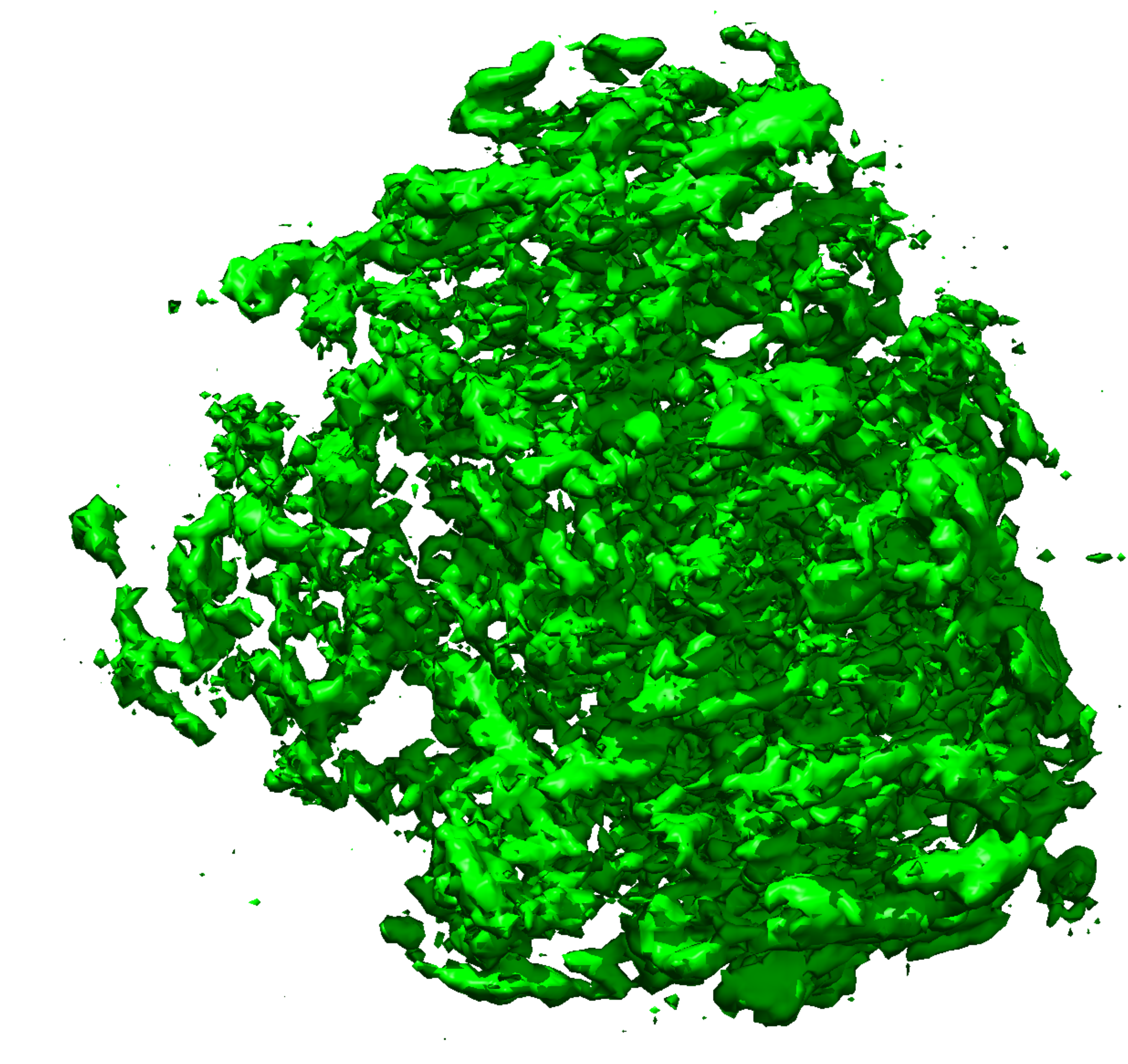}
		}\hspace{0.05cm}
		\subfloat[$K=3$]{
			\label{fig:vol_nn02}
			\includegraphics[width=0.2\textwidth]{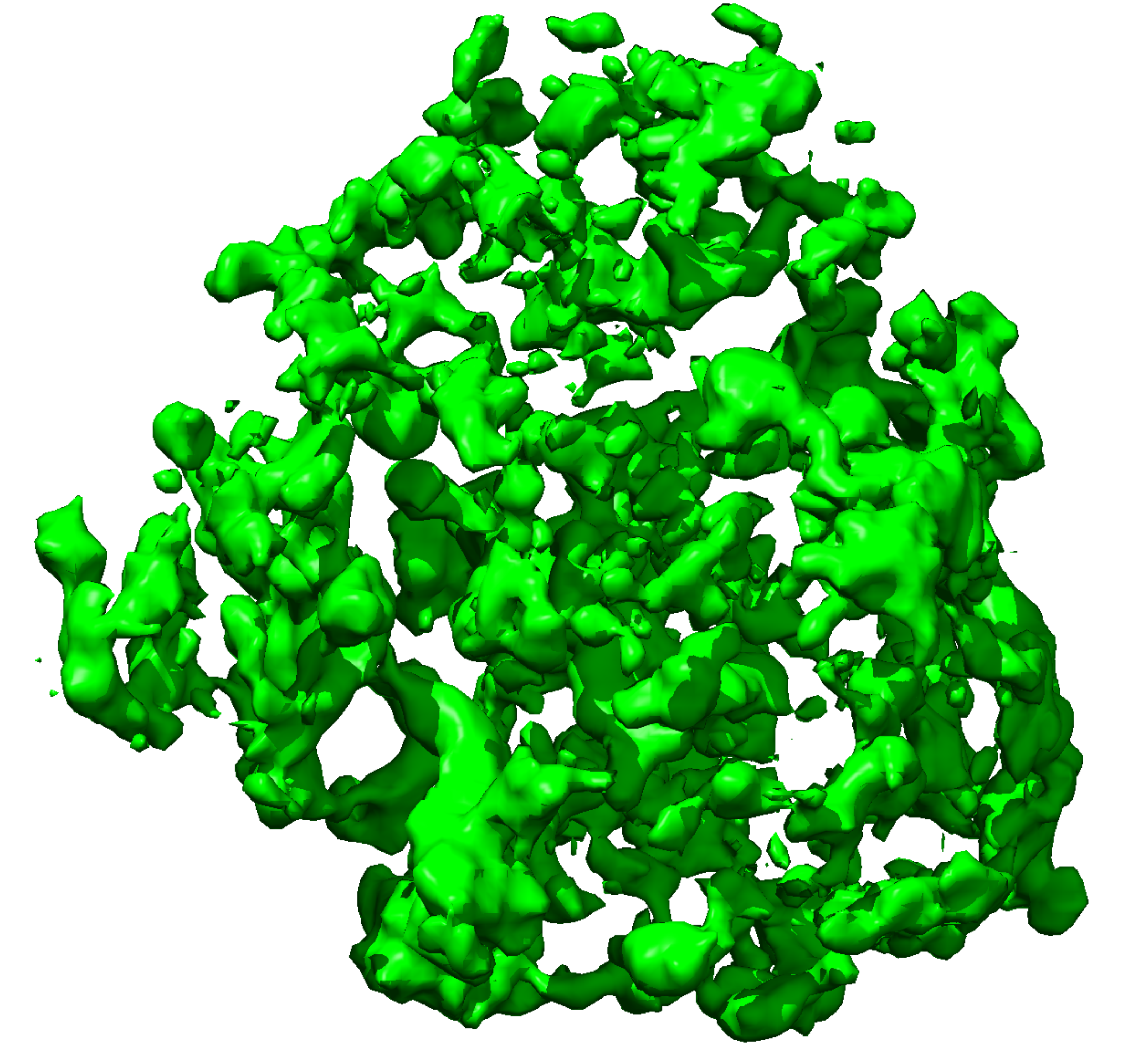}
		} \hspace{0.05cm}
		\subfloat[$K=3$, UW]{
			\label{fig:vol_nn02_old}
			\includegraphics[width=0.2\textwidth]{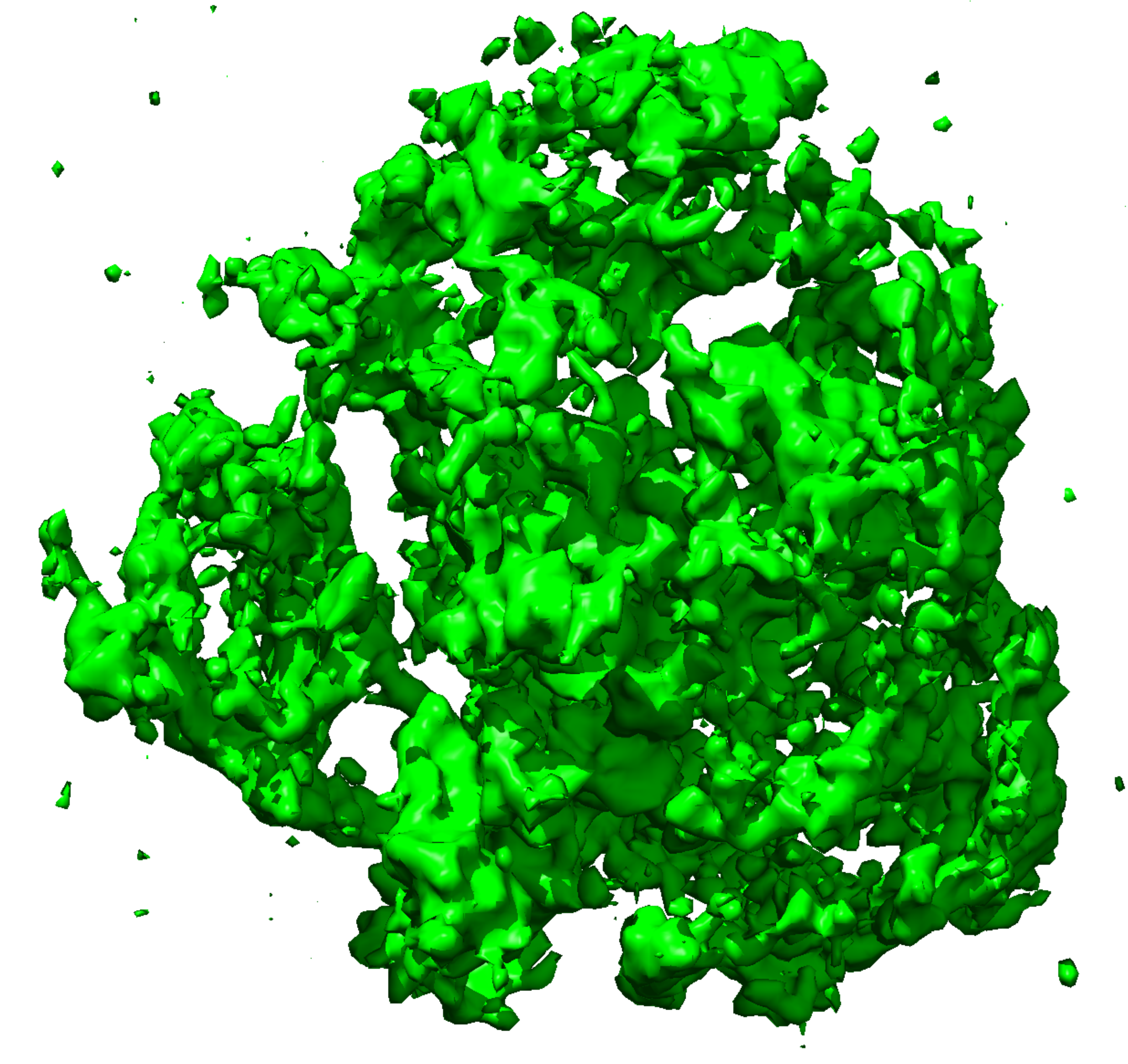}
		}
		\caption{Density maps of the 80S subunit, reconstructed from class averages of $K$ projection-images per class. \textit{UW} stands for the \textit{Un-Weighted} reconstruction algorithm of~\protect\cite{sync3N}. \textit{Reference} stands for the density map EMD-2660 of EMDB~{\cite{EMDB}}.}
		\label{fig:vols80S}
	\end{center}
\end{figure}

Next, we used cross-correlation to find common lines between all class averages, used the algorithm described in Section~\ref{sec:weighting_scheme} to estimate the rotations $R_{1},\ldots,R_{N}$ corresponding the $N$ class averages, and reconstructed a density map using the class averages and their estimated rotations. No refinement was used in the reconstruction, nor CTF correction (except for phase-flipping as described above). To assess the resolutions of our reconstructions, we compared the density map reconstructed from the class averages to the reference density map EMD-2660 available in the Electron Microscopy Data Bank (EMDB)~\cite{EMDB}, which was reconstructed from the same underlying raw images, and is described in detail in~\protect\cite{80S32A}. The comparison was done using the 0.5-criterion of the Fourier shell correlation (FSC) curve~\cite{vanHeel_Schatz}.
The FSC curve is shown in Figure~\ref{fig:fsc_80s_nn49}, and implies a resolution of 8.4\AA. Note that the downsampling during preprocessing removed the higher frequencies of the data, and thus it is possible to have positive correlations throughout the entire range of frequencies. Two-dimensional rendering of the reconstructed density map is shown in Figure~\ref{fig:vol_K50}, along with the reference density map EMD-2660~{\cite{EMDB}}. All the renderings in this section were generated using USCF Chimera~\cite{chimera}.

\subsection{Yeast dataset}\label{sec:yeast}

\begin{figure}
\begin{center}
		
		\subfloat{
			\includegraphics[width=0.23\textwidth]{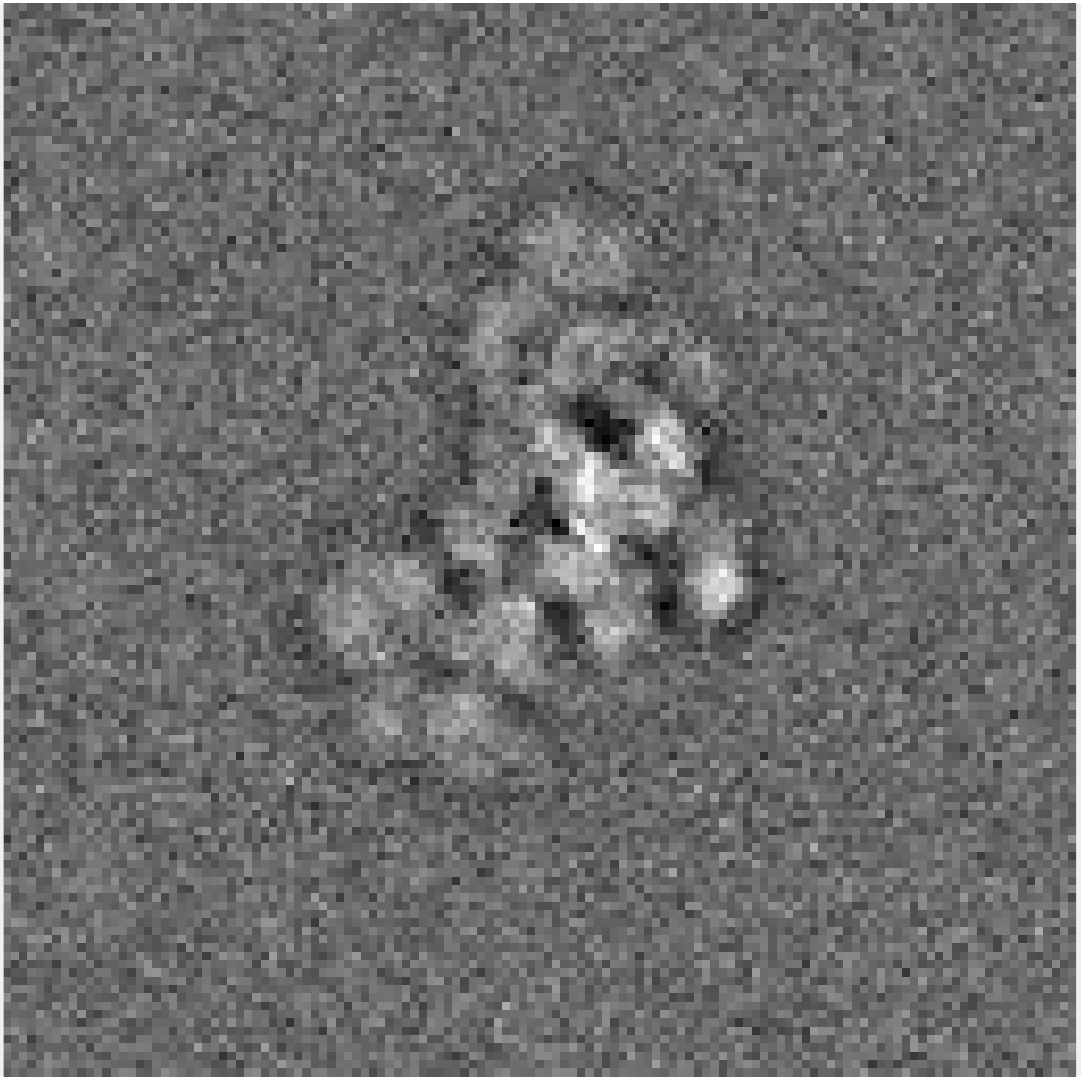}
		} 
		\subfloat{
			\includegraphics[width=0.23\textwidth]{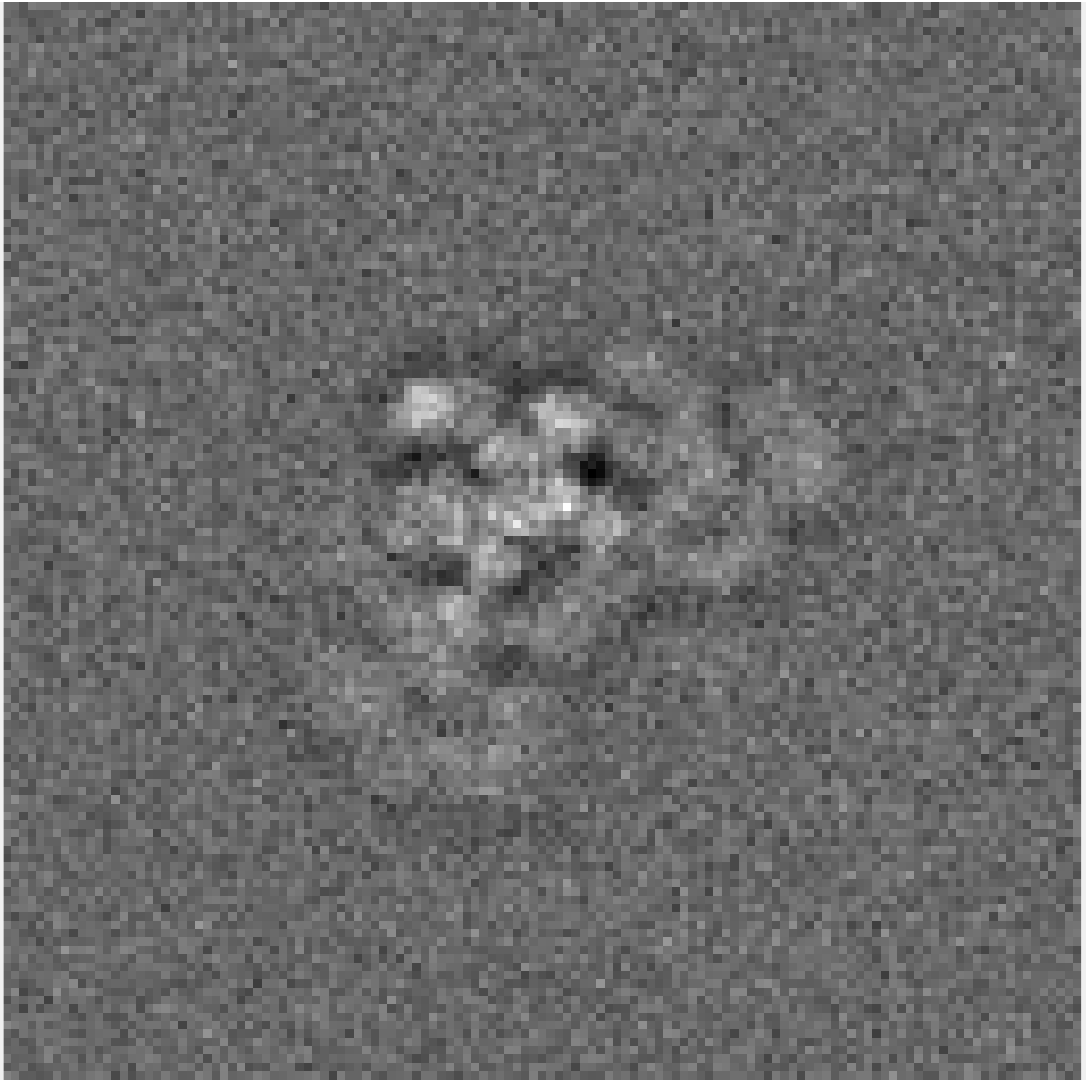}
		} 
		\subfloat{
			\includegraphics[width=0.23\textwidth]{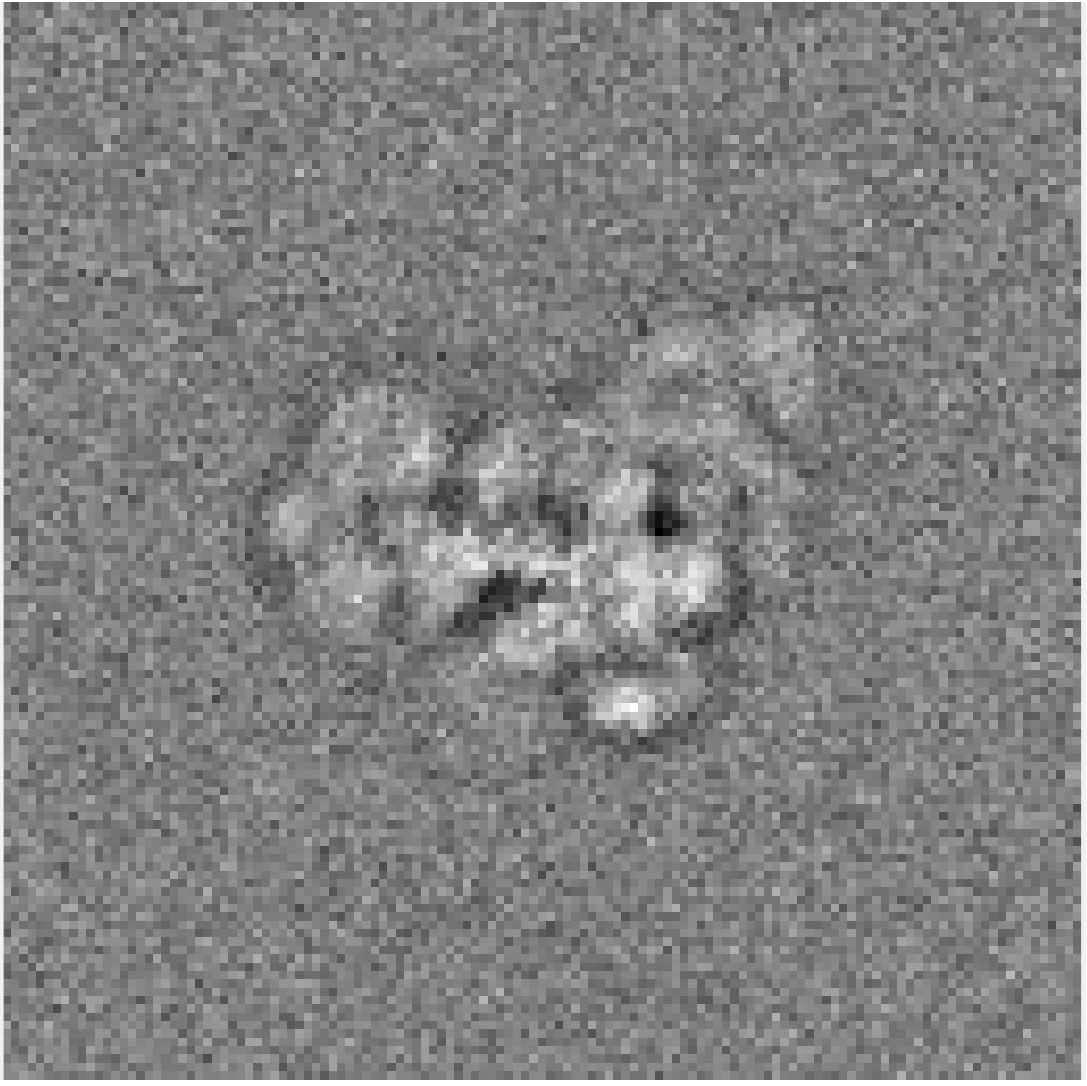}
		} 
		\subfloat{
			\includegraphics[width=0.23\textwidth]{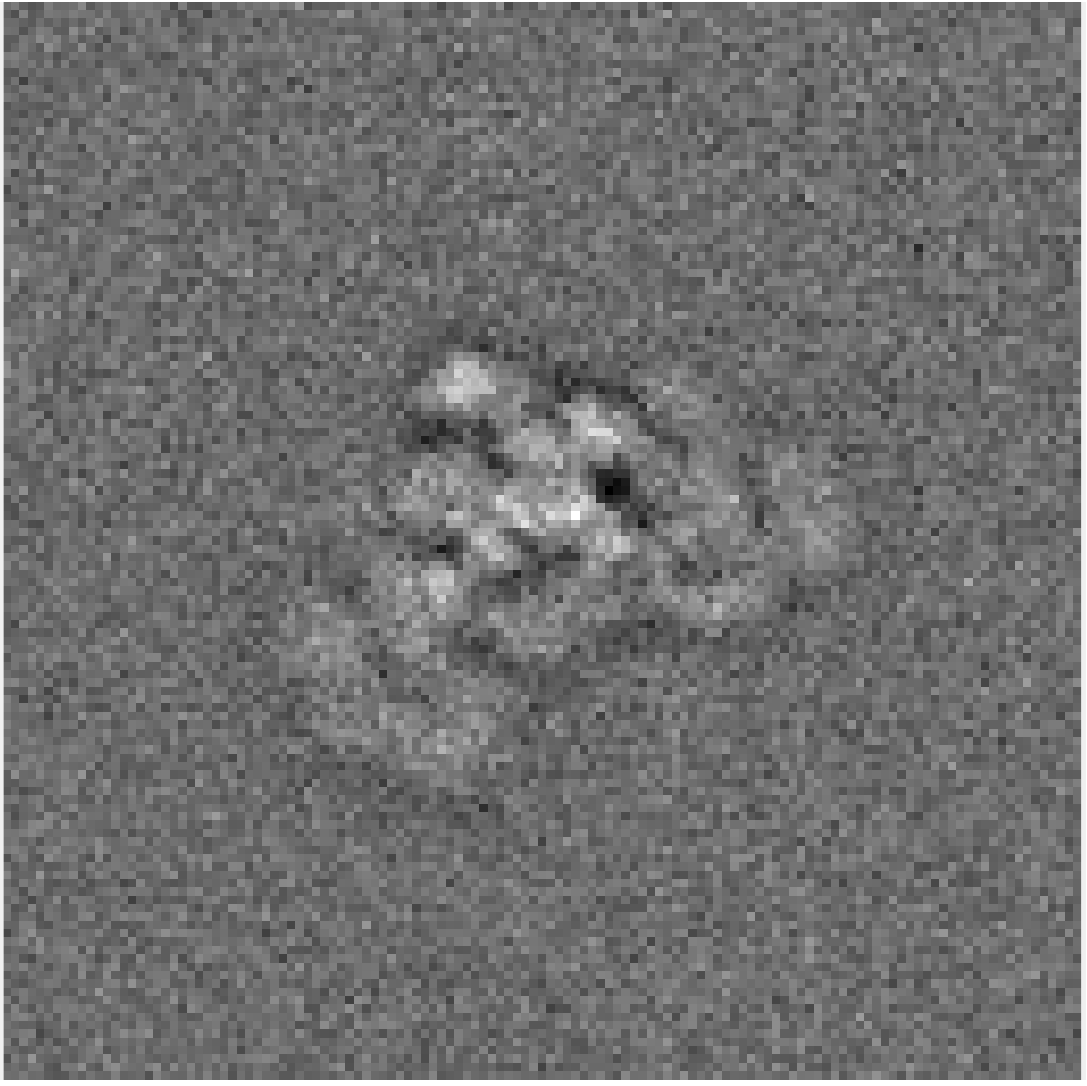}
		}
\end{center}
		\caption{A sample of class averages with $K=50$ raw projection-images per class, generated from $129 \times 129$ images of the yeast data set.}
\label{fig:class_averages yeast}
\end{figure}

Next, we applied the algorithm on class averages of the yeast U4/U6.U5 tri-snRNP, generated from particle images provided in the EMPIAR-10073 data set~\cite{EMPIAR10073}. Our first attempt with this data set consisted of repeating the class averaging procedure exactly as described in Section~\ref{sec:80S}, using the $3000$ highest contrast class averages with $K=50$ images per class, but with the images downsampled to $129 \times 129$. A sample of the class averages is displayed in Figure~\protect\ref{fig:class_averages yeast}.

\begin{figure}
\begin{center}
		\subfloat[Highest contrast]{
			\includegraphics[width=0.3\textwidth]{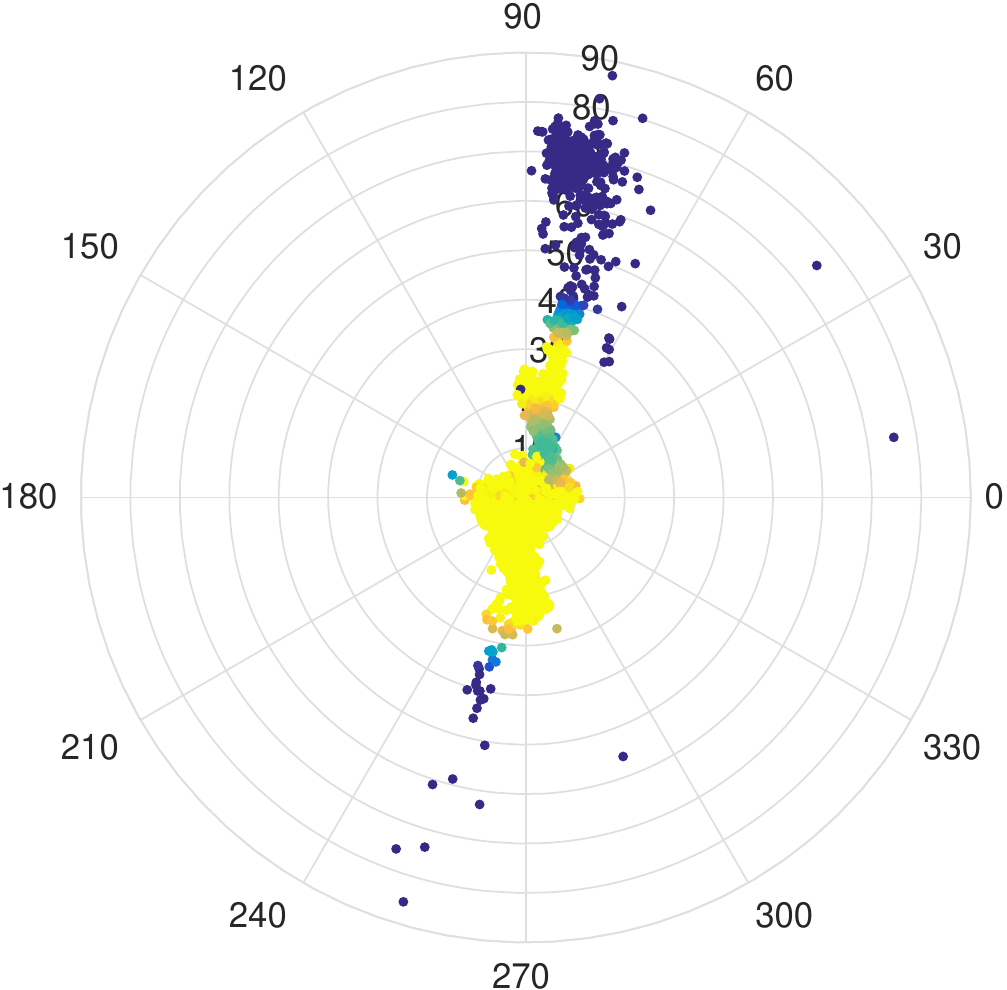}
    \label{fig:yeast highest contrast}
		} \hspace{1cm}
		\subfloat[Uniform sampling]{
			\includegraphics[width=0.3\textwidth]{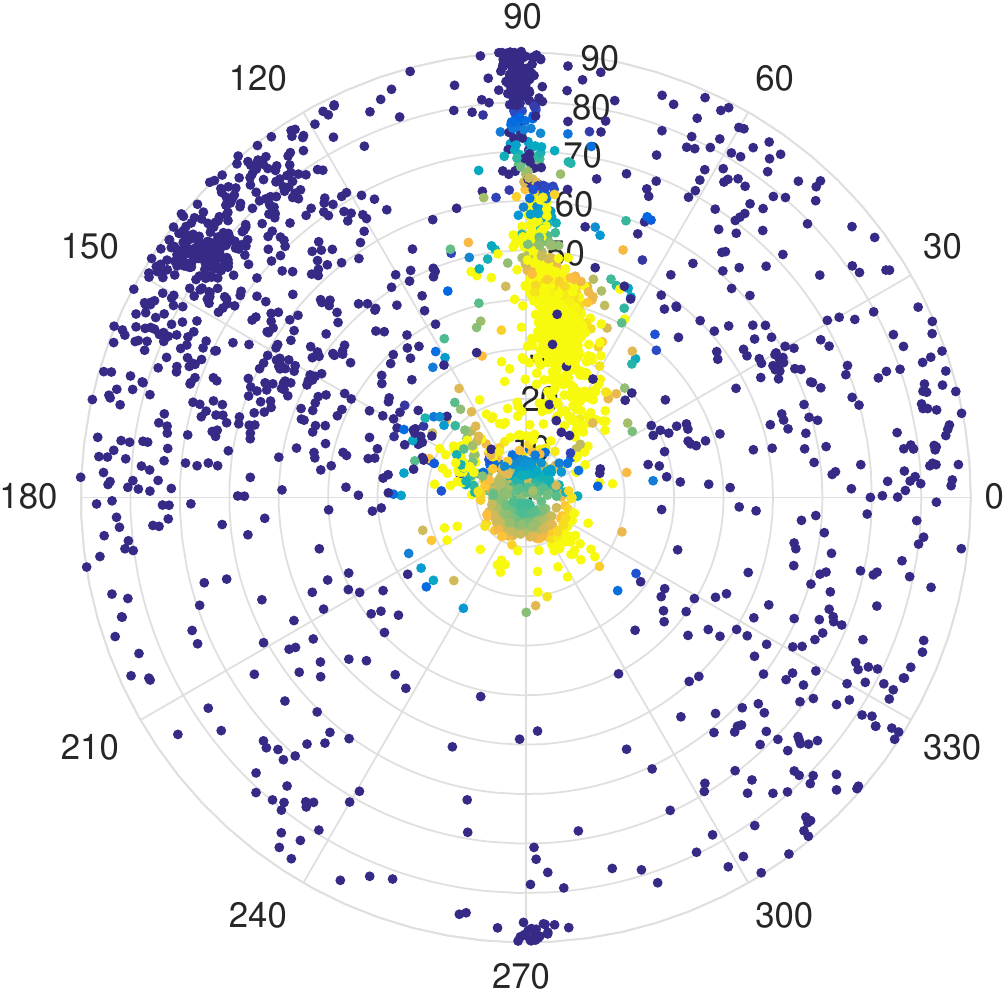}
\label{fig:yeast uniform}
		} 
\caption{Distribution of the estimated orientations for the yeast data set. Each point represents the spherical coordinates of the estimated viewing direction of one of the class averages: the radius represents the polar angle in {$\left[0,90^\circ\right]$}, and the angular direction represents the azimuthal angle in {$\left[0,360^\circ\right)$}. Lighter color corresponds to higher density regions.}
\end{center}
\end{figure}

The ab-initio model generated by our algorithm (and in particular the orientations assigned to the class averages) revealed that the orientations of the 3000 highest contrast averages are restricted to a small subset of all possible orientations. The distribution of the estimated orientations is shown in Figure~\ref{fig:yeast highest contrast}. Note that even though the orientations are highly non-uniform, they were nevertheless estimated accurately according to the reliability measures described in Section~\ref{sec:reliability}. Thus, in this case, instead of choosing the highest contrast averages, we used 3000 averages uniformly sampled from all class averages, in order to better represent all possible viewing directions at the expense of choosing averages of lower quality. The distribution of the estimated orientations for the latter set of class averages is shown in Figure~\ref{fig:yeast uniform}. Note that this change of setup was inferred completely from the outcome of our algorithm, without any prior or external information.

\begin{figure}
	\begin{center}
		\includegraphics[width=0.5\textwidth]{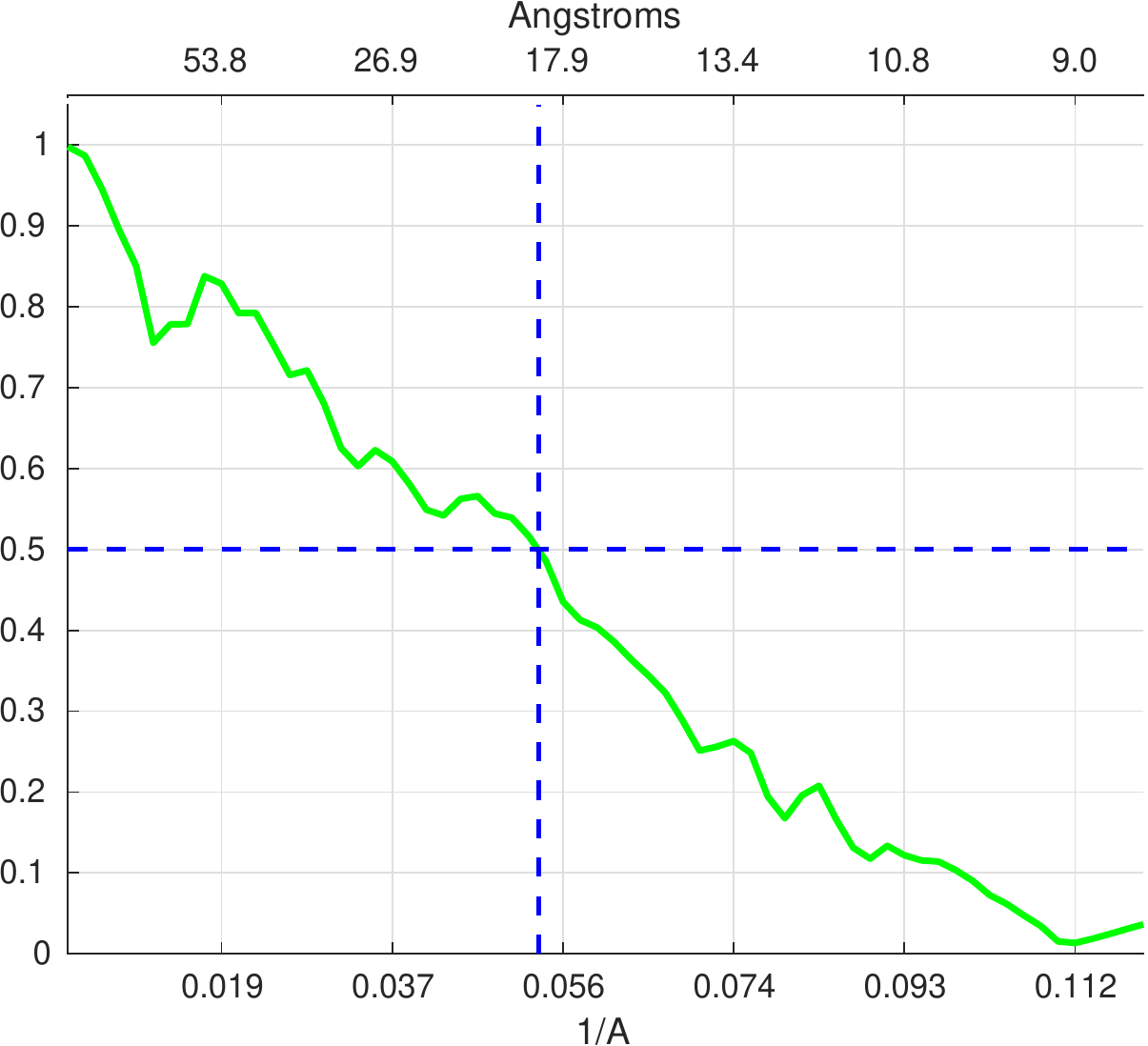}
		\caption{Fourier shell correlation curve for the reconstruction from the yeast data set using class averages of $K=50$ images per class, against the reference density map EMD-8012 of EMDB~{\cite{EMDB}}.}
		\label{fig:fsc_yeast}
	\end{center}
\end{figure}

\begin{figure}
\begin{center}
		\subfloat[Reference]{
			\includegraphics[width=0.2\textwidth]{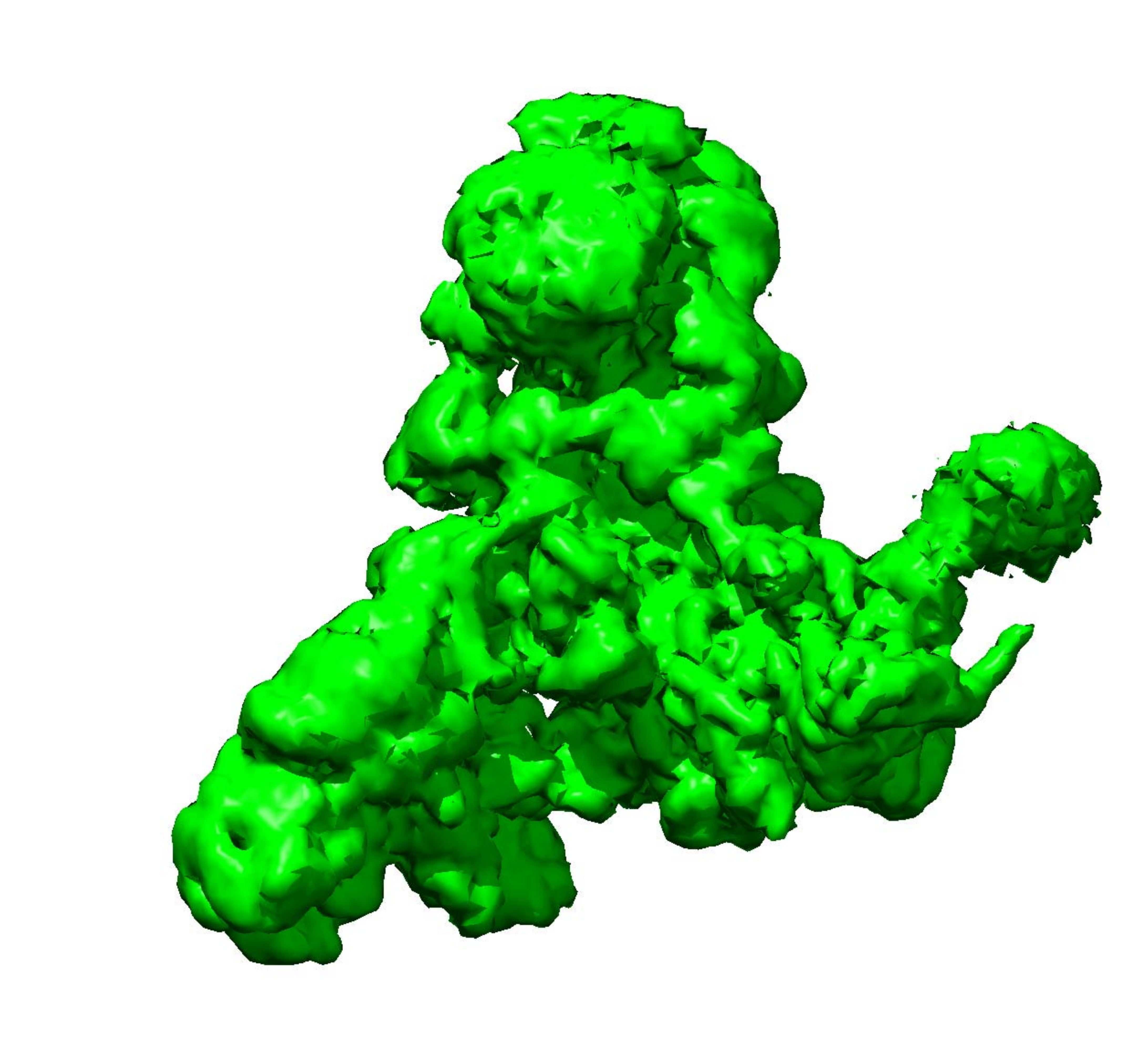}
        \label{fig:vol_yeast_ref}
		} \hspace{0.05\textwidth}
		\subfloat[Ab-initio model]{
			\label{fig:vol_yeast}
			\includegraphics[width=0.2\textwidth]{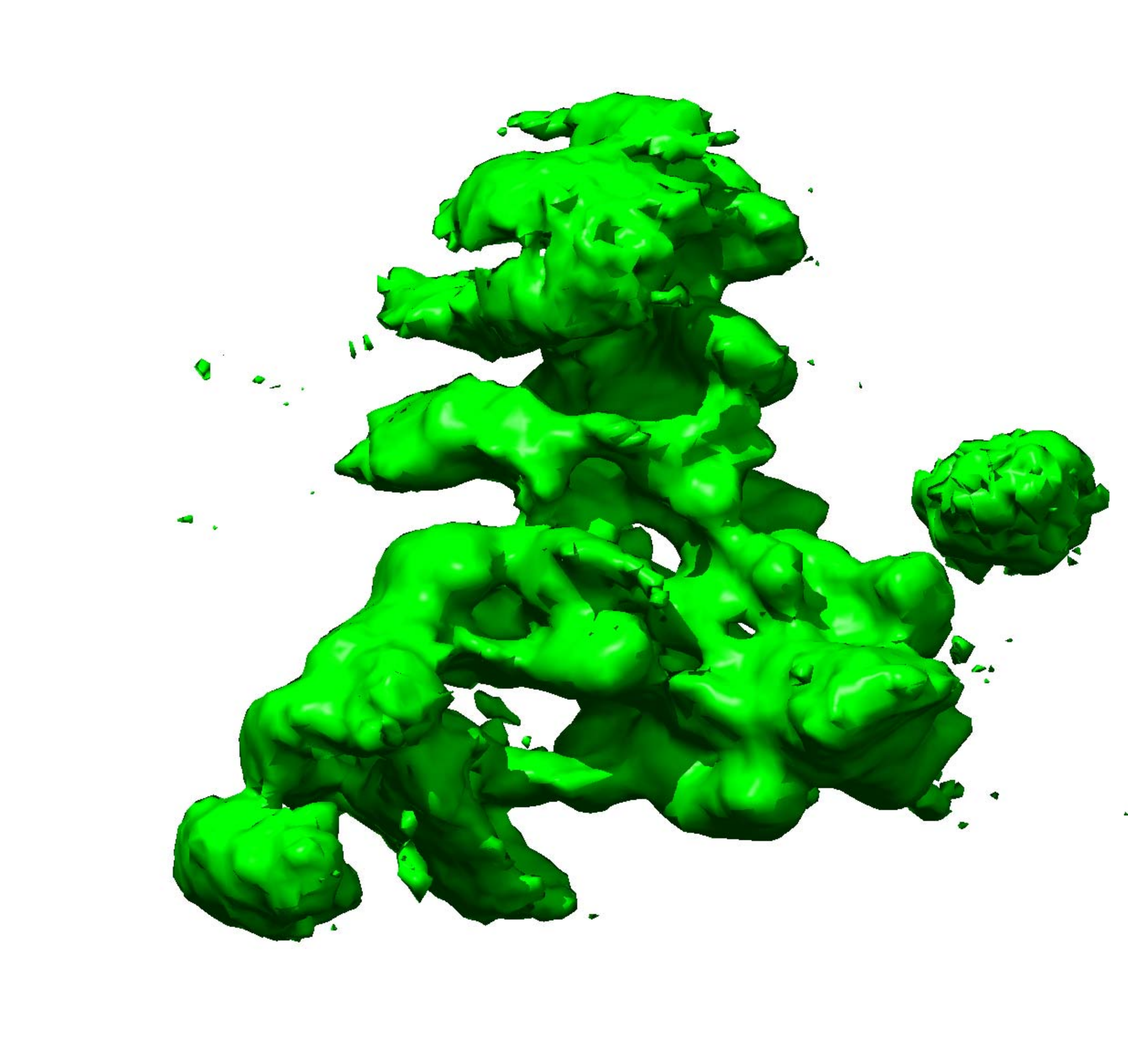}
		}	
		\caption{Density maps of the yeast data set.}
		\label{fig:volsYeast}
\end{center}
\end{figure}

The reconstructed density map was then compared to the reference density map EMD-8012 of (EMDB)~\cite{EMDB}, which was refined from the same set of raw particle images, and is described in detail in~\protect\cite{yeast37A}. The reconstruction achieved a resolution of 18.8{\AA} according to the 0.5-criterion.
Two-dimensional rendering of the density map reconstructed by our algorithm is shown in Figure~\ref{fig:vol_yeast}, along with the reference density map EMD-8012 (Figure~\ref{fig:vol_yeast_ref}). The FSC curves are shown in Figure~\ref{fig:fsc_yeast}.

\subsection{Robustness to noisy class averages}\label{sec:noise robustness}
As discussed in Section~\ref{sec:background}, a major advantage of the simultaneous synchronization of the \rda{relative rotations {$\{R_iR_j^{-1}\}$}}{common lines information}, and in particular of the weighted synchronization suggested in this paper, is improved robustness to noise in the projection-images (or class averages).
To demonstrate the robustness of the algorithm to noise, we applied it on $3000$ class averages of the Plasmodium falciparum 80S ribosome generated as in Section~\ref{sec:80S}, but with as few as $K=3$ images per class, downsampled to $89 \times 89$. A sample of the class averages is displayed \rda{in}{at} the bottom row of Figure~\protect\ref{fig:class_averages 80S}.

\begin{figure}
	\begin{center}
		\includegraphics[width=0.5\textwidth]{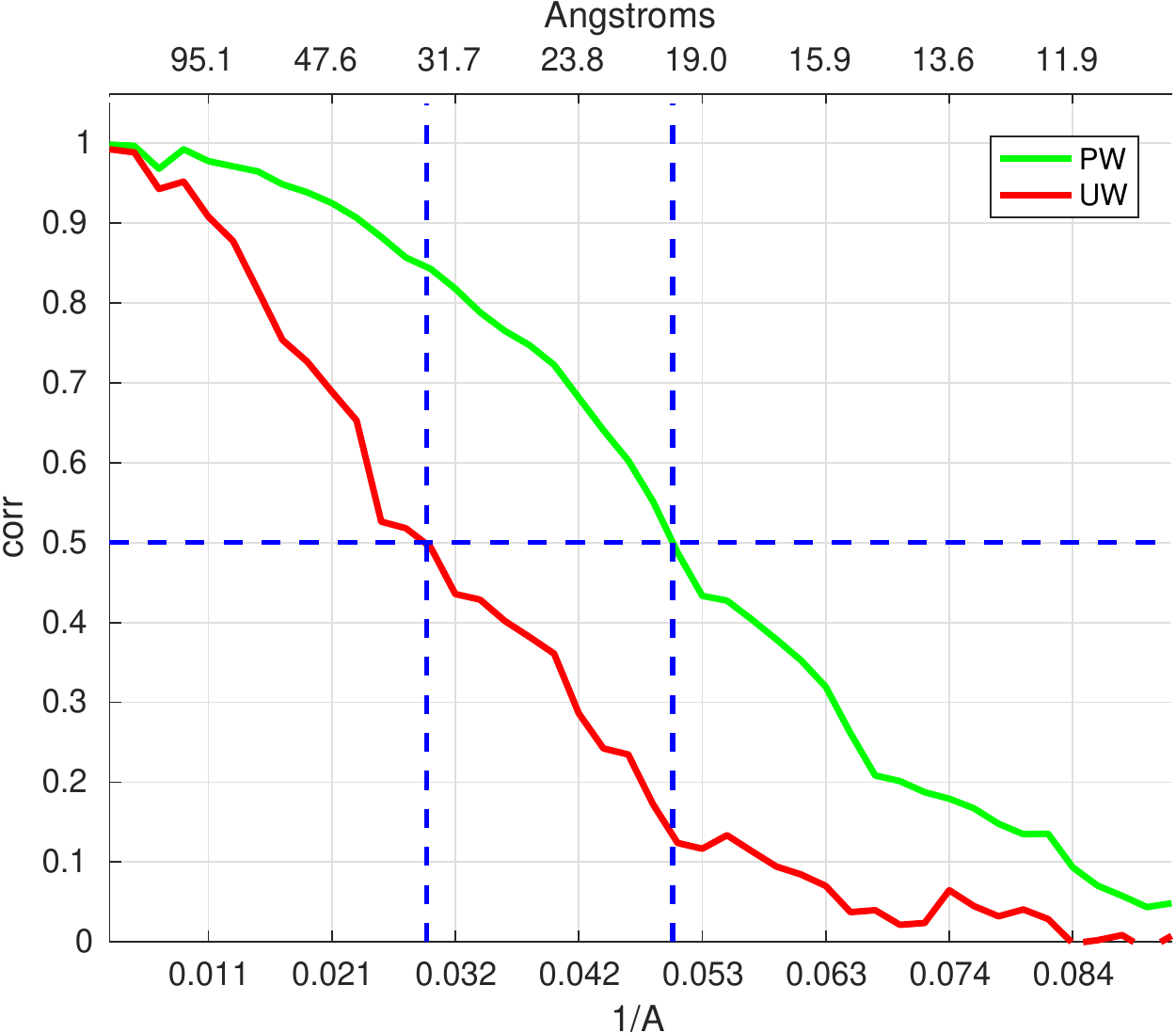}
		\caption{Fourier shell correlation curves for the reconstructions of the 80S subunit from class averages of $K=3$ images per class, against the reference density map EMD-2660 of EMDB~{\cite{EMDB}}. \textit{PW} stands for the \textit{Probability-Weighted} algorithm described in this paper, whereas \textit{UW} stands for the \textit{Un-Weighted} algorithm of \cite{sync3N}.}
		\label{fig:fsc_80s_nn02}
	\end{center}
\end{figure}

Using this setup, our algorithm reconstructed a density map with resolution of 20.0{\AA} (according to the 0.5-criterion of the FSC curve).
By comparison, the algorithm described in \cite{sync3N}, which does not assign weights to the relative rotations (or equivalently, uses $w_{ij} \equiv 1$), resulted in resolution of 34.4{\AA}. Figures~\ref{fig:vol_nn02} and \ref{fig:vol_nn02_old} show the density maps reconstructed from the class averages corresponding to $K=3$. Figure~\ref{fig:fsc_80s_nn02} displays the corresponding Fourier shell correlation curves.

\ra{The robustness of the presented algorithm to noise may even provide robustness to ``bad'' images, which are very common in cryo-EM data sets (due to, for instance, contaminants, crowded structures or wrong class averaging).
In particular, some of the class averages used for the reconstruction discussed in this section seem not to represent any actual projection of the molecule, as demonstrated by the two rightmost images at the bottom row of Figure~{\ref{fig:class_averages 80S}}.}

%
%

\subsection{Assessing map's reliability}\label{sec:reliability}

As discussed in Section~\ref{sec:background}, a major challenge in cryo-EM is to assess the reliability of a reconstructed density map. While the Fourier shell correlation curve describes the consistency between two reconstructed density maps, it cannot verify the correctness of the maps, hence it cannot detect consistent errors. One source of such errors is the bias towards an initial density map \cite{Henderson2013b}, which is avoided completely by the algorithm described in this paper, as it does not assume any initial density map.

A problem with assessing the reliability of an ab-initio model arises when the model is of low resolution. In such cases, it may be unclear whether the reconstruction yielded a low-resolution density map (which can be successfully refined to a high-resolution one), or entirely failed to estimate the viewing directions of the class averages, and consequentially failed to reconstruct a low resolution density map. Thus, it is essential to determine the reliability of the ab-initio reconstruction process directly.

The algorithm described in this paper provides several inherent indicators to assess the success of the reconstruction process.
One such indicator is the estimated rate of the indicative common lines, introduced in Section~\ref{sec:errors_indications}.
For example, this indicator is equal to $P=80\%$ in the reconstruction of the 80S from class averages with $K=50$ images per class; $P=34\%$ for $K=3$; and $P=15\%$ for $K=2$. Note that according to Section~\ref{sec:triplets_scores}, $P=15\%$ is practically indistinguishable from pure noise, which is consistent with the resolution of 54.5{\AA} that was computed for the reconstruction corresponding to $K=2$.

Another indicator is the spectral gap of the matrix $\tilde S$ of \eqref{eq:sync2}.
As explained in Section~\ref{sec:intro_algo}, this matrix encodes the viewing directions of the projection-images.
Roughly speaking, its spectral gap says how reliably can the viewing directions be estimated in the presence of noise. Mathematically, the spectral gap is defined as the ratio between the third and the fourth eigenvalues of the matrix, where the reliability increases with this ratio, and \ra{a} ratio of~1 means that the estimated rotations are completely arbitrary.
This indicator was equal to 14.6 for $K=50$, 3.7 for $K=3$, and 1.1 for $K=2$, hence it also clearly indicates the failure of the ab-initio reconstruction from class averages corresponding to $K=2$.

\subsection{Running time}\label{sec:running time}

All tests were executed on an Intel Xeon CPU running at 3.60GHz (6 cores in total) with 128GB of RAM running Linux. The algorithm was implemented in Matlab. Whenever possible, all $6$ cores were used simultaneously, either explicitly using Matlab's \texttt{parfor}, or implicitly, by employing Matlab's implementation of BLAS, which takes advantage of multi-core computing. Some loop-intensive parts of the algorithm were implemented in \texttt{C} as Matlab \texttt{mex} files. The algorithm for estimating common lines between pairs of class averages was implemented using Matlab's support for GPU computing, running on a single Nvidia GeForce GTX 1080. The total running time of the algorithm for $N=3000$ projection-images was 165 minutes from class averages to a model \--- including detection of the common lines, estimation of the orientations and reconstruction of the density map. Asymptotically, the running time is $\mathcal{O}(N^3)$. The memory usage increases as $\mathcal{O}(N^2)$ and puts no constraints on the algorithm even for tens of thousands of images (given the machine described above).

\section{Summary and future extensions}
\label{sec:summary}

We introduced an improved algorithm for reference-free ab-initio reconstruction of non-symmetrical structures, which is not biased towards any initial model.
This algorithm is essentially a generalization of the angular reconstitution method, which uses thousands of class averages simultaneously, with possibly a small number of raw-images averaged within each class.
We demonstrated several ab-initio reconstructions, all with resulting resolutions of 20{\AA} or better, even from noisy class averages of as few as 3 images per class.
Such an unbiased density map with intermediate resolution is expected to assist refinement algorithms to converge to the global optimum rather than to a local one.
In addition, we demonstrated how to detect a failure of the proposed algorithm. This capability allows to avoid using wrong initial models in the refinement process or in any other consecutive steps of the reconstruction.

In order to reconstruct a reference-free ab-initio model, the algorithm assigns a reliability-based weight to the estimated relative rotation corresponding to each pair of projection-images, automatically damping the contribution of poor estimates (due to, for example, poor class averages). The assigned weights are based on the probability that an estimated common line between a pair of images was correctly identified. This probability is derived using an errors model for correlation-based common lines detection. The accuracy and robustness of the proposed algorithm were demonstrated using experimental data as described above.

While the algorithm in this paper shows promising results, there are several possible directions for further improvement. To start, the weighting scheme proposed in Section~\protect\ref{sec:weighting_scheme} relies on the observation that a misidentified common line necessarily leads to a wrong estimate of a relative rotation. Thus, the probability that a common line is correct is a proxy for the probability that the corresponding relative rotation is correct. Nevertheless, it may be possible to derive weights that are optimal under some criterion, such as minimizing the mean-squared-error of the estimated relative rotations, resulting in further robustness of the algorithm to noise.

Second, the probability $P_{ij}$ of~\eqref{eq:Pij} is in fact an indicator for the ``quality'' of the common line estimated between the projection-images $P_{R_i}$ and $P_{R_j}$. By aggregating all the probabilities $P_{ij}$ corresponding to a fixed $i$, it may be possible to derive an estimate for the quality of the image $P_{R_i}$. This can be used in a mechanism for discarding class averages of low quality (in the sense that they are not consistent with the other class averages).

Finally, the score~\eqref{eq:score} used to derive the probabilities $P_{ij}$ is by no means the only possible option for modeling common lines' reliabilities. It was used due to its observed behavior described in Section~\ref{sec:triplets_scores} and its direct relation to the common lines. It may be possible derive improved indicators for common lines' reliabilities, or even combine several such indicators.

All the improvements proposed above may lead to an even more robust ab-initio reconstruction algorithm, which will require class averages with only very mild averaging, will be applicable to smaller molecules and noisier data sets, and may even allow to reconstruct ab-initio models directly from raw projection-images.

\section*{Acknowledgments}
This research was supported by THE ISRAEL SCIENCE FOUNDATION grant No. 578/14, by Award Number R01GM090200 from the NIGMS, \ra{and} by the European Research Council (ERC) under the European Union's Horizon 2020 research and innovation programme (grant agreement 723991 - CRYOMATH).

\bibliographystyle{plain}
\bibliography{em}

\begin{thebibliography}{10}

\bibitem{aspire}
{ASPIRE} - {Algorithms for Single Particle Reconstruction} software package.
\newblock \url{http://spr.math.princeton.edu/}.

\bibitem{EMPIAR10028}
{Cryo-EM} structure of the {Plasmodium} falciparum {80S} ribosome bound to the
  anti-protozoan drug emetine.
\newblock http://dx.doi.org/10.6019/EMPIAR-10028.

\bibitem{EMPIAR10073}
{Cryo-EM} structure of the yeast {U4/U6.U5 tri-snRNP} at 3.7 angstrom.
\newblock http://dx.doi.org/10.6019/EMPIAR-10073.

\bibitem{EMDB}
{Protein Data Bank in Europe -- EM} resources.
\newblock \url{https://www.ebi.ac.uk/pdbe/emdb/}.

\bibitem{subtomo2016}
{Bharat T.~A.~M.} and {Scheres S.H.W.}
\newblock Resolving macromolecular structures from electron cryo-tomography
  data using subtomogram averaging in {RELION}.
\newblock {\em Nature Protocols}, 11:2054–--2065, 2016.

\bibitem{Walz2015}
{Cheng~Y.}, {Grigorieff~N.}, {Penczek~P.~A.}, and {Walz~T.}
\newblock A primer to single-particle cryo-electron microscopy.
\newblock {\em Cell}, 161(3):438--449, 2015.

\bibitem{Elmlund2012420}
{Elmlund D.} and {Elmlund H.}
\newblock {SIMPLE:} software for ab initio reconstruction of heterogeneous
  single-particles.
\newblock {\em Journal of Structural Biology}, 180(3):420--427, 2012.

\bibitem{PRIME13}
{Elmlund H.}, {Elmlund D.}, and {Bengio S.}
\newblock {PRIME:} probabilistic initial {3D} model generation for
  single-particle cryo-electron microscopy.
\newblock {\em Cell}, 21(8):1299–--1306, 2013.

\bibitem{Henderson2013b}
{Henderson R.}
\newblock Avoiding the pitfalls of single particle cryo-electron microscopy:
  {Einstein} from noise.
\newblock {\em Proc. Natl Acad. Sci. USA}, 110:18 037--18 041, 2013.

\bibitem{Herman09}
{Herman G.~T.}
\newblock {\em Fundamentals of Computerized Tomography: Image Reconstruction
  from Projections}.
\newblock Springer, London, UK, 2nd edition, 2009.

\bibitem{EMPIAR}
{Iudin~A.}, {Korir~P.~K.}, {Salavert-Torres~J.}, {Kleywegt~G.~J.}, and
  {Patwardhan~A.}
\newblock {EMPIAR:} a public archive for raw electron microscopy image data.
\newblock {\em Nature Methods}, 13(5):387--388, 2016.

\bibitem{LYUMKIS2013417}
{Lyumkis D.}, {Vinterbo S.}, {Potter C. S.}, and {Carragher B.}
\newblock Optimod -- an automated approach for constructing and optimizing
  initial models for single-particle electron microscopy.
\newblock {\em Journal of Structural Biology}, 184(3):417--426, 2013.

\bibitem{Subramaniam2016}
{Merk~A.}, {Bartesaghi~A.}, {Banerjee~S.}, {Falconieri~V.}, {Rao~P.},
  {Davis~M.~I.}, {Pragani~R.}, {Boxer~M.~B.}, {Earl L.~A.}, {Milne J.~L.~S.},
  and {Subramaniam S.}
\newblock Breaking {Cryo-EM} resolution barriers to facilitate drug discovery.
\newblock {\em Cell}, 165(7):1698–--1707, 2016.

\bibitem{Natr2001a}
{Natterer F.}
\newblock {\em The Mathematics of Computerized Tomography}.
\newblock Classics in Applied Mathematics. SIAM: Society for Industrial and
  Applied Mathematics, 2001.

\bibitem{yeast37A}
{Nguyen T~.H.~D.}, {Galej W.~P.}, {Bai X.~C.}, {Oubridge C.}, {Newman A.~J.},
  {Scheres S.~H.~W.}, and {Nagai K.}
\newblock {Cryo-EM} structure of the yeast {U4/U6.U5 tri-snRNP} at 3.7 angstrom
  resolution.
\newblock {\em Nature}, 530 298-302, 2016.

\bibitem{viper}
{Penczek P.~A.} and {Asturias F.~J.}
\newblock Ab initio {cryo-EM} structure determination as a validation problem.
\newblock In {\em 2014 IEEE International Conference on Image Processing
  (ICIP)}, pages 2090--2094, 2014.

\bibitem{chimera}
{Pettersen E.~F.}, {Goddard T.~D.}, {Huang C.~C.}, {Couch G.~S.}, {Greenblatt
  D.~M.}, {Meng E.~C.}, and {Ferrin T.~E.}
\newblock {UCSF Chimera}--a visualization system for exploratory research and
  analysis.
\newblock {\em Journal of Computational Chemistry}, 25(13):1605--1612, 2004.

\bibitem{sync3N}
{Pragier G.}, {Greenberg I.}, {Cheng X.}, and {Shkolnisky Y.}
\newblock A graph partitioning approach to simultaneous angular reconstitution.
\newblock {\em IEEE Transactions on Computational Imaging}, 2016.

\bibitem{cryoSPARC}
{Punjani A.}, {Rubinstein J.~L}, {Fleet D.~J.}, and {Brubaker M.~A}.
\newblock {cryoSPARC:} algorithms for rapid unsupervised cryo-{EM} structure
  determination.
\newblock {\em Nature Methods}, 14:290–296, 2017.

\bibitem{Radermacher1}
{Radermacher~M.}, {Wagenknecht~T.}, {Verschoor~A.}, and {Frank J.}
\newblock Three-dimensional reconstruction from a single-exposure random
  conical tilt series applied to the 50s ribosomal subunit of {Escherichia}
  coli.
\newblock {\em Journal of Microscopy}, 146:113--136, 1987.

\bibitem{Scheres2012}
{Scheres S.~H.~W.}
\newblock Relion: Implementation of a {Bayesian} approach to {cryo-EM}
  structure determination.
\newblock {\em Journal of Structural Biology}, 180:519--530, 2012.

\bibitem{SCHERES2016125}
{Scheres S.~H.~W.}
\newblock Processing of structurally heterogeneous cryo-{EM} data in {RELION}.
\newblock {\em Methods in Enzymology}, 579:125--157, 2016.

\bibitem{sync2N}
{Shkolnisky Y.} and {Singer A.}
\newblock Viewing direction estimation in {Cryo-EM} using synchronization.
\newblock {\em SIAM Journal on Imaging Sciences}, 5(3):1088--1110, 2012.

\bibitem{voting}
{Singer A.}, {Coifman R.~R.}, {Sigworth F.~J.}, {Chester D.~W.}, and
  {Shkolnisky Y.}
\newblock Detecting consistent common lines in {cryo-EM} by voting.
\newblock {\em Journal of Structural Biology}, 169(3):312--322, 2010.

\bibitem{Goncharov}
{Vainshtein B.} and {Goncharov A.}
\newblock Determination of the spatial orientation of arbitrarily arranged
  identical particles of an unknown structure from their projections.
\newblock {\em Proc. llth Intern. Congr. on Elec. Mirco.}, pages 459--460,
  1986.

\bibitem{VanHeel1987}
{van Heel M.}
\newblock Angular reconstitution: a posteriori assignment of projection
  directions for {3D} reconstruction.
\newblock {\em Ultramicroscopy}, 21(2):111--123, 1987.

\bibitem{vanHeel_Schatz}
{van Heel M.} and {Schatz M.}
\newblock Fourier shell correlation threshold criteria.
\newblock {\em J. Struct. Biol.}, 151(3):250--262, 2005.

\bibitem{80S32A}
{Wong W.}, {Bai X.~C.}, {Brown A.}, {Fernandez I.~S.}, {Hanssen E.}, {Condron
  M.}, {Tan Y.~H.}, {Baum J.}, and {Scheres S.~H.~W}.
\newblock {Cryo-EM} structure of the {Plasmodium} falciparum {80S} ribosome
  bound to the anti-protozoan drug emetine.
\newblock {\em eLife}, 10.7554/eLife.03080, 2014.

\bibitem{avgVDM}
{Zhao Z.} and {Singer A.}
\newblock Rotationally invariant image representation for viewing direction
  classification in cryo-{EM}.
\newblock {\em Journal of Structural Biology}, 186(1):153--166, 2014.

\end{thebibliography}

\end{document}